\documentclass[12pt,a4wide]{article}
\usepackage{afterpage}
\usepackage[normalem]{ulem}
\usepackage[a4paper]{geometry}
\usepackage{graphicx}	
\usepackage{verbatim} 
\usepackage{enumitem}
\usepackage[dvipsnames]{xcolor}
\newcommand\ignore[1]{}			
\usepackage{amsmath}
\usepackage{amssymb}
\usepackage{amsthm}
\usepackage{array}
\usepackage[urlcolor=black,colorlinks=true,linkcolor  = black]{hyperref}
\usepackage{mathtools}
\usepackage{float}
\usepackage{fullpage}
\usepackage{braket}
\usepackage{url}

\usepackage{tabularx}

\usepackage[section]{placeins}

\usepackage{cite}
\usepackage{subcaption}
\usepackage{upgreek}
\usepackage{outlines}
\usepackage{braket}
\usepackage{appendix}
\def\0{{(0)}}
\def\1{{(1)}}
\usepackage{xspace}

\usepackage{mathrsfs} 
\usepackage{dsfont}

\DeclareSymbolFontAlphabet{\mathbb}{AMSb}

\newcommand{\id}[0]{\mathds{1}}

\def\d{\partial}
\def\del{\nabla}

\def\a{\alpha}

\usepackage{bm}

\def\k{\kappa}
\usepackage{caption}
\usepackage{subcaption}

\def\vol{\operatorname{Vol}}

\def\calC{\mathcal{C}}

\def\t{\tau}

\def\Om{\Omega}

\def\l{\lambda}

\def\calI{\mathcal{I}}

\def\ccccend{\end{array}\right)}

\def\t{\tau}

\newcommand{\calM}{\mathcal{M}}
\newcommand{\calN}{\mathcal{N}}
\newcommand{\calK}{\mathcal{K}}

\renewcommand\Re{\operatorname{Re}}

\def\de{\delta}
\def\De{\Delta}

\usepackage[bottom]{footmisc}

\theoremstyle{definition}

\usepackage{amsthm}
\theoremstyle{theorem}

\def\calN{\mathcal{N}}
\def\calO{\mathcal{O}}

\def\del{\partial}

\def\bbZ{\mathbb{Z}}
\def\bbQ{\mathbb{Q}}

\def\calV{\mathcal{V}}

\def\Ombar{\bar{\Om}}

\def\calT{\mathcal{T}}
\def\calK{\mathcal{K}}
\def\calV{\mathcal{V}}
\def\calC{{\mathcal{C}_{\mathrm{cf}}}}
\def\vol{\operatorname{Vol}}
\def\Xt{\widetilde{X}}
\def\Dec{\De^\circ}

\def\I{i}
\def\calF{\mathcal{F}}
\def\calA{\mathcal{A}}

\newcommand{\Vcal}{\mathcal{V}}
\newcommand{\Vtcal}{\widetilde{\mathcal{V}}}
\newcommand{\Tcal}{\mathcal{T}}
\newcommand{\Kcal}{\mathcal{K}}

\begin{document}

\pagestyle{plain}

\makeatletter
\@addtoreset{equation}{section}
\makeatother
\renewcommand{\theequation}{\thesection.\arabic{equation}}
\pagestyle{empty}

\begin{flushright}
	\text{CERN-TH-2024-090}
\end{flushright}

\begin{center}
\phantom{a}\\
\vspace{0.8cm}
\scalebox{0.90}[0.90]{{\fontsize{24}{30} \bf{Candidate de Sitter Vacua
}}}\\
\end{center}

\vspace{0.4cm}
\begin{center}
\scalebox{0.95}[0.95]{{\fontsize{15}{30}\selectfont  Liam McAllister,$^{a}$ Jakob Moritz,$^{b}$}}\vspace{0.08cm}
\scalebox{0.95}[0.95]{{\fontsize{14}{30}\selectfont Richard Nally,$^{a}$ and Andreas Schachner$^{a,c}$}}
\end{center}

\begin{center}
\vspace{0.25 cm}

\textsl{$^{a}$Department of Physics, Cornell University, Ithaca, NY 14853, USA}\\
\textsl{$^{b}$Department of Theoretical Physics, CERN, 1211 Meyrin, Switzerland}\\
\textsl{$^{c}$ASC for Theoretical Physics, LMU Munich, 80333 Munich, Germany}\\

	 \vspace{1.1cm}
	\normalsize{\bf Abstract} \\[8mm]
\end{center}

\begin{center}
	\begin{minipage}[h]{15.0cm}
        We  construct  compactifications of type IIB string theory that yield,
        at leading order in the $\alpha^\prime$ and $g_s$ expansions,
        de Sitter vacua of the form envisioned by Kachru, Kallosh, Linde, and Trivedi. 
        We specify explicit Calabi-Yau orientifolds and quantized fluxes for which we derive the four-dimensional effective supergravity theories, incorporating the exact flux superpotential, the nonperturbative superpotential from Euclidean D3-branes, and the K\"ahler potential  at tree level in the string loop expansion but to all orders in $\alpha'$.
        Each example includes a Klebanov-Strassler throat region containing a single anti-D3-brane, whose supersymmetry-breaking energy, computed at leading order in $\alpha'$,  causes an uplift to a metastable de Sitter vacuum in which all moduli are stabilized. 
        Finding vacua that demonstrably survive subleading corrections, and in which the quantization conditions are completely understood, is an important open problem for which this work has prepared the foundations.
 
 	\end{minipage}
\end{center}
\vfill
\today
\newpage

\setcounter{page}{1}
\pagestyle{plain}
\renewcommand{\thefootnote}{\arabic{footnote}}
\setcounter{footnote}{0}
%
%
\setcounter{tocdepth}{2}

\tableofcontents

\newpage

\section{Introduction}\label{sec:intro}

The expansion of the Universe is accelerating. The simplest cosmology that is compatible with this fact is de Sitter space, and so to understand the quantization of gravity in our world, and to grapple with the cosmological constant problem, one should study de Sitter vacua of string theory. In spite of the fundamental importance of this problem, and the thousands of papers \cite{Candelas:1985en} on
compactifications of string theory, de Sitter vacua remain elusive.

In this work, we report progress in constructing de Sitter vacua in Calabi-Yau compactifications of type IIB string theory.  
We closely follow the 
lines of a proposal made more than
twenty years ago by Kachru, Kallosh, Linde, and Trivedi (KKLT) \cite{KKLT},
and for the first time we bring together all the necessary components of the KKLT scenario in explicit compactifications.

We begin by specifying  Calabi-Yau orientifolds and choices of quantized fluxes for which we derive the 
leading-order effective 
theories,
which we will precisely define
below. 
In this setting, we construct 33{,}371
compactifications that each contain
a Klebanov-Strassler throat hosting a single anti-D3-brane.
Among these solutions we find five in which
the supersymmetry-breaking energy of the anti-D3-brane,
computed  at leading order  in $\alpha'$, 
causes an uplift to a metastable de Sitter vacuum in which all moduli are stabilized.

Corrections beyond leading order in the $\alpha'$ and $g_s$ expansions are potentially important, but are not fully known.
Moreover, whether the flux quantization conditions in Calabi-Yau orientifolds permit odd integer flux quanta is unclear. 
Thus, our work does not definitively establish the existence of KKLT de Sitter vacua as solutions of string theory.  Even so, it marks an advance 
in the technology for 
investigating   
de Sitter vacua  
in Calabi-Yau compactifications of type IIB string theory,
and
provides a foundation for future research in this area.

\vspace{10pt} 
The outline of this paper is as follows.  
In the remainder of \S\ref{sec:intro}, we give an  overview  of our approach.  
In \S\ref{sec:EFTs}, we lay out the structure of the effective field theories (EFTs) 
that arise in our compactifications, and we explain the approximations that define the leading-order supersymmetric EFT.
In \S\ref{sec:conisec} we describe a procedure for constructing compactifications that contain Klebanov-Strassler throats.
In \S\ref{sec:methods} we explain how we computed the leading-order EFT in a collection of explicit Calabi-Yau  compactifications, 
and then constructed more than 100 million flux vacua therein.   
We present
examples of candidate de Sitter vacua of string theory  in \S\ref{sec:examples}.  
We comment on the prospects for future work in \S\ref{sec:outlook}, and we conclude in \S\ref{sec:conclusion}.
Further technical material appears in the Appendices: in particular, in Appendix \ref{app:corr} we address corrections to the leading-order EFT.

\subsection*{Background on the KKLT scenario}

The setting for this work is type IIB string theory, which is a supersymmetric theory of quantum gravity in ten dimensions. To obtain a four-dimensional universe, we compactify six of the ten dimensions. We choose the compact space to be a Calabi-Yau threefold, i.e., a compact, Ricci-flat K\"ahler manifold of complex dimension three. This choice yields a four-dimensional $\mathcal{N}=2$ supersymmetric EFT, but at the cost of introducing a large number of massless scalar fields, called \textit{moduli}, which parameterize how the geometry of the compact space can vary across the non-compact spacetime. We then quotient the string spectrum by an orientifold action that projects out some, but not all, of these massless degrees of freedom, and yields a four-dimensional $\mathcal{N}=1$  supergravity theory.

The remaining moduli space of massless scalar degrees of freedom is a tree-level accident: quantum corrections to the effective action are believed --- and in some examples, have been shown  --- to lift all flat directions.  By introducing quantized fluxes for the ten-dimensional gauge fields, one induces a classical potential for some of the moduli, while the remaining moduli can be lifted by genuine quantum corrections. Compactifications of this sort are called flux compactifications (on Calabi-Yau orientifolds).\footnote{For a recent review of flux compactifications with references to the original literature, see \cite{McAllister:2023vgy}.}  
The process of computing the most relevant corrections and identifying weakly-coupled local minima of the resulting scalar potential is called
\textit{moduli stabilization}.

The scalar potential at a local minimum with respect to the scalar fields is the vacuum energy, so moduli stabilization sources a cosmological constant. One can therefore hope to obtain a de Sitter vacuum of string theory by stabilizing the moduli with a potential $V$ that has a local minimum with positive energy. A fundamental difficulty in doing so is that de Sitter vacua are necessarily non-supersymmetric. Thus, we must also break the remaining supersymmetries that are preserved by the fluxes and the orientifold projection. 

In their seminal work \cite{KKLT}, Kachru, Kallosh, Linde, and Trivedi (KKLT) proposed a mechanism for obtaining such a potential, i.e., a way to stabilize all moduli in a configuration with positive vacuum energy.\footnote{For alternative approaches, see e.g. \cite{Saltman:2004sn,Balasubramanian:2005zx,Westphal:2006tn,Louis:2012nb,Cicoli:2013cha,Cicoli:2015ylx,Cordova:2018dbb,Crino:2020qwk,DeLuca:2021pej,Cicoli:2021dhg}.}
 
The KKLT mechanism proceeds in two steps. First, one stabilizes all moduli in a supersymmetric anti-de Sitter (AdS) vacuum with negative vacuum energy $V_{\text{AdS}}$. One then considers configurations that include an anti-D3-brane, which breaks all of the remaining supersymmetry and adds a positive contribution $V_{\overline{D3}}$ to the scalar potential.  
 
\begin{figure}[H]
\begin{center} 
\includegraphics[width=\linewidth]{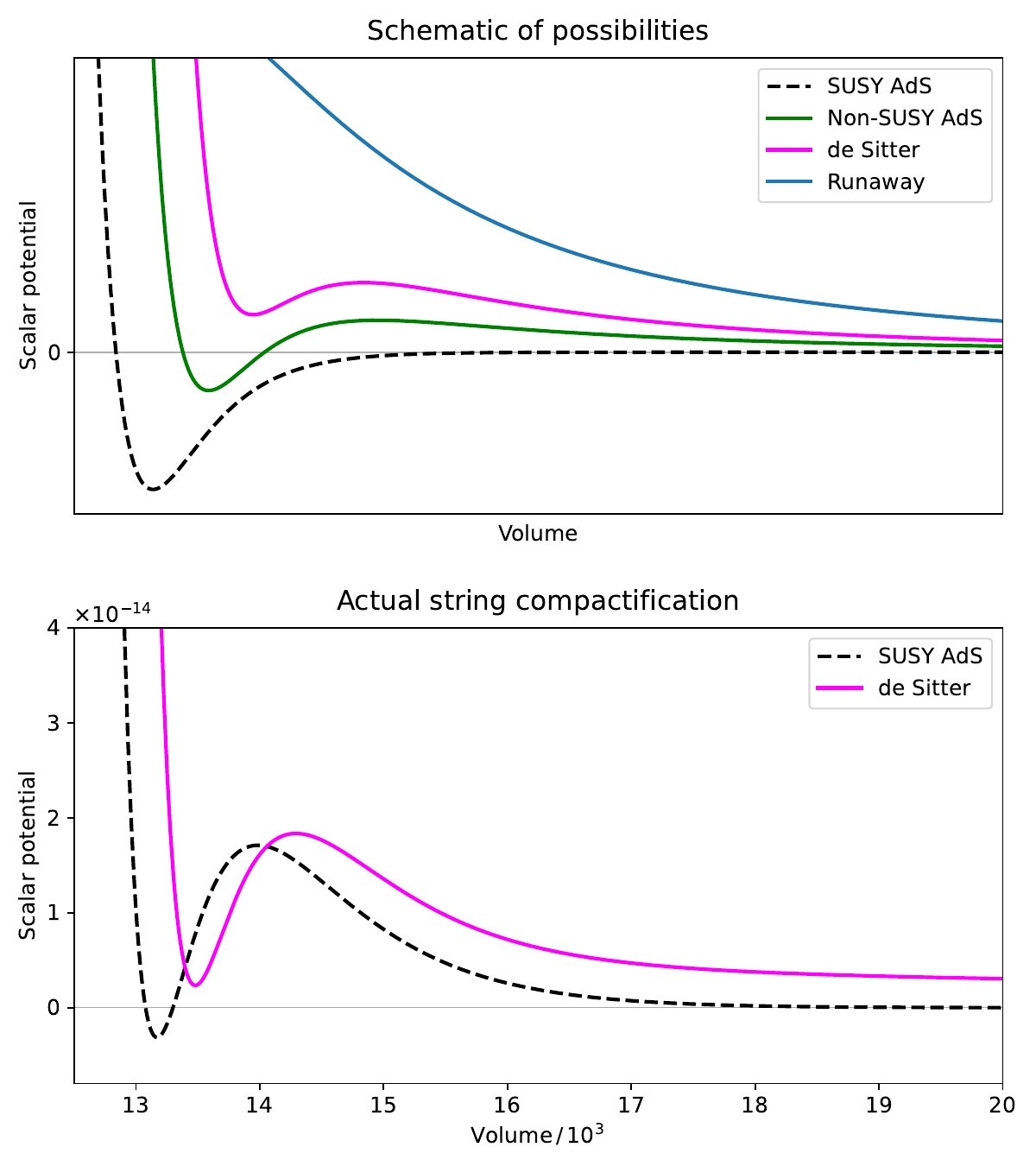}
\caption{\emph{Top:} A sketch of the possible outcomes of anti-D3-brane uplift from a supersymmetric AdS vacuum in the KKLT scenario. \emph{Bottom:} The scalar potentials of the anti-de Sitter vacuum (black) and de Sitter vacuum (pink) in the Calabi-Yau orientifold flux compactification presented in \S\ref{sec:aule}.  
The potentials shown in the lower panel are the result of a complete and explicit computation in the leading-order EFT defined in \S\ref{sec:EFTs} and \S\ref{sec:conisec}. 
The horizontal axis is the Calabi-Yau volume $\mathcal{V}_E$ in Einstein frame and in units of the string length, and the vertical axis is the scalar potential $V$ in Planck units.  
See \eqref{eq:veinsteindef} and \eqref{eq:Vfull}, respectively, for definitions of these quantities. 
The bottom figure is reproduced from Figure \ref{fig:aule_uplift}.
}\label{fig:introExamplePlot}
\end{center}
\end{figure}

Depending on the relative sizes of the AdS vacuum energy $V_{\text{AdS}}$ and the anti-D3-brane contribution $V_{\overline{D3}}$, one of three outcomes can occur: 
\begin{enumerate}
    \item If $V_{\text{AdS}}+V_{\overline{D3}}<0\,,$ 
one obtains a non-supersymmetric AdS vacuum.
    \item If $V_{\overline{D3}} \gg |V_{\text{AdS}}|\,,$
     the volume of the internal space increases without bound in what is called a runaway decompactification.
    \item If $V_{\overline{D3}}$ exceeds $|V_{\text{AdS}}|$ but not by a large factor,
    the existence of a minimum is preserved, and the vacuum energy at the minimum becomes positive: the anti-D3-brane is said to \textit{uplift} the AdS vacuum to a de Sitter vacuum. 
\end{enumerate}
These three possibilities are shown in Figure \ref{fig:introExamplePlot}.

In generic compactifications, the energy of the anti-D3-brane is much larger than the AdS vacuum energy, and one finds a runaway rather than a vacuum. This is because the AdS energy is determined by scales in the EFT, and in particular is below the Kaluza-Klein scale, whereas the energy of the anti-D3-brane is 
proportional to
the much higher string scale. To obtain a de Sitter vacuum, we must therefore arrange for the energy of the anti-D3-brane to be suppressed relative to its na\"ive value. This can be accomplished by embedding into the Calabi-Yau an incarnation of the Randall-Sundrum model \cite{Randall:1999ee}, in which branes at one end of an (approximate) AdS$_5$ throat are exponentially redshifted relative to those at the other end.\footnote{This AdS$_5$ throat is not to be confused with the supersymmetric AdS$_4$ vacuum.} The string theory geometry that reproduces the Randall-Sundrum model in this context is called a Klebanov-Strassler  throat \cite{Klebanov:2000hb}, and arises in Calabi-Yau compactifications that 
contain a type of singular point called a conifold singularity, threaded by three-form fluxes \cite{Giddings:2001yu}.
The potential energy of an anti-D3-brane at the bottom of a Klebanov-Strassler throat was computed by Kachru, Pearson, and Verlinde (KPV) \cite{KPV}, who argued that, in the presence of a sufficiently large amount of flux through the conifold, the anti-D3-brane state at the bottom of the Klebanov-Strassler throat is metastable. Such an anti-D3-brane contributes a positive energy to the scalar potential, and is the origin of the uplifting potential described above.

In summary, to construct a KKLT de Sitter vacuum, one must find a Calabi-Yau orientifold in which one can engineer the following ingredients simultaneously: 
\begin{enumerate}[label=(\alph*)]
    \item\label{it:req1} a supersymmetric AdS vacuum with an exponentially small vacuum energy,
    \item\label{it:req2} a conifold giving rise to a Klebanov-Strassler throat whose redshifted energy scale is commensurate with the AdS vacuum energy, and
    \item\label{it:req3} an anti-D3-brane  whose energy uplifts the AdS vacuum to a de Sitter vacuum.
\end{enumerate}
In this work we fulfill the requirements (a)-(c) in 
explicit flux compactifications on Calabi-Yau orientifolds, at the level of the leading approximation to the EFTs of these compactifications, and so exhibit realizations of KKLT de Sitter vacua at this level.
This is the first time that such vacua have been constructed.

The scalar potential of one example is shown, both before and after uplift, in Figure \ref{fig:introExamplePlot}, and the parameters of the examples are given in Table \ref{tab:summary}.

\subsection*{Scope of this work}
 
The effective theories resulting from string theory are defined by series expansions in two parameters:
the string coupling $g_s$ and the inverse string tension $\a'$. Classic 
references in type IIB flux compactifications such as \cite{KKLT,Klebanov:2000hb,Giddings:2001yu,KPV} worked to leading order in these expansions.
Specifically, the EFT considered by KKLT \cite{KKLT} combines the classical flux superpotential, the nonperturbative superpotential from Euclidean D3-branes, the K\"ahler potential and K\"ahler coordinates at leading order in $g_s$ and $\alpha'$, and the anti-D3-brane potential at leading order in $\alpha'$ \cite{KPV}.

In this work we further incorporate all perturbative and nonperturbative corrections 
in $\alpha'$ to the 
K\"ahler potential and K\"ahler coordinates, at string tree level.  Thus, our treatment of the supersymmetric EFT --- i.e., of the theory before including anti-D3-branes --- is exact in $\alpha'$, but not in $g_s$.

We will see that inclusion of $\alpha'$ corrections is essential.
In the vacua that we find, some curves are small and induce $\alpha'$ corrections to the EFT that cannot be neglected, but that we can consistently incorporate.
In contrast, the string coupling $g_s$ is small enough in our examples so that  truncation to string tree level is arguably consistent.
In both respects, our treatment accords with that of \cite{smallCCs}.
 
The unknown, or incompletely known, effects that we neglect in this work are:
\begin{enumerate}[label=(\roman*)]
    \item $\a'$ corrections to the KPV calculation \cite{KPV} of the anti-D3-brane potential \cite{Junghans:2022exo,Junghans:2022kxg,Hebecker:2022zme,Schreyer:2022len,Schreyer:2024pml}.
    \item string loop corrections to the K\"ahler potential, i.e.~corrections to the K\"ahler potential beyond string tree level (\S\ref{sec:stringloop}).
    \item relevant perturbations of the Klebanov-Strassler solution  that are sourced by 
    quantum effects in the bulk.  The physics of these effects is well-understood \cite{Baumann:2010sx,Gandhi:2011id}, but obtaining precise numerical coefficients of the perturbations, rather than the parametric estimates\footnote{These effects are not formally subleading in the string loop or $\alpha'$ expansion, but they are small when the supersymmetry-breaking energy is small, and so it is consistent to treat them as corrections.}  we give in \S\ref{sec:bulkeffects}, would require data of the Calabi-Yau metric.   
     \item The possibility of restrictive quantization conditions for three-form fluxes.  In toroidal orientifolds, the presence of odd integer flux quanta on certain cycles is consistent only if suitable exotic O3-planes are present \cite{Frey:2002hf}.  Whether corresponding restrictions arise in orientifolds of Calabi-Yau threefolds is not known.  In this work we allow odd and even integer quanta (see \S\ref{sec:even}).   
      \item The normalization of instanton corrections to the nonperturbative superpotential.  We marginalize over this normalization and show that the de Sitter vacua that we find persist over a significant range of values (\S\ref{sec:pfaff}).
\end{enumerate}
We define\footnote{Extensive details of this definition are given in \S\ref{sec:EFTs}, while corrections beyond leading order are studied in Appendix \ref{app:corr}.} the \emph{leading-order EFT} as the EFT obtained in the approximation that the effects (i)-(iv) are negligible, and with a plausible reference value for 
the normalization of the nonperturbative superpotential terms (v).
  
Our principal result is the construction of five totally explicit examples of flux compactifications on Calabi-Yau orientifolds, in each of which the leading-order EFT admits one or more KKLT de Sitter vacua. 
 
We refer to the vacua we find here as \textit{de Sitter vacua at leading order}. It remains unclear if the particular leading-order vacua presented in this work persist as de Sitter solutions to all orders in the expansions that define string theory.  However, even if they do not persist, our solutions serve as templates for a larger-scale search for de Sitter vacua beyond leading order.

\section{Leading-Order Supersymmetric EFT} 
\label{sec:EFTs}

To introduce the leading-order supersymmetric EFT, we begin in \S\ref{sec:known} with an abstract summary of corrections in the $g_s$ and $\alpha'$ expansions.
In \S\ref{sec:flux} we introduce flux vacua and establish  some conventions. 
We then write down the K\"ahler potential and K\"ahler coordinates ---
at string tree level but including corrections to all orders in $\alpha'$ ---
in \S\ref{sec:k},
followed by the superpotential in \S\ref{sec:w}.
We summarize the properties of the leading-order supersymmetric EFT in \S\ref{sec:summarysusyeft}.  A reader who does not wish to be burdened with details can skip to \S\ref{sec:summarysusyeft}.

Unless stated otherwise, we follow the conventions of \cite{Demirtas:2019sip}.  We work in ten-dimensional Einstein frame in units where $\ell_s^2\equiv (2\pi)^2\alpha'=1$, and we set the four-dimensional reduced Planck mass to unity by a Weyl rescaling of the four-dimensional metric.

\subsection{Overview of the EFT}\label{sec:known}
 
We consider type IIB string theory compactified to four dimensions on a Calabi-Yau threefold $X$. The topology of such a manifold is partially characterized\footnote{For recent work on the complete characterization of Calabi-Yau threefold topologies at small $h^{1,1}$, see \cite{Chandra:2023afu,Gendler:2023ujl}.} by two independent Hodge numbers, $h^{1,1}$ and $h^{2,1}$. In the absence of fluxes, compactifying on $X$ yields a four-dimensional effective supergravity theory that preserves $\calN=2$ supersymmetry, and contains the gravity multiplet as well as the following multiplets \cite{Bodner:1989cg,Bohm_2000}:
\begin{enumerate}
    \item $h^{2,1}$ vector multiplets containing massless complex scalars $z^a$ called the \textit{complex structure moduli}. These control the sizes  of odd-dimensional cycles in $X$.
    \item $h^{1,1}$ hypermultiplets containing $4h^{1,1}$ massless real scalars: the $h^{1,1}$ K\"ahler parameters $t^i$ controlling the sizes of even-dimensional cycles of $X$, and the $3h^{1,1}$ Kaluza-Klein zero-modes of the higher-dimensional $p$-form potentials $B_2$, $C_2$, and $C_4$ polarized along the internal directions.
    \item The universal hypermultiplet containing the $C_0$ axion, the axions dual to the two-forms $C_2$ and $B_2$ polarized along the non-compact directions, and the string coupling $g_s$.
\end{enumerate}
Quotienting by an O3/O7 orientifold \cite{Sagnotti:1987tw,Bianchi:1990yu,Dabholkar:1996pc,Sen:1996vd,dabholkar1998lectures,Acharya:2002ag,Brunner:2003zm,Brunner:2004zd} action breaks half the supersymmetries and projects each of these multiplets to an $\mathcal{N}=1$ multiplet, and in particular the light scalar degrees of freedom from the closed string sector reside in chiral multiplets. We will denote these fields as $\Phi^I$.
In all of our examples,  
the chiral multiplets surviving the orientifold projection\footnote{In addition one introduces open string moduli localized on seven-branes. In our compactifications all seven-brane stacks come with gauge algebra $\mathfrak{so}(8)$, and all but one of the stacks yield confining gauge theories without light degrees of freedom.
The normal bundle deformations of the remaining stack are expected to receive a large mass in flux backgrounds, cf.~the analogous discussion in \S3.1 of \cite{smallCCs}, and will not be considered further in this work.} are \cite{Grimm:2004uq}
\begin{enumerate}
    \item $h^{1,1}$ complexified \emph{K\"ahler moduli} $T_i$\,,
    \item $h^{2,1}$ complex structure moduli $z^a$\,, and
    \item the axiodilaton $\tau\coloneqq C_0+\I/g_s$\,.
\end{enumerate}
We will soon make the definition of the K\"ahler moduli more precise ---see \eqref{eq:Ttree} --- but for now it suffices to note that, to leading order, the $\text{Re}(T_i)$ measure Einstein-frame four-cycle volumes, and the $\text{Im}(T_i)$ are the Kaluza-Klein zero modes of the four-form $C_4$.

The moduli experience an F-term potential $V_F$, which is determined by two auxiliary functions: the holomorphic superpotential $W$ and the non-holomorphic K\"ahler potential $\calK$. In terms of $W$ and $\calK$, we define the covariant derivative with respect to the fields $\Phi^I$
\begin{equation}
    D_IW = \d_IW + \d_I\calK \,W\,,\label{eq:DWdef}
\end{equation}
and the K\"ahler metric 
\begin{equation}
    \calK_{I\bar{J}} = \d_I\d_{\bar{J}}\calK\,.
\end{equation} 
We denote the inverse K\"ahler metric, i.e., the matrix inverse of
$\calK_{\bar{I}J}$, as $\calK^{I\bar{J}}$. In terms of these quantities, the F-term potential $V_F$ is given by 
\begin{equation}
    V_F(\Phi,\overline{\Phi}) = e^{\mathcal{K}}\left(\calK^{I\bar{J}}D_IWD_{\bar{J}}\bar{W} - 3\left|W\right|^2\right)\,,\label{eq:VF}
\end{equation}
where the index $I$ runs over the complex structure moduli, the K\"ahler moduli, and the axiodilaton, while the index $\bar{J}$ runs over their complex conjugates. Thus, to compute $V_F$, we need to specify the form of the superpotential and the K\"ahler potential, and evaluate the holomorphic coordinates $\Phi^I$ in terms of data of the compactification.

The superpotential is a holomorphic section of a line bundle over the  moduli space, and thus is highly constrained. At tree level in $\alpha'$, the superpotential is given by the \emph{Gukov-Vafa-Witten flux superpotential} \cite{Gukov:1999ya} --- henceforth $W_{\text{flux}}(z,\tau)$ --- which is independent of the K\"ahler moduli. The axionic shift symmetries of $C_0$ and of the Kaluza-Klein zero-modes of $C_4$ imply that the classical expression is corrected at most nonperturbatively.\footnote{The proof that $W_{\text{flux}}(z,\tau)$ receives no corrections in the string loop expansion is more subtle than in classical nonrenormalization theorems, because the tree-level expression for $W_{\text{flux}}(z,\tau)$ involves $\tau$.  However, the absence of corrections is carefully argued in \cite{Burgess:2005jx}.} 
Therefore, the exact superpotential is given by the sum of two terms,
\begin{equation}
    W(z,\t,T) = W_{\text{flux}}(z,\t) + W_{\text{np}}(z,\t,T)\,.
\end{equation}
The flux superpotential $W_{\text{flux}}$ encodes the contributions of the background fluxes to the scalar potential, and is known exactly \cite{Gukov:1999ya}. The nonperturbative superpotential $W_{\text{np}}$ summarizes the contributions of Euclidean D-branes to the potential. Holomorphy implies that only Euclidean D(-1)-branes and Euclidean D3-branes wrapped on holomorphic four-cycles can contribute \cite{Witten:1996bn}. In addition, strongly coupled gauge dynamics on pure super-Yang-Mills seven-brane gauge groups corrects the infrared superpotential via gaugino condensation; these effects are also controlled by the volumes of four-cycles. In all the configurations studied in this paper, it further turns out that pure Euclidean D(-1) corrections are absent, because all the D7-brane gauge algebras are $\mathfrak{so}(8)$ \cite{Kim:2022jvv}.

We therefore have
\begin{equation}
    \label{eq:wis1} W_{\text{np}} = W_{\l\l} + W_{\text{ED}3}\,,
\end{equation}
where $W_{\l\l}$ represents the effects of gaugino condensation on seven-branes and  $W_{\text{ED}3}$ represents the contribution of Euclidean D3-branes.
In practice we will evaluate \eqref{eq:wis1} by summing the contributions of a finite set of divisors, but the approximation made in doing so is an excellent one: see \S\ref{sec:autoch}.

The K\"ahler potential is a real, non-holomorphic function of all the moduli, and thus its form is much less constrained than that of the superpotential. In particular, unlike the superpotential, the K\"ahler potential receives perturbative and nonperturbative corrections in both the $g_s$ and $\alpha'$ expansions,
\begin{equation}
    \calK =\calK_{\text{tree}} + \de\calK\,.
\end{equation}
Here $\calK_{\text{tree}}$ denotes the K\"ahler potential at leading order in both $g_s$ and $\alpha'$. The corrections $\de\calK$ can be separated into two kinds,
\begin{equation}
    \de\calK = \de\calK_{\calN=2} + \de\calK_{\calN=1}\,,
\end{equation} 
where $\de\calK_{\calN=2}$ contains all corrections in the orientifolded $\calN=1$ theory that are inherited from the $\calN=2$ parent theory (i.e.,~from the theory prior to the orientifold projection), and $\de\calK_{\calN=1}$ contains genuinely $\calN=1$ effects that are not present in the parent theory. Both $\de\calK_{\calN=2}$ and $\de\calK_{\calN=1}$ enjoy double expansions in the string coupling $g_s$ and in $\alpha'$.

The correction $\delta\mathcal{K}_{\mathcal{N}=2}$ is known to all perturbative orders in $g_s$ \cite{Robles-Llana:2006hby,Robles-Llana:2007bbv}, and
we write   
\begin{equation}\label{eq:deltaKN2}
    \delta\mathcal{K}_{\mathcal{N}=2} = \delta\mathcal{K}_{\mathcal{N}=2}^{\text{sphere}}   + \delta\mathcal{K}_{\mathcal{N}=2}^{(g_s)}\,,
\end{equation}
where $\delta\mathcal{K}_{\mathcal{N}=2}^{\text{sphere}}$ is obtained at string tree level, i.e.,~on the sphere, and $\delta\mathcal{K}_{\mathcal{N}=2}^{(g_s)}$ denotes corrections in the string loop expansion that are known exactly, up to the effects of D3-brane and fivebrane instantons \cite{Robles-Llana:2006hby,Robles-Llana:2007bbv}, and that are small enough to be negligible in our examples: see \S\ref{sec:stringloop}. 
Explicitly, we have
\begin{equation}\label{eq:deln2}
    \delta\mathcal{K}_{\mathcal{N}=2}^{\text{sphere}} = \calK_{(\alpha')^3} + \calK_{\text{WSI}}\,,
\end{equation}
where the perturbative correction $\calK_{(\alpha')^3}$ arises at order $(\alpha')^3$ \cite{Grisaru:1986kw,Gross:1986iv,Antoniadis:1997eg,Becker:2002nn}, and the 
worldsheet instanton corrections $\calK_{\text{WSI}}$ \cite{Dine:1986zy,Dine:1987bq} are given in \eqref{eq:Valphap3} and \eqref{eq:VWSI} below.

We can likewise write
\begin{equation}
    \delta\mathcal{K}_{\mathcal{N}=1} = \delta\mathcal{K}_{\mathcal{N}=1}^{\text{sphere}}   + \delta\mathcal{K}_{\mathcal{N}=1}^{(g_s)}\,.
\end{equation}
In flux compactifications the correction term at closed string tree level, $\delta \mathcal{K}_{\mathcal{N}=1}^{\text{sphere}}$, encodes  classical backreaction by (imaginary self-dual) fluxes, and is itself suppressed by a power of $g_s$
in comparison to $\delta\mathcal{K}_{\mathcal{N}=2}^{\text{sphere}}$, as carefully explained in \cite{smallCCs,Cho:2023mhw}.\footnote{In compactifications with seven-branes in general positions, axiodilaton gradients can support further $\mathcal{N}=1$ corrections, as in e.g.~\cite{Minasian:2015bxa}.  In our solutions all seven-branes coincide in $\mathfrak{so}(8)$ stacks, so such effects do not arise at leading order in $g_s$.}  
We therefore omit 
$\delta\mathcal{K}_{\mathcal{N}=1}^{\text{sphere}}$ from our definition of the leading-order EFT, but model its effects in \S\ref{ss:warping}.
 
Rather little is known about $\delta\mathcal{K}_{\mathcal{N}=1}^{(g_s)}$, and
so by necessity we will largely neglect
$\delta\mathcal{K}_{\mathcal{N}=1}^{(g_s)}$ in this work --- though see Appendix \ref{app:corr} for estimates of the sizes of the leading such corrections.
In contrast to the omission of $\delta\mathcal{K}_{\mathcal{N}=2}^{(g_s)}$, which is quite safe \cite{smallCCs}, it is entirely possible that terms in $\delta\mathcal{K}_{\mathcal{N}=1}^{(g_s)}$ spoil the vacuum structure that we report here.  Computing $\delta\mathcal{K}_{\mathcal{N}=1}^{(g_s)}$ is thus an important challenge for future work.

We conclude that at leading order in $g_s$, the K\"ahler potential is known to \emph{all} orders, perturbative and nonperturbative, in $\alpha'$, and we 
\begin{equation}
    \calK_{\text{l.o.}} \coloneqq   \calK_{\text{tree}}  + \delta\mathcal{K}_{\mathcal{N}=2}^{\text{sphere}} = \calK_{\text{tree}} + \calK_{(\alpha')^3} + \calK_{\text{WSI}}\,.
\end{equation}

In summary, we work with the Kähler potential
\begin{subequations}
    \begin{align}
        \calK  &= \calK_{\text{tree}} + \de\calK_{\calN=2} + \de\calK_{\calN=1}\\ 
                &= \calK_{\text{tree}}  + \delta\mathcal{K}_{\mathcal{N}=2}^{\text{sphere}} + \delta \mathcal{K}_{\mathcal{N}=1}^{\text{sphere}} + \de\calK_{\calN=2}^{(g_s)} 
                + \de\calK_{\calN=1}^{(g_s)}\\
                &\approx \calK_{\text{l.o.}} + \de\calK_{\calN=1}^{(g_s)}\label{eq:penultapprox} \\
                &\approx \calK_{\text{l.o.}} \,.\label{eq:lastapprox}
    \end{align}
\end{subequations}
The approximation in \eqref{eq:penultapprox} is fairly robust in our examples: see Table \ref{tab:N2loopSummary}.
Without a direct computation of $\delta\mathcal{K}_{\mathcal{N}=1}^{(g_s)}$,
any assessment of the approximation in \eqref{eq:lastapprox} 
is necessarily incomplete.  In Table \ref{tab:loopSummary} we give na\"ive dimensional analysis estimates of corrections to \eqref{eq:lastapprox}.

The K\"ahler moduli $T_i$ that appear above are defined to be the natural K\"ahler coordinates, i.e., those in which the K\"ahler nature of the metric $\calK_{I\bar{J}}$ is manifest. One can equally well parameterize the K\"ahler moduli space by curve volumes, denoted by $t^i$. At leading order, there is a simple relationship between the $T_i$ and the $t^i$. However, because $\calK$ receives perturbative corrections, so too does $\calK_{I\bar{J}}$, and therefore the relationship between the K\"ahler coordinates $T_i$ and the curve volumes $t^i$ does as well. We therefore have
\begin{equation}\label{eq:firstdefT}
    T_i(t) = T^{\text{tree}}_i + \de T^{\calN=2}_i + \de T^{\calN=1}_i\,,
\end{equation} 
where $T^{\text{tree}}_i(t)$ is defined in \eqref{eq:Ttree}.
Writing
\begin{equation}
    \de T^{\calN=2}_i(t) = \de T^{\calN=2,\text{tree}}_i + \de T^{\calN=2,(g_s)}_i\,,
\end{equation}
the $\mathcal{N}=2$ corrections 
to the K\"ahler coordinates 
at (open string) tree level, $\de T^{\calN=2,\text{tree}}_i$, are known exactly through the c-map \cite{Cecotti:1988qn} (see also e.g.~\cite{Baume:2019sry,Marchesano:2019ifh}), and take the schematic form
\begin{equation}
\label{eq:Kahlercoordinatessplitpreview}
T_i^{\text{tree}}  +\de T^{\calN=2,\text{tree}}_i = T_i^{\text{tree}} + T_i^{(\alpha')^2} + T_i^{\text{WSI}} \,,
\end{equation}
with $T_i^{\text{tree}}$,
$T_i^{(\alpha')^2}$, and $T_i^{\text{WSI}}$ given
in \eqref{eq:Kahlercoordinatessplit}.
As with $\delta\mathcal{K}_{\mathcal{N}=2}^{(g_s)}$,
$\de T^{\calN=2,(g_s)}_i$ can be obtained from \cite{Robles-Llana:2006hby,Robles-Llana:2007bbv}: see \S\ref{sec:stringloop}.  
However, $\delta T^{\calN=1}_i$ is less well-understood (see, however, \cite{Witten:2012bh, Haack:2018ufg,Sen:2024nfd}).

We can thus summarize the superpotential $W$, 
the leading-order K\"ahler 
potential $\calK_{\text{l.o.}}$, and the leading-order K\"ahler coordinates $T^{\text{l.o.}}_i$ 
as follows:
\begin{subequations}
    \begin{align}
        W &= W_{\text{flux}} + W_{\text{ED}3} + W_{\l\l}\,,  \label{eq:wlo} \\
        \calK &\approx \calK_{\text{l.o.}} \coloneqq  \calK_{\text{tree}} + \calK_{(\alpha')^3} + \calK_{\text{WSI}}\,, \label{eq:klo}  \\
        T_i &\approx T^{\text{l.o.}}_i \coloneqq   T^{\text{tree}}_i + \de T^{\calN=2,\text{tree}}_i
        =T_i^{\text{tree}} + T_i^{(\alpha')^2} + T_i^{\text{WSI}}\,. \label{eq:tlo} 
    \end{align}
\end{subequations}
The terms that we retain in  \eqref{eq:klo} and \eqref{eq:tlo} are exact in $\alpha'$, at tree level in $g_s$.\footnote{As remarked in \S\ref{sec:intro}, the disparity in our treatment of $\alpha'$ and $g_s$ corrections is dictated by the vacua that we find: these have $g_s \ll 1$ but include some small curves, so that truncation to string tree level is arguably consistent, while omission of $\alpha'$ corrections is not.}
The approximations made in \eqref{eq:klo} and \eqref{eq:tlo} are further examined in Appendix \ref{app:corr}.

In the presence of an anti-D3-brane, the complete scalar potential is given by 
\begin{equation}
    V = V_F + V_{\overline{D3}}\,,
\end{equation}
where $V_F$ is the $F$-term potential defined in \eqref{eq:VF} and $V_{\overline{D3}}$ is the anti-D3-brane potential.  Once again, $V_{\overline{D3}}$ has both leading and subleading terms, 
\begin{equation}
    V^{\text{up}} = V^{\overline{D3}}_{\text{KPV}} + \Delta V^{\overline{D3}}_{(\alpha')^2} + \ldots\,,
\end{equation} where
$V^{\overline{D3}}_{\text{KPV}}$
is the potential derived by KPV \cite{KPV} and used by KKLT \cite{KKLT}, and certain corrections at order $(\alpha')^2$ have been obtained by Junghans \cite{Junghans:2022exo,Junghans:2022kxg}, and by Hebecker, Schreyer, and Venken \cite{Hebecker:2022zme,Schreyer:2022len,Schreyer:2024pml}.
 
In summary, to specify the leading contributions to the effective potential of a flux compactification, we need to compute $W_\text{flux}$, $W_{\text{ED}3}$, $W_{\l\l}$, $\calK_{\text{tree}}$, 
$\calK_{(\alpha')^3}$, $\calK_{\text{WSI}}$, $T^{\text{tree}}_i$, 
$T_i^{(\alpha')^2}$, $T_i^{\text{WSI}}$, and $V^{\overline{D3}}_{\text{KPV}}$.  In \S\ref{sec:flux}-\S\ref{sec:KS} we will explain 
how these computations were accomplished. 

\subsection{Flux compactifications}\label{sec:flux}

We now turn to a more detailed study of flux compactifications of type IIB string theory on O3/O7 orientifolds of Calabi-Yau threefolds. Taking $(X,\Xt)$ to be a pair of mirror dual Calabi-Yau threefolds, we begin by compactifying on $X$ without any fluxes or orientifolds. As explained in \S\ref{sec:known}, Type IIB string theory compactified on $X$ gives $\mathcal{N}=2$ supergravity in four dimensions coupled to $h^{1,1}(X)+1$ hypermultiplets and $h^{2,1}(X)$ vector multiplets. 

We introduce a symplectic basis of $H_3(X,\mathbb{Z})$, which we denote by $\{\Sigma_{(3)A},\Sigma_{(3)}^A\}$, and corresponding Poincaré dual forms, denoted by $\{\alpha^A,\beta_A\}$. By symplectic, we mean that  
\begin{equation}\label{eq:symplectic}
\int_{X} \alpha^{A}\wedge \beta_B=\delta^A_{~B}\, ,\quad \int_{X}\alpha^{A}\wedge \alpha^B=\int_{X}\beta_{A}\wedge \beta_B=0\, ,\quad A,B=0,\ldots,h^{2,1}(X)\, .
\end{equation}
By integrating the holomorphic three-form $\Omega$ over these cycles, we obtain the \textit{periods}
\begin{equation}
z^A=\int_{\Sigma_{(3)A}}\Omega=\int_{X} \Omega\wedge \alpha^A\, ,\quad \mathcal{F}_A=\int_{\Sigma_{(3)}^A}\Omega=\int_{X} \Omega \wedge \beta_A,
\end{equation}
which we collect in the period vector
\begin{equation}
   \vec{\Pi}=\left ( \begin{array}{c}
        \mathcal{F}_A \\
        z^A
   \end{array}\right ) \, .
\end{equation}
The periods $z^A$ serve as  homogeneous  complex coordinates on a local patch of the complex structure moduli space of $X$. Away from the locus $z^0=0$, we normalize $\Omega$ such that $z^0=1$. 
In doing so, we define local affine coordinates $z^a$, $a=1,\ldots,h^{2,1}(X)$, on the complex structure moduli space; locally, we then have that $\mathcal{F}_A=\mathcal{F}_A(z)$.
Further, the dual periods $\mathcal{F}_A(z)$ are determined by a prepotential $\mathcal{F}(z)$ through
\begin{equation}
    \mathcal{F}_a(z)=\partial_{z^a} \mathcal{F}(z) \, ,\quad \mathcal{F}_0 =2\mathcal{F}-z^a\del_{z^a}\mathcal{F}\, .
\end{equation}
Now, let ${\mathcal{I}}:X\rightarrow X$ be a holomorphic and isometric involution of $X$ under which the holomorphic three-form transforms as $\Omega\mapsto -\Omega$, and denote by $X/\mathcal{I}$ the corresponding O3/O7 orientifold.

Under the orientifold action, the cohomology groups split into even and odd eigenspaces,
\begin{equation}
H^{p,q}(X,\mathbb{Q})=H^{p,q}_+(X,\mathbb{Q})\oplus H^{p,q}_-(X,\mathbb{Q})\, .
\end{equation}
The complex structure moduli surviving this projection come in $\mathcal{N}=1$ chiral multiplets counted by $h^{2,1}_-(X,{\mathcal{I}})= \text{dim}\, H^{2,1}_-(X,\mathbb{Q})$ and will be denoted by $z^a$, $a=1,\ldots,h^{2,1}_-(X,{\mathcal{I}})$.
Similarly, the surviving K\"ahler moduli $T_i$ are counted by $h^{1,1}_+(X,{\mathcal{I}})$.
In this paper we work exclusively with involutions $\mathcal{I}$ such that $h^{1,1}_-(X,{\mathcal{I}})=h^{2,1}_+(X,{\mathcal{I}})=0$, $h^{2,1}_-(X,{\mathcal{I}})=h^{2,1}(X)$,
and $h^{1,1}_+(X,{\mathcal{I}})=h^{1,1}(X)$.  For more general involutions there would be additional axionic chiral multiplets and $U(1)$ vector multiplets counted by $h^{1,1}_-(X,{\mathcal{I}})$ and $h^{2,1}_+(X,{\mathcal{I}})$, respectively.

The fixed locus of ${\mathcal{I}}$ defines the locations of O3-planes and O7-planes. O3-planes fill the four non-compact spacetime dimensions and are localized at points in the orientifold; O7-planes fill the non-compact spacetime and additionally wrap four-cycles inside the compact space.  For the involutions ${\mathcal{I}}$ considered in this work, 
there is a stack of four D7-branes on top of each O7-plane, so that the D7-brane tadpole is locally cancelled. Thus, any four-cycle in $X$ that hosts an O7-plane also hosts an $\mathfrak{so}(8)$ $\mathcal{N}=1$ super Yang-Mills theory. If the cycle is furthermore rigid, then the gauge theory has no matter multiplets and thus confines in the infrared.

The ten-dimensional B-field is forced to take discrete values in $H^2(X,\mathbb{Z}/2)$ modulo integer classes, and by setting
\begin{equation}
    B_2=\frac{1}{2}[\text{O7}]\, ,
\end{equation}
with $[\text{O7}]$ the total class of the O7-planes, one cancels all Freed-Witten anomalies on seven-brane stacks, as in \cite{smallCCs}.

Having introduced an orientifold, we can turn on background fluxes for the ten-dimensional gauge fields.  We preserve four-dimensional Poincar\'{e} invariance by choosing the fluxes to be nonvanishing only along the compact directions. In the symplectic basis introduced above, the fluxes $F_3$ and $H_3$ are characterized by integral flux vectors 
\begin{equation}
    \vec{f}=(f_A,f^A)\, ,\quad\vec{h}=(h_A,h^A) \in H^3(X,\bbZ) \cong \bbZ^{2h^{2,1}(X)+2}\,,
\end{equation}
in terms of the flux quanta
\begin{equation}
f^A=\int_{\Sigma_{(3)A}}F_3\, ,\quad f_A=\int_{\Sigma_{(3)}^A}F_3\, ,\quad h^A=\int_{\Sigma_{(3)A}}H_3\, ,\quad h_A=\int_{\Sigma_{(3)}^A}H_3\, .
\end{equation} 
These are constrained by Gauss's law for the ten-dimensional gauge fields, which requires that the total D3-brane charge in the compact space must vanish. D3-brane charge is sourced by orientifolds, spacetime-filling D3 and D7-branes, and fluxes.  In conventions where each spacetime-filling D3-brane carries a single unit of D3-brane charge and each D3-brane frozen onto the orientifold fixed locus of ${\mathcal{I}}$ carries charge $1/2$, the orientifold and 7-brane contribution to the D3-brane charge is given by $-\frac{1}{4}\chi_f$, where $\chi_f$ is the Euler character of the fixed locus of $\mathcal{I}$ in $X$, and the contribution from the fluxes is given by $\frac{1}{2}\int H_3 \wedge F_3$. 
Thus, defining
\begin{equation}\label{eq:D3-charges}
     Q_{\text{O}} \coloneqq  \frac{1}{2}\chi_f\, ,\quad Q_{\text{flux}} \coloneqq  \int_X H_3 \wedge F_3 = \vec{f}\,^\top \Sigma \vec{h}\,,
\end{equation} 
with
\begin{equation}
    \Sigma \coloneqq  \begin{pmatrix} 0 & \id\\
                -\id & 0
                \end{pmatrix}\,,
\end{equation}
Gauss's law reads 
\begin{equation}
     2\left(N_{\text{D3}}-N_{\overline{\text{D3}}}\right) + Q_{\text{flux}} - Q_{\text{O}} = 0\,,\label{eq:gaussLaw}
\end{equation} 
where $N_{\text{D3}}$ ($N_{\overline{\text{D3}}}$) is the number of spacetime-filling (anti-)D3-branes in the system.

The tadpole constraint \eqref{eq:gaussLaw} must be satisfied in any consistent solution of string theory. Thus, given fluxes such that $Q_{\text{flux}}<Q_{\text{O}}$, spacetime-filling D3-branes must be present. Conversely, if the fluxes are such that $Q_{\text{flux}}>Q_{\text{O}}$, then anti-D3-branes must be present, and supersymmetry is necessarily broken spontaneously.

We will mainly consider flux choices that obey \eqref{eq:gaussLaw} with 
$N_{\overline{\text{D3}}}=1$ and $N_{\text{D3}}=0$: in particular, the examples presented in \S\ref{sec:examples} will all obey
\begin{equation}\label{eq:qqplus2}
 Q_{\text{flux}} = Q_{\text{O}} + 2 \,.
\end{equation}

\subsection{K\"ahler potential and K\"ahler coordinates}\label{sec:k}

We now describe the K\"ahler potential in some detail. 
Recall from \eqref{eq:klo} that we work with the K\"ahler potential at closed string tree level, but incorporate perturbative and nonperturbative $\alpha'$ corrections. 
 
We introduce a basis  $\{\omega^i\}_{i=1}^{h^{1,1}(X)}$
of $H^4(X,\mathbb{Z})$, together with its dual basis $\{\omega_i\}_{i=1}^{h^{1,1}(X)}$ of $H^2(X,\mathbb{Z})$, such that $\int_X\omega^i\wedge  \omega_j={\delta^i}_j$.
In our notation, a $p$-form class and its Poincar\'e-dual cycle class are represented by the same symbol.

The K\"ahler cone of $X$, which we denote by $\mathcal{K}_X\subset H^{1,1}(X,\mathbb{R})$, is parameterized by the K\"ahler parameters $\{t^i\}_{i=1}^{h^{1,1}(X)}$. The string-frame K\"ahler class $J$ of $X$ is then
\begin{equation}
J=\sum_i t^i\,\omega_i\, ,
\end{equation}
and we denote by $\kappa_{ijk}\coloneqq \int_X \omega_i\wedge \omega_j\wedge \omega_k$ the triple intersection form on $X$. The cone dual to $\mathcal{K}_X$ is the Mori cone $\mathcal{M}(X) \subset H_{2}(X,\mathbb{R})$; any $\vec{q}\in\calM(X)$ satisfies $q_it^i>0$ for all $\vec{t}$ in the K\"ahler cone.

In terms of the string-frame K\"ahler parameters $t^i$ and the triple intersection numbers $\k_{ijk}$ of $X$, the string-frame volume 
of $X$, before including any corrections in the $\alpha'$ expansion, is given by 
\begin{equation}
    \calV^{(0)} = \frac{1}{6}\k_{ijk}t_it_jt_k\,,\label{eq:Vst}
\end{equation}
and the Einstein-frame volume is 
\begin{equation}\label{eq:VEtree}
    \calV_E^{(0)} = \frac{\calV^{(0)}}{g_s^{3/2}}\, .
\end{equation}
The K\"ahler potential obtained from classical dimensional reduction on $X$ is given by 
\begin{equation}
    \calK_{\text{tree}} = -2\log\Bigl(2^{3/2}\calV_E^{(0)}\Bigr) -\log\bigl(-\I\left(\t-\bar{\t}\right)\bigr) - \log\left(-i\int_{X}\Om\wedge\Ombar\right)\, ,\label{eq:Ktree}
\end{equation}
where  
\begin{equation}
    \int_{X}\Om\wedge\Ombar = \vec{\Pi}^\dagger\cdot\Sigma\cdot\vec{\Pi}\,.
\end{equation}
At leading order in $\alpha'$ the holomorphic K\"ahler moduli $T_i$ are given by \cite{Grimm:2004uq}
\begin{equation}
    T_{i}^{\text{tree}} = g_s^{-1}\frac{1}{2}\k_{ijk}t^jt^k+i\int_X C_4\wedge \omega_i\, .\label{eq:Ttree}
\end{equation}

The above expressions are corrected order by order in the string loop and $\alpha'$ expansions. We write
\begin{equation}
    e^{-\frac{\calK}{2}}=\sum_{k=0}^{\infty} g_s^{k-2} \Phi^{[k]}+\mathcal{O}\left(e^{-\frac{2\pi}{g_s}}\right)\, ,
\end{equation}
where $k=0$ denotes closed string tree level, $k=1$ denotes open-string tree level, and so on. While not much is known about the $\Phi^{[k>0]}$, the term at closed string tree level, $\Phi^{[0]}$, is inherited from the $\mathcal{N}=2$ parent theory \cite{smallCCs}, and is known exactly due to mirror symmetry and the c-map \cite{Cecotti:1988qn}.

We have   
\begin{equation}
    \Phi^{[0]}=4\left(-i\int_X \Omega\wedge \overline{\Omega}\right)^{\frac{1}{2}}\times \mathcal{V} \, ,
\end{equation}
with the $\alpha'$-corrected, string-frame volume $\mathcal{V}$ given by
\begin{align}\label{eq:Kahlerpotential}
\mathcal{V} &=\calV^{(0)}+\delta\calV_{(\alpha')^3} +\delta\calV_{\text{WSI}}\,,
\end{align}
in terms of the tree level $(\alpha')^3$ correction \cite{Grisaru:1986kw,Gross:1986iv,Antoniadis:1997eg,Becker:2002nn}
\begin{align}\label{eq:Valphap3}
    \delta\calV_{(\alpha')^3} &= -\frac{\zeta(3)\chi(X)}{4(2\pi)^3}\, , 
\end{align}
and worldsheet instanton corrections \cite{Dine:1986zy,Dine:1987bq,Grimm_2007}
\begin{equation}\label{eq:VWSI}
     \delta\calV_{\text{WSI}} = \frac{1}{2(2\pi)^3}
\sum_{\mathbf{q}\in \mathcal{M}(X)}\, \mathscr{N}_{\mathbf{q}}\,\Biggl( \text{Li}_3\Bigl((-1)^{\mathbf{\gamma}\cdot \mathbf{q}}e^{-2\pi \mathbf{q}\cdot \mathbf{t}}\Bigr) + 2\pi \mathbf{q}\cdot \mathbf{t}\,\,\text{Li}_2\Bigl((-1)^{\mathbf{\gamma}\cdot \mathbf{q}}e^{-2\pi \mathbf{q} \cdot \mathbf{t}}\Bigr)\Biggr)\,,
\end{equation}
where $\gamma^i\coloneqq \int_X [\text{O7}]\wedge \omega^i$ is twice the class of the ten-dimensional B-field, and the coefficients $\mathscr{N}_{\mathbf{q}}$ are the genus-zero Gopakumar-Vafa (GV) invariants \cite{Gopakumar:1998ii,Gopakumar:1998jq} of $X$.
Recall that polylogarithms of arbitrary index $k$ are defined for $|z|<1$ as \begin{equation}\label{eq:polylog}
    \text{Li}_k(z)=\sum_{n=1}^\infty \frac{z^n}{n^k}\, ,
\end{equation}
and by analytic continuation to the rest of the complex plane. 
We define the Einstein-frame volume including $\alpha'$ corrections as
\begin{equation}\label{eq:veinsteindef}
    \mathcal{V}_E := \frac{\mathcal{V}}{g_s^{3/2}}\,.
\end{equation}

At tree level in $g_s$ the coordinates $(\tau,z^a)$ remain uncorrected, while 
the holomorphic K\"ahler moduli are corrected as in \eqref{eq:Kahlercoordinatessplitpreview}, 
\begin{equation}\label{eq:firstdefcalT}
    T_i \rightarrow \frac{1}{g_s}\mathcal{T}_i^{\text{l.o.}}+i\int_X C_4\wedge \omega_i\, ,
\end{equation}
with (see e.g.~\cite{Cecotti:1988qn,Grimm:2007xm})
\begin{equation}\label{eq:Kahlercoordinates}
\mathcal{T}_i^{\text{l.o.}}=\frac{1}{2}\kappa_{ijk}t^jt^k-\frac{\chi(D_i)}{24}+\frac{1}{(2\pi)^2}\sum_{\mathbf{q}\in \mathcal{M}(X)}q_i\, \mathscr{N}_{\mathbf{q}} \,\text{Li}_2\Bigl((-1)^{\mathbf{\gamma}\cdot \mathbf{q}}e^{-2\pi \mathbf{q}\cdot \mathbf{t}}\Bigr)\,, 
\end{equation}
where we decompose $\mathcal{T}_i^{\text{l.o.}}$ as
\begin{equation}\label{eq:Kahlercoordinatessplit}
\mathcal{T}_i^{\text{l.o.}}= \mathcal{T}_i^{\text{tree}} + \mathcal{T}_i^{(\alpha')^2} +  \mathcal{T}_i^{\text{WSI}}\,.
\end{equation}

Corrections beyond tree level in $g_s$, involving genuinely $\mathcal{N}=1$ corrections to the K\"ahler potential and K\"ahler coordinates, are discussed in Appendix \ref{app:corr}.

\subsection{Superpotential}\label{sec:w}

Let us now describe the structure of the superpotential that arises in our constructions. 
As explained above, we consider the contributions of fluxes, Euclidean D3-branes, and gaugino condensates on $\mathfrak{so}(8)$ stacks of seven-branes, and so arrive at 
\cite{Gukov:1999ya,Giddings:2001yu}
\begin{equation}
    W(z,\tau,T) = W_{\text{flux}}(z,\t) + W_{\text{np}}\left(z,\tau,T\right)\,.
\end{equation}
We now examine each term in detail.

\subsubsection{The flux superpotential}

The classical flux superpotential is given by \cite{Gukov:1999ya,Giddings:2001yu},
\begin{align}\label{eq:flux_superpotential}
     W_{\text{flux}}(\tau,z^a)=\sqrt{\tfrac{2}{\pi}}\int_X (F_3-\tau H_3)\wedge \Omega(z)=\sqrt{\tfrac{2}{\pi}}\,\vec{\Pi}^\top \,{\cdot}\,\Sigma\,{\cdot}\, (\vec{f}-\tau \vec{h})\, .
\end{align}
Given a choice of a Calabi-Yau orientifold $X/\calI$ and flux vectors $\vec{f}$, $\vec{h}\in H^3(X,\bbZ)$, to compute the flux superpotential one must compute the period vector $\vec{\Pi}$ as a function of the complex structure moduli.  

To achieve this,
we begin by working in the large complex structure (LCS) patch; we defer to \S\ref{sec:coniPFV} the discussion of how to extend this analysis to the vicinity of a conifold singularity. At LCS, the prepotential can be expanded in terms of the topological data of the mirror Calabi-Yau $\Xt$ as the sum of a polynomial piece and an exponential piece \cite{Hosono:1994av}:
\begin{align}\label{eq:prepotential}
\mathcal{F}(z)=\mathcal{F}_{\text{poly}}(z)+\mathcal{F}_{\text{inst}}(z)\,,
\end{align}
where
\begin{align}
    \label{eq:Fpoly}\mathcal{F}_{\text{poly}}(z)&=-\frac{1}{3!}\widetilde{\kappa}_{abc}z^az^bz^c+\frac{1}{4}\mathbb{A}_{ab}z^az^b+\frac{1}{24}\tilde{c}_a z^a+\frac{\zeta(3)\chi(\widetilde{X})}{2(2\pi \I)^3}\,
    ,\\[0.5em]
    \label{eq:Finst}\mathcal{F}_{\text{inst}}(z)&=-\frac{1}{(2\pi \I)^3}\sum_{\tilde{\mathbf{q}}\in \mathcal{M}(\widetilde{X})}\mathscr{N}_{\tilde{\mathbf{q}}}\,\text{Li}_3\Bigl(e^{2\pi \I\,\tilde{\mathbf{q}}\cdot \mathbf{z}}\Bigr)\, .
\end{align}
In \eqref{eq:Fpoly}, $\widetilde{\kappa}_{abc}$ denotes the triple intersection numbers of $\widetilde{X}$, while
\begin{equation}
\tilde{c}_a=\int_{\widetilde{X}}c_2(\widetilde{X})\wedge \tilde{\beta}_a\, ,\quad
\mathbb{A}_{ab}\equiv \begin{cases}
\widetilde{\kappa}_{aab} & a\geq b\\
\widetilde{\kappa}_{abb} & a<b
\end{cases}\, , \quad \text{and} \quad \chi(\widetilde{X})=\int_{\widetilde{X}} c_3(\widetilde{X})\, . \label{eq:amatrix}
\end{equation}
We have introduced a basis $\{\tilde{\beta}_a\}_{a=1}^{h^{2,1}(X)}$ of $H^2(\widetilde{X},\mathbb{Z})$ that is mirror dual to the set of three-forms $\beta_a\in H^3(X,\mathbb{Z})$, cf.~\eqref{eq:symplectic}, and $c_2(\widetilde{X})$ and $c_3(\widetilde{X})$ denote the second and third Chern classes of $\widetilde{X}$.

The instantonic piece $\calF_{\text{inst}}$ is generated by type IIA worldsheet instantons, and takes the form of a sum over effective curve classes $\tilde{\mathbf{q}}$ in $H^4(\widetilde{X},\mathbb{Z})\simeq H_2(\widetilde{X},\mathbb{Z})$; here we work in a  basis $\left\{\tilde{\alpha}^a\right\}_{a=1}^{h^{2,1}(X)}$ that is mirror dual to the set of three-forms $\alpha^a\in H^3(X,\mathbb{Z})$.  
The coefficients $\mathscr{N}_{\tilde{\mathbf{q}}}$ are the genus-zero Gopakumar-Vafa (GV) invariants \cite{Gopakumar:1998ii,Gopakumar:1998jq} of $\widetilde{X}$.

With these definitions, it is straightforward to compute that  
\begin{align}
\mathcal{F}_a(z)&=\del_a\mathcal{F}_{\text{poly}}-\frac{1}{(2\pi \I)^2}\sum_{\tilde{\mathbf{q}}\in \mathcal{M}(\widetilde{X})}\mathscr{N}_{\tilde{\mathbf{q}}}\, \, \tilde{q}_a\, 
\,\text{Li}_2\Bigl(e^{2\pi \I\,\tilde{\mathbf{q}}\cdot \mathbf{z}}\Bigr)\,,\\
\calF_0(z) &= 2\mathcal{F}-z^a\mathcal{F}_a\,.
\end{align}
We have thus specified the entire period vector at LCS, from which the definition in \eqref{eq:flux_superpotential} suffices to compute the flux superpotential. Because $\calF$ enjoys a series expansion, so too does $W_{\text{flux}}$. We therefore write 
\begin{equation}\label{eq:wpolywinst}
W_{\text{flux}}(z,\t) = W_{\text{poly}}(z,\t) + W_{\text{inst}}(z,\t)\,,    
\end{equation} where $W_{\text{poly}}$ and $W_{\text{inst}}$ are the contributions to the flux superpotential of $\calF_{\text{poly}}$ and $\calF_{\text{inst}},$ respectively.

\subsubsection{The nonperturbative superpotential}\label{sec:wnp}

Having laid out the form of $W_\text{flux}$, we proceed to define the nonperturbative superpotential. As explained \S\ref{sec:known}, we only need to consider the contributions of Euclidean D3-branes and gaugino condensates.  
 
A Euclidean D3-brane wrapped on a four-cycle $D\in H_4(X,\bbZ)$ contributes to the superpotential as
\begin{equation}
    W_{\text{np}}\supset \calA_{D}(z,\t)\,e^{-2\pi T_D}\, ,
\end{equation}
where at leading order $T_D$ is the Einstein-frame volume of the divisor $D$, which can be written as a linear combination of the K\"ahler moduli $T_i$. 
The Pfaffian $\calA_D$ can in general depend on the complex structure moduli, the axiodilaton, and the positions of D3-branes and D7-branes. 
For divisors $D$ that support more than two fermion zero modes, $\calA_D$ vanishes identically.
 
We call a divisor $D$ \textit{rigid} if 
\begin{equation}\label{eq:drigid}
    h^{\bullet}_+(D,\mathcal{O}_D)=(1,0,0)\,, \qquad h^{\bullet}_-(D,\mathcal{O}_D)=0\,.
\end{equation}
Witten showed that if a divisor $D$ is rigid and smooth,\footnote{The smoothness condition is not always necessary: see e.g.~\cite{Gendler:2022qof} for conditions under which singular divisors can contribute to the superpotential.} then $D$ hosts exactly two zero modes, and so Euclidean D3-branes wrapped on $D$ contribute to the superpotential with a Pfaffian $\calA_D$ that does not vanish identically \cite{Witten:1996bn}.\footnote{See \cite{Blumenhagen:2009qh} for a review on D-instanton calculus.}

To avoid complications from the moduli dependence of the unknown Pfaffian prefactors, it will be useful to impose a stronger condition than rigidity. Type IIB flux compactifications of the kind considered here are dual to F-theory compactified on elliptically-fibered Calabi-Yau fourfolds. Let $\widehat{X}$ be the F-theory geometry dual to type IIB on $X/\calI$. A Euclidean D3-brane wrapping a cycle $D \subset X$ can be realized in F-theory as a Euclidean M5-brane wrapping a vertical divisor $\widehat{D}\subset\widehat{X}$. To understand the Pfaffian $\calA_D$, one can compute the intermediate Jacobian of $\widehat{D}$: in particular, $\calA_D$ is a section of a line bundle over the complex structure moduli space of this auxiliary variety \cite{Witten:1996hc}. We call a rigid divisor $D$ \textit{pure rigid} if its F-theory uplift $\widehat{D}$ satisfies \cite{smallCCs} 
\begin{equation}\label{eq:dpurerigid}
    h^{2,1}\big(\widehat{D}\big)=0\,.
\end{equation}
The F-theory uplifts of pure rigid divisors have trivial intermediate Jacobian, and so their Pfaffian prefactors are simply numbers.\footnote{This holds up to corrections from warping effects sourced by D3-brane charge \cite{Ganor_1997,Berg:2004ek,Giddings:2005ff,Shiu:2008ry,Baumann:2006th,Frey:2013bha,Martucci:2016pzt}, which can be thought of as part of $\delta T_i^{\mathcal{N}=2}$ in \eqref{eq:firstdefT}.} Thus, a smooth and pure rigid divisor $D$ contributes to the superpotential as 
\begin{equation}
    W_{\text{np}} \supset \calA_D\,e^{-2\pi T_D}\, ,
\end{equation}
where $\calA_D$ is a nonzero constant that we will refer to as a \emph{Pfaffian number}. 
 
We now proceed to the contributions of gaugino condensates.
The orientifold involutions in each of our compactifications, defined by \eqref{eq:odef} below, are such that each O7-plane coincides with four D7-branes, leading to an
$\mathfrak{so}(8)$ gauge algebra on each such stack.
Thus, the contributions of gaugino condensation are indexed by the divisors $D$ hosting O7-planes. 
Gaugino condensation on such a divisor contributes to the nonperturbative superpotential as
\begin{equation}
    W_{\text{np}}(z,\t,T) \supset \calA_D(z,\t)\,e^{-2\pi T_D/6}\, ,
\end{equation}
where the numerical factor in the exponent occurs because the dual Coxeter number of $\mathfrak{so}(8)$ is six.  As before, a smooth and pure rigid $D$ has $\calA_D$ constant and nonvanishing. 

A divisor $D$ contributes to the superpotential if it is smooth and rigid. However, in general, other divisors $D$ can contribute as well. This is especially true in the presence of flux, which can ``rigidify" nonrigid divisors by freezing out certain deformations \cite{Bianchi:2011qh,Bianchi:2012pn,Bianchi:2012kt}. We can therefore write the general expression
\begin{equation}
    W_{\text{np}}(z,\t,T) = 
     \sum_{D,\,\text{pure rigid}} \calA_D \, e^{-\frac{2\pi}{c_D}T_D}
     +  \sum_{D,\,\text{rigid}} \calA_D(z,\t)\, e^{-\frac{2\pi}{c_D}T_D}
     +  \sum_{D,\,\text{rigidified}} \calA_D(z,\t)\, e^{-\frac{2\pi}{c_D}T_D}\,,
    \label{eq:Wnpgeneral}
\end{equation} 
where $c_D=6$ if $D$ hosts an O7-plane and $c_D=1$ otherwise. 

If $D$ is a rigid divisor, then we do not expect contributions to the superpotential from Euclidean D3-branes wrapping $nD$ with $2\le n \in \mathbb{Z}$:
the unique representative of the class $[nD]$ has too many zero modes to contribute \cite{Blumenhagen:2009qh}.\footnote{If $D$ is instead ample, then $nD$ may have a smooth connected representative whose moduli could be lifted by fluxes.  However, all ample divisors are very large in our examples, so this effect can be neglected.}
Moreover, if $D_1$ and $D_2$ are rigid, then the unique representative of the class $[D_1]+[D_2]$ is simply the union of the two rigid divisors, intersecting each other normally. This configuration has the zero modes of $D_1$ and those of $D_2$, plus potentially additional zero modes localized along the intersection, and so a contribution to $W$ from $[D_1]+[D_2]$ is likewise forbidden.\footnote{For a general divisor $D$, one should compute $h^\bullet(\overline{D},\mathcal{O}_{\overline{D}})$, with $\overline{D}$ the \emph{normalization} of $D$ \cite{Gendler:2022qof}. This prescription matches the simplified picture stated above.} 
Thus, we can restrict attention to a finite set of rigid divisors, omitting any sums thereof.  It will suffice to consider prime toric divisors, i.e.,~the intersections with the Calabi-Yau $X$ of irreducible toric divisors of the ambient toric variety,\footnote{Background on toric varieties is provided in \S\ref{sec:polytopeselection}.}
as any effective divisor can be expressed as a sum of prime toric divisors with (not necessarily positive) integer coefficients.

In  this work we will only incorporate the contributions of \emph{pure rigid prime toric divisors}, omitting the sums in \eqref{eq:Wnpgeneral} that range over rigid (but not pure rigid) and rigidified divisors, and likewise omitting divisors that are pure rigid but are not prime toric. We will validate this restriction \emph{a posteriori} in each example by demonstrating that the pure rigid prime toric divisors have significantly smaller  volumes than all other effective divisors that we can identify, and are therefore dominant in the superpotential, regardless of whether or not fluxes rigidify other divisors: see \S\ref{sec:autoch}.\footnote{For the computation of pure rigidity, we use a conjecture formulated in \cite{Jefferson:2022ssj} for correcting the Euler characteristic of a divisor by contributions from O3-planes. We tested our computation on the examples of \cite{smallCCs,Jefferson:2022ssj}, finding exact agreement.  Specifically, the computation arrives at the same results for pure rigidity, in all the examples of \cite{smallCCs,Jefferson:2022ssj}, as those found by hand by M.~Kim in a direct computation that does not rely on the conjecture of \cite{Jefferson:2022ssj}.}

To stabilize $h^{1,1}(X)$ K\"ahler moduli $T_i$ by the mechanism proposed by \cite{KKLT}, one needs at least $h^{1,1}(X)$ terms in the nonperturbative superpotential.\footnote{More precisely, the divisor classes corresponding to the nonperturbative superpotential terms need to span all of $H^2(X,\mathbb{R})$.} We therefore restrict ourselves to Calabi-Yau threefolds $X$ that have at least $h^{1,1}(X)$ rigid prime toric divisors. We do not impose at the outset the stronger requirement that all rigid prime toric divisors that make non-negligible contributions are in fact pure rigid, but this nonetheless turns out to be true in all of our examples: see \S\ref{sec:autoch} and \S\ref{sec:examples}.

The final form of the nonperturbative superpotential that we will work with is 
\begin{equation}
    W_{\text{np}} = \sum_D \calA_D\, e^{-\frac{2\pi}{c_D}T_D}\,,
    \label{eq:Wnppure}
\end{equation}
where the sum on $D$ ranges only over pure rigid prime toric divisors. The full superpotential is therefore 
\begin{equation}
    W(z,\t,T) = \sqrt{\frac{2}{\pi}}\, \, \vec{\Pi}^\top \,{\cdot}\,\Sigma \,{\cdot}\,\big(\vec{f}-\t\vec{h}\big) + \sum_D \calA_D\, e^{-\frac{2\pi}{c_D}T_D}\,.\label{eq:superpotential}
\end{equation}
Let us briefly recapitulate how the quantities appearing in \eqref{eq:superpotential} are obtained.
The choice of a Calabi-Yau threefold and of an orientifold projection dictates which divisors $D$ contribute to the sum in \eqref{eq:superpotential}, either with $c_D=1$ or with $c_D=6$, and similarly determines how the complexified volumes $T_D$ can be expanded as linear combinations of the K\"ahler moduli $T_i$.  The fluxes $\vec{f}$, $\vec{h}$ are vectors of integers that can be chosen to specify a flux compactification.
The dependence of \eqref{eq:superpotential} on the complex structure moduli $z_a$ is encoded in the periods $\vec{\Pi}$; the dependence on $\tau$ is shown explicitly; and the dependence on the K\"ahler moduli $T_i$ occurs via $T_D$.

The only quantities that remain to be specified in \eqref{eq:superpotential} are the values of the Pfaffian numbers $\calA_D$. These are presently unknown, but have been the subject of significant recent work \cite{Kim:2022jvv,Alexandrov:2022mmy,Kim:2022uni,Kim:2023cbh}.
We adopt the form
\begin{equation}
    \calA_D = \sqrt{\frac{2}{\pi}}\frac{n_D}{4\pi^2}\,,\label{eq:K0}
\end{equation}
where the constant $n_D$ is related to an integral over worldsheet modes \cite{Alexandrov:2022mmy}.  
No complete computation of $n_D$ has yet appeared in the literature, but based on the results of \cite{Kim:2023cbh} it is reasonable to expect that $n_D$ is an order-one number.\footnote{Indeed, one might expect that $n_D\in \mathbb{Z}$ for the following reason: the O3/O7 orientifolds considered here are mirror dual to O6 orientifolds in type IIA string theory, which at strong coupling become M-theory compactifications on $G_2$ manifolds. In this context, both the exponential corrections in $W_{\text{flux}}$ and Euclidean D3-brane contributions to $W_{\text{np}}$ are reinterpreted as resulting from Euclidean M2-branes wrapping three-cycles. The common origin of these corrections to the superpotential, and the fact that the exponential corrections to $W_{\text{flux}}$ have integer coefficients in our above normalization --- cf.~\eqref{eq:Widef} --- suggests that $n_D$ is integral as well, and 
supports our choice of numerical factors in \eqref{eq:K0}.}  
For definiteness, in \S\ref{sec:methods} and \S\ref{sec:examples} we take 
\begin{equation}\label{eq:pfaffvalue}
    n_{D} =    1 \implies \calA_D = \sqrt{\frac{2}{\pi}}\frac{1}{4\pi^2}\,,
\end{equation}
for all divisors $D$. 
   
In \S\ref{sec:pfaff} we show  
that each of the de Sitter vacua exhibited in  \S\ref{sec:examples} persists throughout the wide range
\begin{equation}\label{eq:ndrange}
10^{-3} \le n_D \le 10^4\,.
\end{equation} 
Persistence outside the range \eqref{eq:ndrange} is also common, but varies depending on the example, and so we do not characterize a wider range here. 

In view of \eqref{eq:ndrange},
the fact that the values of the Pfaffian numbers $\calA_D$ are unknown does not appear to be a significant weakness of our constructions.  Even so, it would be   worthwhile to compute the $\calA_D$ directly.

\subsection{Summary of the supersymmetric EFT}\label{sec:summarysusyeft}

Collecting the above results, the data of the leading-order supersymmetric EFT is as follows.
Adopting the normalization \eqref{eq:pfaffvalue}, the superpotential is 
\begin{align}
W &= W_{\text{flux}} + W_{\text{ED}3} + W_{\l\l}\,  \label{eq:wlo2} \\
&=\sqrt{\frac{2}{\pi}} \, \, \vec{\Pi}^\top \,{\cdot}\,\Sigma \,{\cdot}\,\big(\vec{f}-\t\vec{h}\big) + \sqrt{\frac{2}{\pi}}\frac{1}{4\pi^2}\sum_D\, e^{-\frac{2\pi}{c_D}T_D}\,,
\label{eq:wlo3}    
\end{align}
where the sum runs over pure rigid prime toric divisors.

The K\"ahler potential obtained in \S\ref{sec:k} is
\begin{align}
\calK \approx \calK_{\text{l.o.}} \coloneqq  &\,\,\calK_{\text{tree}} + \calK_{(\alpha')^3} + \calK_{\text{WSI}}\, \\
= & -2\log\Bigl(2^{3/2}g_s^{-3/2}\calV \Bigr) -\log\bigl(-\I\left(\t-\bar{\t}\right)\bigr) - \log\Bigl(-\I\int_{X}\Om\wedge\Ombar\Bigr)\,, \label{eq:detailedform2}
\end{align}
in terms of the $\alpha'$-corrected, string tree level, string-frame volume $\calV$ given in \eqref{eq:Kahlerpotential},
\begin{align}
\mathcal{V} & =  \frac{1}{6}\kappa_{ijk}t^it^jt^k-\frac{\zeta(3)\chi(X)}{4(2\pi)^3}\nonumber\\
&+\frac{1}{2(2\pi)^3}
\sum_{\mathbf{q}\in \mathcal{M}(X)}\, \mathscr{N}_{\mathbf{q}}\,\Biggl( \text{Li}_3\Bigl((-1)^{\mathbf{\gamma}\cdot \mathbf{q}}e^{-2\pi \mathbf{q}\cdot \mathbf{t}}\Bigr)+ 2\pi \mathbf{q}\cdot \mathbf{t}\,\,\text{Li}_2\Bigl((-1)^{\mathbf{\gamma}\cdot \mathbf{q}}e^{-2\pi \mathbf{q} \cdot \mathbf{t}}\Bigr)\Biggr)\,. \label{eq:detailedform3}
\end{align}
The K\"ahler coordinates are
\begin{align}
T_i \approx T^{\text{l.o.}}_i
\coloneqq \frac{1}{g_s}\Bigl(\mathcal{T}^{\text{tree}}_i+ 
\mathcal{T}_i^{(\alpha')^2}    
+
\mathcal{T}_i^{\text{WSI}}\Bigr) +i\int_X C_4\wedge \omega_i\,,
\end{align}
with 
\begin{align}
\mathcal{T}^{\text{tree}}_i+ 
\mathcal{T}_i^{(\alpha')^2}   
+
\mathcal{T}_i^{\text{WSI}} 
&=\frac{1}{2}\kappa_{ijk}t^jt^k-\frac{\chi(D_i)}{24}\nonumber\\
&\hphantom{=}+\frac{1}{(2\pi)^2}\sum_{\mathbf{q}\in \mathcal{M}(X)}q_i\, \mathscr{N}_{\mathbf{q}} \,\text{Li}_2\Bigl((-1)^{\mathbf{\gamma}\cdot \mathbf{q}}e^{-2\pi \mathbf{q}\cdot \mathbf{t}}\Bigr)\,. \label{eq:detailedform4}
\end{align}

In total, the superpotential $W$ \eqref{eq:wlo3}, the K\"ahler potential $\mathcal{K}_{\text{l.o.}}$ \eqref{eq:detailedform2}, 
\eqref{eq:detailedform3}, and the K\"ahler coordinates 
$\tau$, $z^a$, and $T^{\text{l.o.}}_i$ \eqref{eq:detailedform4} specify an $\mathcal{N}=1$ supersymmetric supergravity theory, 
which we call the \emph{leading-order supersymmetric EFT}.
The F-term potential of this theory is
\begin{equation}\label{eq:vfsum}
    V_F = V_F(W;\mathcal{K}_{\text{l.o}};\tau, z^a, T^{\text{l.o.}}_i)\,.
\end{equation}

Let us briefly recapitulate the approximations that have been made 
in reaching \eqref{eq:vfsum}.  
We are aware of no significant corrections to the form \eqref{eq:wlo3} for the superpotential, except for deviations from the 
reference normalization \eqref{eq:pfaffvalue}; we will marginalize over this unknown in our solutions.
The expressions
\eqref{eq:detailedform2}, \eqref{eq:detailedform3}, \eqref{eq:detailedform4}
for $\mathcal{K}_{\text{l.o}}$ and $T^{\text{l.o.}}_i$, respectively, 
incorporate corrections at string tree level, to all orders --- perturbative and nonperturbative --- in $\alpha'$.  
Perturbative corrections in the string loop expansion have not been included in
\eqref{eq:detailedform2}, \eqref{eq:detailedform3}, \eqref{eq:detailedform4}, and are discussed in Appendix \ref{app:corr}.

The potential $V_F$ in \eqref{eq:vfsum} depends on all the moduli, and can be computed \textit{explicitly} in any example, in terms of the topological data of a Calabi-Yau $X$ and its mirror $\widetilde{X}$, an orientifold involution  $\mathcal{I}$ of $X$, and a choice of flux quanta.  
By computing $V_F$ and finding supersymmetric minima, one can thus construct
supersymmetric AdS$_4$ vacua of type IIB string theory \cite{smallCCs}. 
The cosmological constant of such a solution
is governed by the value of the superpotential at the minimum, and will be small in magnitude for
choices of flux such that $W_{\text{flux}}$ has an extremely small vev, 
\begin{equation} \label{eq:W0def}
\langle |W_{\text{flux}}| \rangle \equiv W_0 \ll 1\,.
\end{equation}
It was shown in \cite{Demirtas:2019sip,smallCCs} that, by choosing fluxes such that the term $W_{\text{poly}}$ appearing in \eqref{eq:wpolywinst} vanishes identically, one can generate exponentially small flux superpotentials.\footnote{The flux superpotential vevs engineered in \cite{Demirtas:2019sip,smallCCs} are much smaller than counting arguments suggest is generically possible \cite{Denef:2004ze}: the integer nature of the flux vectors is essential to ensuring the smallness of the superpotential in \cite{Demirtas:2019sip,smallCCs}, and so these configurations are invisible in the continuous flux approximation made in \cite{Denef:2004ze}.}

Our goal is instead to find non-supersymmetric minima in which anti-D3-branes in Klebanov-Strassler throats break supersymmetry.  We now turn to the task of engineering such configurations.

\section{Conifolds and Supersymmetry Breaking}\label{sec:conisec}

In this section  
we explain how 
to construct supersymmetric AdS vacua with Klebanov-Strassler throats, and to break supersymmetry via anti-D3-branes therein.\footnote{Supersymmetric AdS vacua without Klebanov-Strassler throats were constructed in \cite{smallCCs}, and supersymmetric configurations with Klebanov-Strassler throats but without all moduli stabilized were obtained in \cite{Alvarez-Garcia:2020pxd,coniLCS}, building on the framework of \cite{Demirtas:2019sip}; we employ much of the technology used in those works.
Earlier work on supersymmetric AdS vacua of KKLT type appears in \cite{Denef:2004dm,Denef:2005mm}.}

\subsection{Engineering conifolds}\label{sec:coniPFV}

To construct a Klebanov-Strassler throat region in a flux compactification, we must stabilize the complex structure moduli near a conifold locus.  This will be our first major departure from the framework of \cite{smallCCs}, which we have otherwise followed thus far. 
To engineer small cosmological constants in the presence of a Klebanov-Strassler throat, we would like to adapt the formalism of
\cite{Demirtas:2019sip}
to work near a boundary of moduli space where a conifold curve shrinks. This was done in \cite{coniLCS}, whose approach we will follow; similar results appeared simultaneously in \cite{Alvarez-Garcia:2020pxd}. Here we will just sketch the construction, referring the reader to the original work for more details.

A \textit{conifold singularity} is a locus in the complex structure moduli space of a Calabi-Yau threefold $X$ where a set of $n_{\mathrm{cf}}$ three-cycles, all of which lie in the same homology class $[\mathcal{C}] \in H_3(X,\mathbb{Z})$, shrink to zero volume. In an LCS patch, we can identify the complex structure moduli space of  $X$ with the  complexified  K\"ahler cone 
$\calK(\Xt)$
of the mirror threefold $\Xt$. In this picture, the conifold locus is identified with the facet of 
$\calK(\Xt)$ where a fixed set of curves $\calC$ in some effective curve class  $\tilde{\mathbf{q}}^{\mathrm{cf}} \in \mathcal{M}(\widetilde{X})\cap H_2(\Xt,\mathbb{Z})$, which we call the \textit{conifold class}, shrink to zero size. We will defer discussion of how to actually find conifolds and how to compute $n_{\mathrm{cf}}$ to \S\ref{sec:polytopeselection}, and for now just assume the existence of some conifold class $\tilde{\mathbf{q}}^{\mathrm{cf}}$.
The volume of the curves $\calC$ is measured by the absolute value of
\begin{equation}
    z_{\text{cf}}\coloneqq z^a \tilde{\mathbf{q}}^{\mathrm{cf}}_a\, ,
\end{equation}
which we will refer to as the \emph{conifold modulus}. Denoting by ${\omega^a}_\alpha$ the generators of the lattice orthogonal to 
$\tilde{\mathbf{q}}^{\mathrm{cf}}$, it will be useful to parameterize moduli space according to
\begin{equation}
    z^a={\omega^a}_{\alpha} z^\alpha+ \xi^a z_{\text{cf}}\, ,
\end{equation}
with $a \in 1,\ldots,h^{2,1}$ and
$\alpha \in 1,\ldots,h^{2,1}-1$, in terms of the conifold modulus $z_{\mathrm{cf}}$ and the \emph{bulk moduli} $z^\alpha$. Here, $\xi^a$ is an arbitrary constant vector  satisfying $\tilde{\mathbf{q}}^{\mathrm{cf}}_a\xi^a=1$.

We want to engineer a long but finite Klebanov-Strassler throat, so we need to stabilize the moduli such that $z_{\mathrm{cf}}$ is exponentially small but still nonzero, 
and we are otherwise at LCS. 
In this regime, we need to evaluate the flux superpotential and the K\"ahler potential systematically, order by order in $z_{\mathrm{cf}}$. As in \cite{coniLCS},  if one defines  the following  quantized fluxes,
\begin{subequations}
    \begin{align}
        \vec{f} &= \left(P_0,P_a,0,M^a\right)^\top \, , \label{eq:PFVfflux} \\
        \vec{h} &= \left(0,K_a,0,0^a\right)\,, \label{eq:PFVhflux}
    \end{align}
\end{subequations}
one can continue the flux superpotential \eqref{eq:superpotential} to small $z_{\text{cf}}$, resulting in
\begin{equation}
    \sqrt{\tfrac{\pi}{2}} \cdot W(z^\alpha,z_{\mathrm{cf}},\tau)=W_{\mathrm{bulk}}(z^\alpha,\tau)+z_{\mathrm{cf}} W^{(1)}(z^\alpha,z_{\mathrm{cf}},\tau)+\mathcal{O}(z_{\mathrm{cf}}^2)\, ,
\end{equation}
with
\begin{align}
    W_{\mathrm{bulk}}(z^\alpha,\tau)=&\frac{1}{2}M^a \widetilde{\kappa}_{a \beta\gamma}z^\beta z^\gamma-\tau K_\alpha z^\alpha+\left(P_\beta-\frac{1}{2}M^a \mathbb{A}_{a\beta}\right)z^\beta+\left(P_0-\frac{1}{24}M^a \tilde{c}'_a\right)  \nonumber \\
    &-\frac{1}{(2\pi)^2}\sum_{\tilde{\mathbf{q}}\neq \tilde{\mathbf{q}}^{\mathrm{cf}}} \mathscr{N}_{\tilde{\mathbf{q}}}\,\tilde{\mathbf{q}}_a M^a\,\text{Li}_2(e^{2\pi i\tilde{\mathbf{q}}_\alpha z^\alpha})\,  ,
\end{align}
where we have defined $\tilde{c}'_a\coloneqq \tilde{c}_a+n_{\mathrm{cf}}\tilde{\mathbf{q}}^{\mathrm{cf}}_a$, and lower Greek indices are obtained from Latin ones by contracting with the projection ${\omega^a}_{\alpha}$.

If $P_\beta= \frac{1}{2}M^a \mathbb{A}_{a\beta}$
and $P_0=\frac{1}{24}M^a \tilde{c}'_a$, the polynomial part of the bulk superpotential becomes homogenous and quadratic in the bulk moduli $z^\alpha$ and $\tau$, and therefore vacua with small superpotential can be found as in \cite{Demirtas:2019sip}.

At linear order in $z_{\mathrm{cf}}$ we have
\begin{align}
    W^{(1)}(z^\alpha,z_{\mathrm{cf}},\tau)=&-M\frac{n_{\mathrm{cf}}}{2\pi i}\Bigl(\log(-2\pi i z_{\mathrm{cf}})-1\Bigr)+M^a \widetilde{\kappa}_{ab\gamma}\xi^b z^\gamma+\left(P_a-\frac{1}{2}\mathbb{A}_{ab}M^b\right)\xi^a \nonumber\\
    &+\frac{1}{2\pi i }\sum_{\mathbf{q}\neq \tilde{\mathbf{q}}^{\mathrm{cf}}}
    \tilde{\mathbf{q}}_a \tilde{\mathbf{q}}_b \,M^a  \xi^b\, \mathscr{N}_{\tilde{\mathbf{q}}}\,\text{Li}_1(e^{2\pi i \tilde{\mathbf{q}}_\alpha z^\alpha})\, ,
\end{align}
where we have defined
\begin{equation}\label{eq:coniMdef}    
M\coloneqq \tilde{\mathbf{q}}^{\mathrm{cf}}_a M^a\,,
\end{equation} 
and the number of conifolds is given by $n_{\mathrm{cf}}=\mathscr{N}_{\tilde{\mathbf{q}}}$; this expression
involves the famous logarithmic branch cut of the conifold. 

In the compactification geometry $X$, rather than its mirror $\Xt$, the conifold   singularity involves a shrinking three-cycle that we denote by $A_{\text{coni}}$, and refer to as `the conifold A-cycle'.  The three-form flux threading the conifold A-cycle is equal to $M$ defined in \eqref{eq:coniMdef}.

We now make the convenient choice $\xi^a=M^a/M$ and define the quantities
\begin{equation}\label{eq:Mnorm}
\mathcal{K}_{a\bar{b}}\coloneqq \widetilde{\kappa}_{ab\gamma}\text{Im}(z)^\gamma-\frac{1}{2\pi} \sum_{\tilde{\mathbf{q}}\neq \tilde{\mathbf{q}}^{\mathrm{cf}}}
    \tilde{\mathbf{q}}_a \tilde{\mathbf{q}}_b \,\mathscr{N}_{\tilde{\mathbf{q}}}\,\text{Li}_1(e^{2\pi i \tilde{\mathbf{q}}_\alpha z^\alpha})\, ,
\end{equation}
\begin{equation}\label{eq:mnormdef}
||\vec{M}||^2\coloneqq -M^a M^b \mathcal{K}_{a\bar{b}}\,,
\end{equation}
and
\begin{equation}\label{eq:D3_charge_in_throats}
    Q^{\text{throat}}_{\text{flux}}\equiv Q_{\text{flux}}-g_s ||\vec{M}||^2\,,
\end{equation}
where the D3-brane charge in fluxes, $Q_{\text{flux}}$, was defined in \eqref{eq:D3-charges}. 
Even though $\mathcal{K}_{a\bar{b}}$ has indefinite signature $(+1,-1,\ldots, -1)$, one can show that $||\vec{M}||>0$ on the space of admissible fluxes.

The F-terms of the conifold modulus are solved at the vev
\begin{equation}
\label{eq:conifold_vev}
    \langle|z_{\mathrm{cf}}|\rangle= \frac{1}{2\pi}\exp\Biggl(-\frac{2\pi}{g_s M^2 \, n_{\mathrm{cf}}}Q^{\text{throat}}_{\text{flux}}\Biggr)\, ,
\end{equation}
and $Q^{\text{throat}}_{\text{flux}}$ can be thought of as the part of the overall D3-brane charge in fluxes that resides in the $n_{\mathrm{cf}}$ local conifold regions. Conversely, the quantity
\begin{equation}\label{eq:D3_charge_in_bulk}
    Q_{\text{flux}}^{\text{bulk}}:= g_s ||\vec{M}||^2
\end{equation}
can be thought of as the part residing in the bulk.

Importantly --- provided that $Q^{\text{throat}}_{\text{flux}}>0$ --- the conifold modulus gets stabilized exponentially close to $z_{\mathrm{cf}}=0$, leading to a warped throat region with hierarchy of scales proportional\footnote{The precise numerical relation is derived in Appendix \ref{sec:alignment}: see \eqref{eq:warpz}.
} to $\langle|z_{\mathrm{cf}}|\rangle^{\frac{1}{3}}$.

\subsection{Perturbatively flat vacua with conifolds}\label{sss:cpfv}

As long as the vev of the conifold modulus is very small, the F-terms of the bulk moduli $(z^\alpha,\tau)$ are insensitive to $z_{\mathrm{cf}}$, and we can analyze their stabilization using $W_{\text{bulk}}(z^\alpha,\tau)$, along the lines of \cite{Demirtas:2019sip}.

In terms of $N_{\alpha\beta}\coloneqq M^a \kappa_{a\alpha\beta}$, we define $p^\alpha\coloneqq  N^{\alpha\beta} K_\alpha$, and $p^a:={w^a}_{\alpha}p^\alpha$, and the conditions for solutions as in \cite{Demirtas:2019sip,coniLCS} become
\begin{align}
    \label{eq:detN_conition}
    &\det N\neq 0\, ,\\
    \label{eq:p_in_Kcf}
    &\vec{p}\in \mathcal{K}_{\mathrm{cf}}\, ,\\
    \label{eq:Diophantine_eq}
    &K_{\alpha}p^\alpha=0\, ,\\
    \label{eq:integrality_condition_I}
    & \mathbb{A}_{\alpha b}M^b\in 2\mathbb{Z}\, ,\\
    \label{eq:integrality_condition_II}
    & \tilde{c}'_a M^a\in 24\mathbb{Z}\, ,
\end{align}
where $\mathcal{K}_{\mathrm{cf}}$ denotes the (interior of) the facet of the K\"ahler cone $\calK(\Xt)$ of the mirror threefold $\widetilde{X}$ along which the conifold curve shrinks.
In general, $p^\alpha$ is a rational vector, but it is occasionally useful to rescale it by a positive integer $\mathtt{r}$ to obtain a vector with integer entries, which we denote by $p_{\text{int}}^{\alpha}\equiv \mathtt{r}\, p^{\alpha}$.
We remark that the conditions \eqref{eq:detN_conition}-\eqref{eq:integrality_condition_II}
are invariant under  
$K_a\rightarrow K_a-\delta K \, \tilde{\mathbf{q}}^{\mathrm{cf}}_a$ for $\delta K\in \mathbb{Z}$, under which the vev of the conifold modulus transforms as
\begin{equation}
    \langle |z_{\mathrm{cf}}|\rangle\longrightarrow e^{-2\pi\frac{\delta K}{g_s M}} \langle |z_{\mathrm{cf}}|\rangle\, .
\end{equation}

If the conditions \eqref{eq:detN_conition}-\eqref{eq:integrality_condition_II} are met, the F-term conditions for the bulk moduli $z^\alpha$ and the axiodilaton $\tau$ are solved along the one-dimensional locus in field space where
\begin{equation}
    z^\alpha=p^\alpha \tau\, ,
\end{equation}
provided that exponential terms in $W_{\text{bulk}}$ are neglected. 
Such a locus is called a perturbatively flat vacuum (PFV) \cite{Demirtas:2019sip}, and we refer to a perturbatively flat vacuum with a conifold as a \emph{conifold PFV}.

Integrating out all but the single light degree of freedom,\footnote{Recall that in this section we are not yet considering K\"ahler moduli.} 
one obtains an effective theory that we will refer to as the PFV effective theory, with effective superpotential
\begin{equation}
    W^{\text{eff}}_{\text{bulk}}(\tau)=-\frac{1}{(2\pi)^2}\sum_{\tilde{\mathbf{q}}\neq \tilde{\mathbf{q}}^{\mathrm{cf}}} \mathscr{N}_{\tilde{\mathbf{q}}}\, \tilde{\mathbf{q}}_a M^a\,\text{Li}_2\left(e^{2\pi i \tilde{\mathbf{q}}_\alpha p^\alpha \tau}\right)\, .
\end{equation}
This can be evaluated order by order in the worldsheet instanton expansion of the mirror dual $\widetilde{X}$. For later convenience, we write
\begin{equation}
    W^{\text{eff}}_{\text{bulk}}(\tau) = \sum_{N=1}^{\infty}\, W_N
\end{equation}
where  
\begin{equation} 
   W_N \coloneqq -\frac{1}{(2\pi)^2} 
\sum_{\mathbf{p}_{\text{int}}\cdot\tilde{\mathbf{q}}=N}\mathscr{N}_{\tilde{\mathbf{q}}}\,\tilde{\mathbf{q}}_a M^a\,\text{Li}_2\Bigl(e^{\frac{2\pi \I}{\mathtt{r}} N \t}\Bigr). \label{eq:Widef}
\end{equation}

An F-term solution for the light degree of freedom $\tau$ frequently arises from a competition between consecutive terms with somewhat hierarchical coefficients: this effect is called the \emph{racetrack mechanism}. 
For example, if the superpotential can be well-approximated by the two leading terms,
\begin{equation}\label{eq:twoterm}
    W^{\text{eff}}_{\text{bulk}}(\tau) \approx C_1 \exp\bigl(2\pi \I q_1 \t\bigr) + C_2 \exp\bigl(2\pi \I q_2\t\bigr)\,,
\end{equation}
then an F-term solution arises for 
\begin{equation}
    \langle\t\rangle  = \frac{\I}{2\pi\left(q_1-q_2\right)}\log\left(-\frac{C_1q_1}{C_2q_2}\right)\, ,
\end{equation}
and the superpotential has expectation value
\begin{equation}
    \langle W^{\text{eff}}_{\text{bulk}} \rangle = C_2\frac{q_2-q_1}{q_1}\left(-\frac{C_2q_2}{C_1q_1}\right)^{q_2/(q_2-q_1)}\,,
\end{equation}
which is exponentially small in $1/(q_2-q_1)$, provided that $C_2q_2> C_1 q_1$. 

It is important to note that the effective superpotential in general receives corrections of order $\left(W^{\text{eff}}_{\text{bulk}}(\tau)\right)^2$, and so care must be taken when considering solutions for which the vev of the effective superpotential is not extremely small. All of our solutions will feature superpotentials that are somewhat small, but not extremely small, and so we will take the following approach:
\begin{enumerate}
    \item Identify a vacuum $\langle \tau \rangle_{\mathrm{PFV}}$ of the PFV effective theory.
    \item Use the point $\tau = \langle \tau \rangle_{\mathrm{PFV}}$, $z^\alpha=p^\alpha \langle \tau \rangle_{\mathrm{PFV}}$ as an initial guess for a numerical search for a root of the full F-terms, no longer making the approximations that led to the PFV effective theory.
\end{enumerate}
If a solution can be found in step 2 it is reliable, and is usually found very quickly.

We thus arrive at expectation values for the axiodilaton, bulk complex structure moduli, and conifold modulus, at the level of the full theory of the complex structure moduli and axiodilaton, but not yet including the K\"ahler moduli or any source of supersymmetry breaking:
\begin{align}\label{eq:promotedvevs}
 \tau &\rightarrow  \langle \tau \rangle_{\text{F}}\,,\\
z^{\alpha} &\rightarrow \langle z^{\alpha} \rangle_{\text{F}}\,,\\
 z_{\text{cf}}  &\rightarrow  \langle z_{\text{cf}} \rangle_{\text{F}}\,,
\end{align}
and we write 
\begin{equation}
 W_0\coloneqq  \langle |W_{\text{flux}}|\rangle_{\text{F}}
\equiv \Bigl|W_{\text{flux}}\Bigl(\langle \tau \rangle_{\text{F}}, \langle z^{\alpha} \rangle_{\text{F}}, \langle z_{\text{cf}} \rangle_{\text{F}}\Bigr)\Bigr|\,.
\end{equation}

\subsection{Anti-D3-branes in Klebanov-Strassler throats}\label{sec:KS} 

Given any vacuum in complex structure moduli space that lies sufficiently close to a conifold singularity --- i.e.~with $\langle|z_{\mathrm{cf}}|\rangle \ll 1$, cf. \eqref{eq:conifold_vev} --- a warped Klebanov-Strassler (KS) throat emerges \cite{Klebanov:2000hb,Giddings:2001yu}. The radius of curvature near the tip is of order $\sqrt{g_sM\a'}$ \cite{Klebanov:2000hb}. Thus, in addition to the vev of the flux superpotential, the 't Hooft coupling $g_sM$ is a key control parameter in our analysis. 

There are two interesting regimes. In the first regime, where $g_sM\ll1$, the supergravity description of the throat is not controlled, and instead one should work with the dual description in terms of a confining gauge theory \cite{Klebanov:2000hb}; we refer to this as the gauge theory regime. In contrast, for $g_sM\gg1$, the supergravity description is the correct one and we correspondingly refer to this as the supergravity regime.  
In this work we will focus on the supergravity regime, where supersymmetry breaking is much better understood; we defer discussion of the gauge theory regime to future work.\footnote{Nevertheless, we present an example in the gauge theory regime in Appendix \ref{sec:KSInKS}.}

The infrared region of the Klebanov-Strassler throat provides a weakly curved background for a spacetime-filling 
anti-D3-brane, which is the source of supersymmetry breaking in the KKLT proposal \cite{KKLT}. Two properties of the throat are essential for this purpose.

First, an anti-D3-brane state at the bottom of the throat can be metastable. At leading order in the $\alpha'$ expansion, this was studied by Kachru, Pearson, and Verlinde (KPV) \cite{KPV}, who found that $p$ anti-D3-branes in a Klebanov-Strassler throat can form a metastable  state if 
\begin{equation}\label{eq:KPV_constraint}
    \dfrac{M}{p}\gtrsim 12\, .
\end{equation}
We will study configurations with a single anti-D3-brane, i.e.~with $p=1$, and so we will consider configurations with $M>12$. The constraint \eqref{eq:KPV_constraint} is generally subject to corrections in $\alpha'$, i.e., in $1/(g_s M)$, which were studied in a series of recent works \cite{Junghans:2022exo,Junghans:2022kxg,Hebecker:2022zme,Schreyer:2022len,Schreyer:2024pml}. These results indicate that $g_s M$ has to be rather large before $\alpha'$ corrections to the computation of KPV can be safely disregarded.\footnote{See also \cite{Lust:2022xoq}.} But by the same token,  
the results of \cite{Junghans:2022exo,Junghans:2022kxg,Hebecker:2022zme,Schreyer:2022len,Schreyer:2024pml} cannot rule out the existence of metastable anti-D3 brane states in the regime $g_s M\sim 1$ because the series of $\alpha'$ corrections is not controlled when the leading correction starts to matter. In this work we focus on finding explicit solutions in which $g_s M\gtrsim 1$,  leaving the important question of what the actual range of $g_sM$ for which the anti-D3-brane state remains metastable is for future work. 
We emphasize that the existence of a metastable supersymmetry-breaking state for small values of the 't Hooft coupling, even for arbitrarily small values, is certainly plausible.  

Second, the positive tension of a single metastable anti-D3-brane state contributes to the four-dimensional vacuum energy. The potential $V_{\overline{D3}}$ of an anti-D3-brane in a Klebanov-Strassler throat scales with the overall  Einstein-frame  volume $\mathcal{V}_E$ of the Calabi-Yau as
\cite{KPV,KKLT,Kachru:2003sx}
\begin{equation}
\label{eq:anti-D3-potential0}
    V_{\overline{D3}} = \frac{c}{\calV_E^{4/3}}\, ,
\end{equation}
where the constant $c$ is suppressed by the redshift of the throat. In Appendix \ref{sec:alignment} we find
\begin{equation}
    c = \eta\,\frac{z_{\mathrm{cf}}^{4/3}}{g_sM^2 \widetilde{\calV}^{2/3}}
    \; ,\quad \eta \approx 2.6727\, ,\label{eq:cdef}
\end{equation}
where $\widetilde{\calV}$ is the string-frame volume of the mirror Calabi-Yau.

Thus, in the presence of an anti-D3-brane, the full scalar potential is given by
\begin{equation}\label{eq:Vfull}
    V = V_F + V_{\overline{D3}}\,,
\end{equation}
where $V_F$ is the F-term potential \eqref{eq:vfsum}, and 
$V_{\overline{D3}}$ is specified by \eqref{eq:anti-D3-potential0} and \eqref{eq:cdef}.

In order for the term $V_{\overline{D3}}$ to lift the overall vacuum energy to a positive value but not to induce a runaway instability, one requires that the uplift is approximately of the same order as 
$|V_F|$ in an F-term minimum, which leads to the constraint
\begin{equation}\label{eq:throat_tuning_maintext}
\Xi := \dfrac{V_{\overline{D3}}}{|V_F|} \approx \frac{|z_{\text{cf}}|^{\frac{4}{3}}}{|W_0|^2}\frac{\mathcal{V}_E^{\frac{2}{3}}\widetilde{\mathcal{V}}^{\frac{1}{3}}}{(g_s M)^2}\cdot \zeta \sim 1
\,, 
\end{equation}   
where the constant $\zeta\approx 114$ is derived in Appendix \ref{sec:alignment}. We call a vacuum satisfying the relation \eqref{eq:throat_tuning_maintext} \textit{well-aligned}, and our task in the rest of the paper is to construct well-aligned vacua.

\subsection{Constraints on control parameters}
 
In practice, the alignment condition  \eqref{eq:throat_tuning_maintext}
requires the conifold modulus to have a small vev, $|z_{\text{cf}}| \ll 1$. 
Through \eqref{eq:conifold_vev},
this
vev is governed by the D3-brane charge in the throat as 
defined in \eqref{eq:D3_charge_in_throats},
which obeys
\begin{equation}\label{eq:throatisless}
Q_{\text{flux}}^{\text{throat}} \le  Q_{\text{flux}} = Q_{\text{O}} + 2\,,
\end{equation} where 
the D3-brane tadpole from orientifolds $Q_{\text{O}}$ and the total D3-brane charge in flux $Q_{\text{flux}}$ were defined in \eqref{eq:D3-charges},
and we have used \eqref{eq:qqplus2}.

Combining \eqref{eq:throatisless} with \eqref{eq:conifold_vev}, we find  
\begin{equation}\label{eq:zbound}
    \langle |z_{\mathrm{cf}}| \rangle \gtrsim \frac{1}{2\pi} \exp \left(-\frac{2\pi Q_{\text{O}}}{g_s M^2\, n_{\mathrm{cf}}}\right)\, .
\end{equation}
The tadpole $Q_\text{O}$ is determined by the Hodge numbers of the Calabi-Yau, which are believed to be bounded in general.
For Calabi-Yau hypersurfaces in toric varieties, which are the setting for this work, we have\footnote{Explicit orientifolds that saturate this bound appear in \cite{Crino:2022zjk,orientifolds}.}
\begin{equation}
Q_{\text{O}}  \le 2 + h^{1,1} + h^{2,1} \le 2+11+491=504\,. 
\end{equation}

Using \eqref{eq:zbound}, we can write \eqref{eq:throat_tuning_maintext} as a \emph{lower bound on the flux superpotential} \cite{Bena:2018fqc,Bena:2020xrh,Gao:2022fdi}. 
Generically, for a (conifold-)PFV the quantity $\widetilde{\mathcal{V}}$ is of order $g_s^{-3}$, and empirically we typically find\footnote{This finding implies that control over warping corrections \cite{Giddings:2005ff,Martucci:2016pzt,Carta:2019rhx,Gao:2020xqh,Carta:2021lqg} is good for small $g_s$, cf.~\cite{smallCCs}.} $\mathcal{V}_E^{2/3}\sim Q_{\text{O}}/g_s$. Using these estimates   
and taking $n_{\text{cf}}=2$,
we find 
\begin{equation}\label{eq:cookedness}
    W_0 \gtrsim \frac{1}{g_s^2M} \sqrt{\frac{\zeta Q_{\text{O}}}{(2\pi)^{4/3}}}\, \exp\left(-\frac{2\pi Q_{\text{O}}}{3g_sM^2}\right)\, .
\end{equation}
When the 't Hooft coupling $g_s M$ is large,\footnote{If a metastable supersymmetry breaking state could be established directly in the Klebanov-Strassler gauge theory \cite{Klebanov:2000hb}, i.e., for small 't Hooft coupling $g_s M$, then one could search for de Sitter vacua with exponentially small superpotentials. Identifying such a state in the gauge theory regime is an important (and formidable!) task for future research.} \eqref{eq:cookedness} presents
a severe constraint that sharply bounds the best control
parameters one can even in principle hope to find.
For example, taking $g_s=1/(g_s M)=0.2$ and using\footnote{
The much larger D3-brane tadpoles that arise in generic F-theory models could significantly ameliorate this problem; however, see e.g.~\cite{Bena:2020xrh,Grana:2022dfw,Lust:2022mhk}.} $Q_{\text{O}} \le 504$, we find $W_0>0.015$. 

Moreover, even with the very conservative values $g_sM\sim1$ and $M=14$, 
we find that $W_0>10^{-31}$.
This limit on the vacuum energy in a KKLT de Sitter vacuum stands in sharp contrast to the case of supersymmetric AdS vacua,  
where there does not appear to be a fundamental limit on how small a $W_0$ one can hope to find \cite{smallCCs}.\footnote{Indeed, in forthcoming work \cite{evenSmaller} we will present a supersymmetric AdS vacuum with $W_0<10^{-500}$.}

The bound \eqref{eq:cookedness}  implies that control over the string loop expansion and over corrections in $1/(g_sM)$ will necessarily be somewhat limited in our solutions, and indeed
none of the de Sitter candidates presented in this work will have extremely small flux superpotential. 
Nonetheless, we have been able to construct well-aligned vacua that uplift to metastable de Sitter vacua of the leading-order theory.

\section{Search Procedure}
\label{sec:methods}

At this point the leading-order EFT has been fully specified in terms of computable data. Our task is now to find incarnations of this EFT in explicit Calabi-Yau compactifications, and to stabilize the moduli in these theories. 

In this section we explain the process of identifying candidate KKLT de Sitter vacua, from the first step of specifying topological data, down to the final step of exhibiting a metastable minimum of the potential $V$ defined in \eqref{eq:Vfull}.

\subsection{Overview of requirements}\label{sec:summofreq} 

We will select Calabi-Yau orientifolds and three-form flux quanta as follows:
\begin{enumerate}[label=\roman*.)]
    \item We begin with a Calabi-Yau threefold hypersurface $X$ in a toric variety obtained from a triangulation of a reflexive polytope.  We select $X$ for which a conifold singularity can arise from shrinking a toric flop curve in the mirror $\widetilde{X}$.\footnote{A set of further criteria we impose on the polytope are enumerated and explained in \S\ref{sec:polytopeselection}.}
    \item We compactify on a particular orientifold of $X$, inherited from an involution of the toric variety, for which $h^{1,1}_-=h_{+}^{2,1}=0$ and $Q_{\text{O}} =  2+h^{1,1}+h^{2,1}$. 
    \item We check that the conifold and orientifolds do not intersect.
   \item We check that at least $h^{1,1}(X)$ pure rigid divisors that do not intersect the conifold contribute to the nonperturbative superpotential \eqref{eq:Wnppure}.
    \item We choose quantized fluxes with $Q_{\text{flux}} = Q_{\text{O}}+2,$ so that Gauss's law \eqref{eq:gaussLaw} requires the presence of a single anti-D3-brane.\footnote{One could in addition include an arbitrary number of D3-brane/anti-D3-brane pairs, but in general such pairs quickly annihilate.} 
    \item We further require that the quantized fluxes are such that the complex structure moduli and axiodilaton are stabilized
    near a conifold locus, and at weak string coupling, and such that the vev of the flux superpotential is small, i.e.~we require  
    $\langle|z_{\mathrm{cf}}|\rangle_{\text{F}}\ll1$, $(\text{Im}\langle\tau\rangle_{\text{F}})^{-1} \equiv g_s \ll 1$, 
    and  
    $W_0\ll1$, cf.~\eqref{eq:promotedvevs}.
    \item We select cases in which $M>12$, with $M$ the flux on the conifold A-cycle, cf.~\eqref{eq:coniMdef}, 
    so that the anti-D3-brane metastability condition \eqref{eq:KPV_constraint} obtained in \cite{KPV} is met.

    \item We select cases in which the full scalar potential \eqref{eq:Vfull}, $V=V_F+V_{\overline{D3}}$, viewed as a function of all the moduli fields $\Phi$ --- the complex structure moduli, axiodilaton, and K\"ahler moduli --- has a de Sitter minimum, i.e.~a point $\Phi_{\star}$ where $V>0$, $V'=0$, and the Hessian of $V$ is positive definite.
    \item We check that the truncations made are consistent at $\Phi_{\star}$: the contributions of all divisors omitted from \eqref{eq:Wnppure} must be small compared to the terms retained in \eqref{eq:Wnppure}, and the contributions of all curves omitted from \eqref{eq:detailedform3} and \eqref{eq:detailedform4}  must be small compared to the terms retained in 
\eqref{eq:detailedform3} and \eqref{eq:detailedform4}.  
\end{enumerate}
A vacuum meeting all of these requirements is a KKLT de Sitter vacuum in the leading order EFT.
In the following sections we explain in more detail the algorithm we used to find such configurations. 
 
We collect  the various conditions being imposed, as well as the number of configurations satisfying each constraint, in Table~\ref{tab:scan_summary}.

\begin{table}
    \centering
    \begin{tabular}{c|c|c}
   \textbf{Condition}      & \textbf{Number of configurations} & \textbf{Explanation} \\ \hline 
    & & \\[-1.3em]
   $3\le h^{2,1} \le 8$      &   202{,}073 polytopes  & \S\ref{sec:polytopeselection} \\[0.2em]
   trilayer, $\Delta$ and $\Delta^\circ$ favorable     &  3187 polytopes   & \S\ref{sec:polytopeselection} \\[0.2em]  
   Hodge number cuts & 322   polytopes   & \S\ref{sec:polytopeselection} \\[0.2em]
    $\ge h^{1,1}$ rigid divisors     &  322 polytopes   & \S\ref{sec:polytopeselection} \\[0.2em]
   conifold disjoint from O-planes  &  2669 conifolds  & \S\ref{sec:conifoldselection} \\[0.2em]
   conifold consistent with KKLT point  &  416 conifolds  & \S\ref{sec:conifoldselection} \\[0.2em]
   fluxes giving conifold PFV   &  240{,}480{,}253 conifold PFVs  & \S\ref{sec:pfv} \\[0.2em]
    two-term racetrack    &  141{,}594{,}222  racetrack PFVs  & \S\ref{sec:pfv}, \eqref{eq:twoterm} \\[0.2em]
   $M>12$; one anti-D3-brane     & 33{,}371  anti-D3-brane PFVs  & \S\ref{sec:pfv}, \eqref{eq:qqplus2} \\[0.2em]
   de Sitter vacuum & 30 de Sitter vacua & \S\ref{sec:examples}
   
    \end{tabular}
    \caption{Number of configurations found at each stage of the selection process. The conditions are cumulative: in a given row the requirements of all preceding rows are imposed.  At the level of polytopes, the configurations found exhaust the possibilities in the Kreuzer-Skarke list that meet the stated conditions.  
    In the corresponding Calabi-Yau threefolds, all conifolds arising from toric flop curves were considered, but non-toric flop curves were not.  
    We do not quote here the exponentially large number of inequivalent triangulations of $\Delta^{\circ}$, but do allow exploration of any such Calabi-Yau phase.  
    At the level of flux choices yielding PFVs, the search was extensive but not exhaustive.  }
    \label{tab:scan_summary}
\end{table} 

\subsection{Selection of polytopes}\label{sec:polytopeselection}

We begin by setting notation and terminology for Calabi-Yau threefold hypersurfaces in toric varieties, which we will obtain from triangulations of  
four-dimensional reflexive polytopes from the Kreuzer-Skarke list \cite{Kreuzer:2000xy}.

Suppose that $\Delta \subset \mathbb{Z}^4$ is a four-dimensional reflexive polytope, and denote by $\Delta^{\circ}$ its polar dual.
We call a dual pair $(\Delta,\Delta^{\circ})$ of polytopes $\Delta^{\circ}$-favorable
if every two-face of $\Delta^{\circ}$ that has interior points is dual to a one-face of $\Delta$ that has no interior points, and similarly we call the pair $\Delta$-favorable if every two-face of $\Delta $ that has interior points is dual to a one-face of $\Delta^{\circ}$ that has no interior points.  Both conditions are convenient, and we will only consider pairs of polytopes that are both $\Delta$-favorable and $\Delta^{\circ}$-favorable.

Let $\mathscr{T}$ be a regular, star triangulation of $\Delta^{\circ}$
in which every point of $\Delta^{\circ}$ that is not interior to a facet is a vertex of a simplex of $\mathscr{T}$.\footnote{Such a $\mathscr{T}$ is not in general a \emph{fine}, regular, star triangulation (FRST), because of the omission of points interior to facets, but the associated subvarieties of $V$ do not intersect a generic hypersurface $X$, and so are immaterial for our analysis.  With this understanding, we refer to such triangulations $\mathscr{T}$ as FRSTs, because they are fine enough for our needs.} 
The toric fan associated to $\mathscr{T}$  defines a four-dimensional toric variety $V$ in which the generic anticanonical hypersurface $X$ is a smooth Calabi-Yau threefold.  

A consequence of $\Delta^{\circ}$-favorability is that
\begin{equation}
    h^{1,1}(V) = h^{1,1}(X)\,,
\end{equation}
with  
$h^{1,1}(V)+4$ toric coordinates $x_I$ generating the Cox ring.  We define the prime toric divisors of $V$ as
\begin{equation}
    \mathscr{D}_I \coloneqq  \{x_I = 0\}\,,
\end{equation} 
and we refer to 
\begin{equation}\label{eq:primetoric}
    D_I \coloneqq  \mathscr{D}_I \cap X
\end{equation}
as the prime toric divisors of $X$.
The $D_I$ are all effective divisors, and (again using $\Delta^{\circ}$-favorability) they generate $H_4(X,\mathbb{Z})$.  Even so, in general there exist effective divisors $D$ that are non-positive integer linear combinations of the $D_I$, which we term \emph{autochthonous divisors}.

We say that a polytope $\Delta^{\circ}$ is \emph{trilayer} if the points of $\Delta^{\circ}$ lie in exactly three distinct affine sub-lattices of codimension one, in a sense made precise in \cite{orientifolds}.
Calabi-Yau threefold hypersurfaces in toric varieties $V$ resulting from triangulations of trilayer polytopes admit very convenient orientifold actions \cite{orientifolds}.
In each case there exists a certain toric coordinate, which we denote by $x_1$, such that the involution of $V$ defined by
\begin{equation}\label{eq:odef}
    x_1 \to -x_1\,
\end{equation}
yields, when restricted to the generic invariant hypersurface $X \subset V$, an orientifold with $h^{1,1}_-=h^{2,1}_+=0$, which we will refer to as a \emph{trilayer orientifold}. All  orientifolds considered in this work will be of this type.

Given a Calabi-Yau orientifold $X/\calI$, the D3-brane tadpole, defined in \eqref{eq:D3-charges}, is a useful measure of the richness of flux vacua that one can expect in compactification on $X/\calI$. For the orientifolds considered here,  we have 
\begin{equation}\label{eq:qdef}
Q_{\text{O}} =  2+h^{1,1}+h^{2,1}\, .
\end{equation}
Thus, the D3-brane tadpole is large if either  
Hodge number is large. 
However,  the construction of large ensembles of PFVs becomes expensive for $h^{2,1} \gtrsim 10$, so we restrict ourselves to the range $3 \le h^{2,1} \le 8$. The D3-brane tadpole can then still be large for sufficiently large values of $h^{1,1}$. In practice, we restrict to polytopes with either $3\leq h^{2,1}\leq 5$ and $Q_{\text{O}}\ge 100$, or $6\leq h^{2,1}\leq 8$ and $Q_{\text{O}}\ge 150$. 

Furthermore, we will only consider polytopes in which at least $h^{1,1}$ of the prime toric divisors $D_I$ are rigid, as defined in \eqref{eq:drigid}, and so generate nonperturbative superpotential terms.\footnote{The prime toric divisors of a smooth Calabi-Yau threefold hypersurface $X$ are themselves smooth at generic points in the complex structure moduli space of $X$.}
 
Finally, for technical reasons related to our construction of conifold limits in Calabi-Yau hypersurfaces, which we will explain in \S\ref{sec:conifoldselection}, we impose that $\Delta^\circ$ has one or more one-faces with exactly one interior point, such that the dual two-face admits at least two distinct fine regular triangulations (FRTs).

In summary, we select Calabi-Yau hypersurfaces from pairs of reflexive polytopes $(\Delta, \Delta^{\circ})$ that  
\begin{enumerate}
    \item are $\Delta$-favorable,
    \item are $\Delta^{\circ}$-favorable,
    \item have $Q_{\text{O}} \ge 100$ and $3\leq h^{2,1} 
    \leq 5$, \emph{or} $Q_{\text{O}} \ge 150$ and $6\leq h^{2,1} 
    \leq 8$,
    \item have at least $h^{1,1}$ rigid prime toric divisors,
    \item are such that $\Delta^\circ$ has a one-face with a unique interior point, with the dual two-face of $\Delta$ admitting at least two FRTs.
\end{enumerate}
As summarized in Table~\ref{tab:scan_summary}, there are  $202{,}073$ polytopes in the Kreuzer-Skarke list with $3 \le h^{2,1} \le 8$, of which 322 fulfill all our requirements. These 322 polytopes are the arena for this work.

\subsection{Selection of Calabi-Yau orientifolds with conifolds}\label{sec:conifoldselection}
 
Next, let us describe how to construct Calabi-Yau orientifolds with conifolds, while avoiding intersections of the conifold singularities with the orientifold planes.
First, we recall that the anti-canonical polynomial for our Calabi-Yau hypersurface $X$ is given by
\begin{align}\label{eq:EquationHypS} 
f = \sum_{q\in \Delta\cap \mathbb{Z}^4} \psi_q s_q\, ,\quad s_q\coloneqq  \prod_{p_I} x_I^{\langle q,p_I\rangle +1}
\end{align}
where $p_I$ are the lattice points of $\Delta^\circ$ not interior to facets and not the origin, and the sum runs over all integral points $q$ in the dual polytope $\Delta$.

In order for the hypersurface to be invariant under the involution $\mathcal{I}$ in \eqref{eq:odef} all coefficients $\psi_q$ in \eqref{eq:EquationHypS} with $\langle q, p_1\rangle\in 2\mathbb{Z}$ are set to zero, making the otherwise generic hypersurface orientifold-invariant. 

We would like to engineer a situation in which \eqref{eq:EquationHypS} develops conifold singularities along a pair of points. This can be achieved in the following way: one considers a one-face of $\Delta^\circ$ with a unique interior point $p_{\mathrm{cf}}$. We will denote by $x_{\mathrm{cf}}$ the homogeneous coordinate associated to this point. Upon setting $x_{\mathrm{cf}}=0$, the anti-canonical polynomial $f$ degenerates to $g\coloneqq f|_{x_{\mathrm{cf}}=0}$, which is a linear combination of monomials $s_q$ with $q$ in the dual two-face of $\Delta$. We assume that this two-face admits at least two FRTs, and so in the simplest case it is a quadrilateral with four vertices $q^{1,\ldots,4}$. If this is so, one chooses local coordinates such that $f$ can be expanded as
\begin{equation}\label{eq:f_conifold}
    f = x_{\mathrm{cf}}\cdot  h+g+\mathcal{O}(x_{\mathrm{cf}}^2)= x_{\mathrm{cf}}\cdot  h+s_{q^1}\left(\psi_{q^1}+\psi_{q^2}\mathtt{t_1} +\psi_{q^3}\mathtt{t_2} +\psi_{q^4}\mathtt{t_1} \mathtt{t_2} \right)+\mathcal{O}(x_{\mathrm{cf}}^2)\, ,
\end{equation}
where the pair $(\mathtt{t_1}, \mathtt{t_2})$ parameterizes an algebraic torus $(\mathbb{C}^*)^2\subset (\mathbb{C}^*)^4\subset V$, and $h$ is a generic global section of the appropriate line bundle. 

For 
\begin{equation}\label{eq:conifold_singularity}
    \frac{\psi_{q_2}\psi_{q_3}}{\psi_{q_1}\psi_{q_4}}=1\, ,
\end{equation}
conifold singularities arise along
\begin{equation}\label{eq:conifold_locations}
    x_{\mathrm{cf}}=h=0\, ,\quad \mathtt{t_1}=\mathtt{t_2}=-\frac{\psi_{q_2}}{\psi_{q_4}}\, ,
\end{equation}
which is a set of two nodal points in $X$. The generalization to cases where the relevant two-face of the dual polytope $\Delta$ is not a quadrilateral is straightforward, and leads to higher-order corrections in $\mathtt{t}_{1,2}$ in \eqref{eq:f_conifold}, which can be neglected in the (partial) LCS limit.

Using the mirror map \cite{Hosono:1994av}, one reads off from \eqref{eq:conifold_singularity} that the resulting pair of shrinking three-spheres are mirror dual to a pair of isolated shrinking $\mathbb{P}^1$s in a class $[\mathcal{C}_{\mathrm{cf}}]$, shrinking likewise to conifold singularities. Thus, the methods of \cite{coniLCS} apply. In order to reach the conifold limit at a facet of the large complex structure cone, one has to further ensure that the Calabi-Yau periods are expanded around the appropriate LCS cone, i.e., one that is dual to a K\"ahler cone of the mirror threefold that has a facet dual to the class $[\mathcal{C}_{\mathrm{cf}}]$. 
Any such LCS cone yields a model with an appropriate conifold limit, and thus we generate models  by iterating over all FRSTs of the dual polytope $\Delta$.\footnote{More generally, one could use the results of \cite{Gendler:2022ztv} to work with conifold singularities that do not have a simple combinatorial description in terms of neighboring triangulations of the dual polytope. We leave to future work the task of generating models along these lines.}

Finally, given any candidate conifold, we impose a further condition: the orientifold symmetry \eqref{eq:odef} must exchange the two conifold singularities \eqref{eq:conifold_locations}, rather than mapping each of them to itself, which would force the presence of orientifold planes in  the infrared region of the resulting Klebanov-Strassler throats. In particular, we will impose that the divisor $x_{\mathrm{cf}}=0$ does not host an O7-plane.

As an aside, one could attempt to find models with only a single throat in the Calabi-Yau threefold that gets mapped to itself by the orientifold \eqref{eq:odef}. The only way this can occur without O7-planes reaching to the tip of the throat is if a pair of O3-planes reside on opposite poles of the shrinking $S^3$.\footnote{Such configurations were studied in a different context in \cite{AbdusSalam:2022krp}.} At first sight, this seems easy to engineer, by considering a pair of O3-planes that arise as the intersection with the Calabi-Yau hypersurface of an orientifold-invariant curve $\mathcal{C}_{IJK}\coloneqq \{x_I=x_J=x_K=0\}$ of the ambient variety.  In this case, the anti-canonical polynomial $f$ degenerates to a quadratic polynomial in a single $\mathbb{C}^*$ coordinate $\mathtt{t}$, 
\begin{equation}
    f|_{\mathcal{C}_{IJK}}\propto \psi_{q_1}+\psi_{q_2}\mathtt{t}+\psi_{q_3} \mathtt{t}^2\, ,
\end{equation}
with vanishing discriminant for tuned coefficients $\frac{\psi_{q_1}\psi_{q_3}}{ \psi_{q_2}^2}=\frac{1}{4}$. Here, $q_{1,2,3}$ are the three points of the one-face of $\Delta$ dual to the two-face of $\Delta^\circ$ that hosts the points $p_{I,J,K}$. Again using the mirror map, this identifies the singular limit as mirror dual to a facet of the K\"ahler cone where a divisor degenerates to a rational curve, yielding non-abelian $\mathfrak{su}(2)$ enhancement \cite{Aspinwall_1996,Katz_1996}. Unfortunately, the methods of \cite{coniLCS} are not directly applicable in this regime, and so we leave its exploration for future work.

A final restriction on the conifold results from considering K\"ahler moduli stabilization.
A Euclidean D3-brane on a divisor $D$ that intersects the conifold in the singular limit will pass through a highly warped region of the Klebanov-Strassler throat that arises after flux compactification. The corresponding Euclidean D3-brane contribution to the nonperturbative superpotential will be exponentially suppressed compared to those from unwarped Euclidean D3-branes,\footnote{See e.g.~\cite{Baumann:2006th} for background on Euclidean D3-branes and gaugino condensates in warped throats.} and cannot materially contribute to K\"ahler moduli stabilization.  
Thus, \eqref{eq:conifold_locations} implies that we should omit from the sum  in \eqref{eq:wlo3} the prime toric divisor $D_{\mathrm{cf}}\coloneqq \{x_{\mathrm{cf}}=0\}$ that intersects the $n_{\text{cf}}=2$ conifolds.  
We therefore require that of the prime toric divisors \emph{excluding} $D_{\mathrm{cf}}$, at least $h^{1,1}$ are pure rigid and allow for the existence of a KKLT point, as will be explained in \S\ref{ss:modstab}.

From the list of 322 polytopes meeting our topological criteria, there are 416 conifolds in Calabi-Yau orientifolds that meet all of our criteria: see Table~\ref{tab:scan_summary}.

\subsection{Selection of fluxes}\label{sec:pfv}

Equipped with a Calabi-Yau orientifold $X/\calI$ and a choice of conifold curve $\calC$, our next task is to select flux vectors that stabilize the complex structure moduli exponentially close to a conifold singularity with small $W_0$. Following \cite{coniLCS}, we will do this by engineering  conifold PFVs, as defined in  \S\ref{sss:cpfv}. 

We recall that the data of a conifold PFV consists of vectors $\vec{M}\in\bbZ^{h^{2,1}(X)}$, $\vec{K}\in\bbZ^{h^{2,1}(X)}$, $\vec{p}\in\bbQ^{h^{2,1}(X)}$ satisfying \eqref{eq:detN_conition}-\eqref{eq:integrality_condition_II}. In particular, 
\eqref{eq:p_in_Kcf} requires that $\vec{p}$ lies in a facet $\mathcal{K}_{\mathrm{cf}}$ of the K\"ahler cone of the mirror threefold $\Xt$ along which the conifold curve $\calC$ shrinks; the entries of $\vec{M}$ must satisfy the  integrality conditions \eqref{eq:integrality_condition_I} and \eqref{eq:integrality_condition_II}; and a Diophantine equation \eqref{eq:Diophantine_eq} in the entries of $\vec{K}$ and $\vec{M}$ must be satisfied.

We now describe an efficient algorithm for enumerating solutions to these constraints.\footnote{We present a variant of an algorithm discovered in the course of related work by M. Demirtas and implemented by A. Rios-Tascon; we thank them for letting us use their algorithm here.}

The simplest conditions are \eqref{eq:integrality_condition_I} and \eqref{eq:integrality_condition_II}, which constrain $\vec{M}$ to lie in some full-dimensional sub-lattice. Its lattice generators are easily found, for example using the Euclidean Algorithm. Expanding $\vec{M}$ as an integral combination of lattice generators identically solves the constraints \eqref{eq:integrality_condition_I} and \eqref{eq:integrality_condition_II}.

The starting point for what follows is a set of lattice points $p$ interior to the facet $\mathcal{K}_{\mathrm{cf}}$. This is a convex, locally polyhedral cone of dimension $h^{2,1}(X)-1$. Given its generators, or the dual hyperplanes, enumerating lattice points is an exercise in integer linear programming. 

Given any such lattice point $p$ we can write the Diophantine equation \eqref{eq:Diophantine_eq} as
\begin{equation}
    M^a \mathcal{T}_a = 0 \, ,\quad \mathcal{T}_a(p)\coloneqq \frac{1}{2}\kappa_{a\alpha\beta}p^\alpha p^\beta\, .
\end{equation}
This is a linear integer constraint on $\vec{M}$, and so again we can compute the sub-lattice along which this constraint is satisfied identically. Denoting by ${e_\mu}^a$, with $\mu=1,\ldots, h^{2,1}(X)-1$, a lattice basis for the sub-lattice along which \eqref{eq:Diophantine_eq}, \eqref{eq:integrality_condition_I}, and \eqref{eq:integrality_condition_II} are satisfied, we can expand
\begin{equation}
    M^a = \sum_{\mu} m^\mu {e_\mu}^a\, ,\quad \vec{m}\in \mathbb{Z}^{h^{2,1}(X)-1}\, ,   
\end{equation}  
and for each $\Vec{m}$ we can set $K_{\alpha}\rightarrow N_{\alpha\beta}p^\beta = M^a \kappa_{a\alpha\beta} p^\beta$. As $K_{\alpha}$ is obtained from the flux vector $K_a$ via the projection $\pi:\,K_a\mapsto K_\alpha=K_a {\omega^a}_\alpha$, we thus obtain a one-parameter family of solutions
\begin{equation}\label{eq:Kprime_def}
    K_a = \kappa_{ab\gamma}M^bp^\gamma-K' \tilde{\mathbf{q}}^{\mathrm{cf}}_a\, ,
\end{equation}
where $K'\in \mathbb{Z}$. As a consequence, the D3-brane charge in fluxes is given by
\begin{equation}\label{eq:D3_flux_splitting}
    Q_{\mathrm{flux}}=-M^a M^b \kappa_{ab\gamma}p^\gamma+M K' \, ,
\end{equation}   
and we recall that Gauss's law enforces \eqref{eq:gaussLaw} on the D3-brane charge in fluxes.

Dropping instanton corrections from \eqref{eq:Mnorm} and using $\text{Im}(z^\alpha)= p^\alpha/g_s$, we may approximately identify the first term in \eqref{eq:D3_flux_splitting} with $Q_{\text{flux}}^{\text{bulk}}\equiv g_s||M||^2$, cf. \eqref{eq:D3_charge_in_bulk}, and likewise $Q_{\text{flux}}^{\text{throat}}\approx M K'$.
Thus, \eqref{eq:conifold_vev} can be written
\begin{equation}\label{eq:zvev_coni_PFV}
    \langle |z_{\mathrm{cf}}| \rangle\simeq \frac{1}{2\pi}\exp\left(-\frac{2\pi K'}{g_s M}\right)\, .
\end{equation}
As a consequence of \eqref{eq:zvev_coni_PFV}, stabilizing near the conifold requires that $ M K'>0$. Thus, \eqref{eq:gaussLaw} implies that
\begin{equation}\label{eq:flux_ellipsoid}
    Q_{\text{flux}}^{\text{bulk}}\approx - m^\mu m^\nu \kappa_{\mu\nu\gamma}p^\gamma\overset{!}{=}Q_O+2(N_{\overline{\text{D3}}}-N_{\text{D3}})- M K'\, .
\end{equation}
The Weil-Petersson metric on complex structure moduli space is proportional to the quantity 
$-\tilde{\kappa}_{ab\gamma}\text{Im}(z^\gamma)$, up to an additive term $\propto \mathcal{T}_a \overline{\mathcal{T}_b}$ that is annihilated by the generators ${e_\mu}^a$. Therefore, \eqref{eq:Mnorm} defines a positive definite norm on the space of quantized fluxes $m^\mu$. For any given integral $p^\alpha$ we can hence find \emph{all} conifold PFVs by enumerating all points in the ellipsoid defined by
\begin{equation}\label{eq:flux_ellipsoid2}
    Q_{\text{flux}}^{\text{bulk}}\approx - m^\mu m^\nu \kappa_{\mu\nu\gamma}p^\gamma\leq Q_O+2(N_{\overline{\text{D3}}}-N_{\text{D3}})\, ,
\end{equation}
and dialing $ K'$ such that \eqref{eq:flux_ellipsoid} is solved.

Furthermore, for any integral $p^\alpha$ one can instead consider a rational multiple $p^\alpha\rightarrow p^\alpha/k$ for $1<k \in \mathbb{Z}$. This is equivalent to enlarging the right-hand side of \eqref{eq:flux_ellipsoid} by a factor $k$ and post-selecting those solutions for which $K_a$ is divisible by $k$ (the flux number $K'$ need not be integer if $k\neq 1$). We know of no analytic bound for the largest admissible $k$, so in practice we fix 
some reasonably large integer $k_{\mathrm{max}}$ and
seek solutions with $k\leq k_{\mathrm{max}}$.

\begin{figure}[!t]
\centering
\includegraphics[width=1\linewidth]{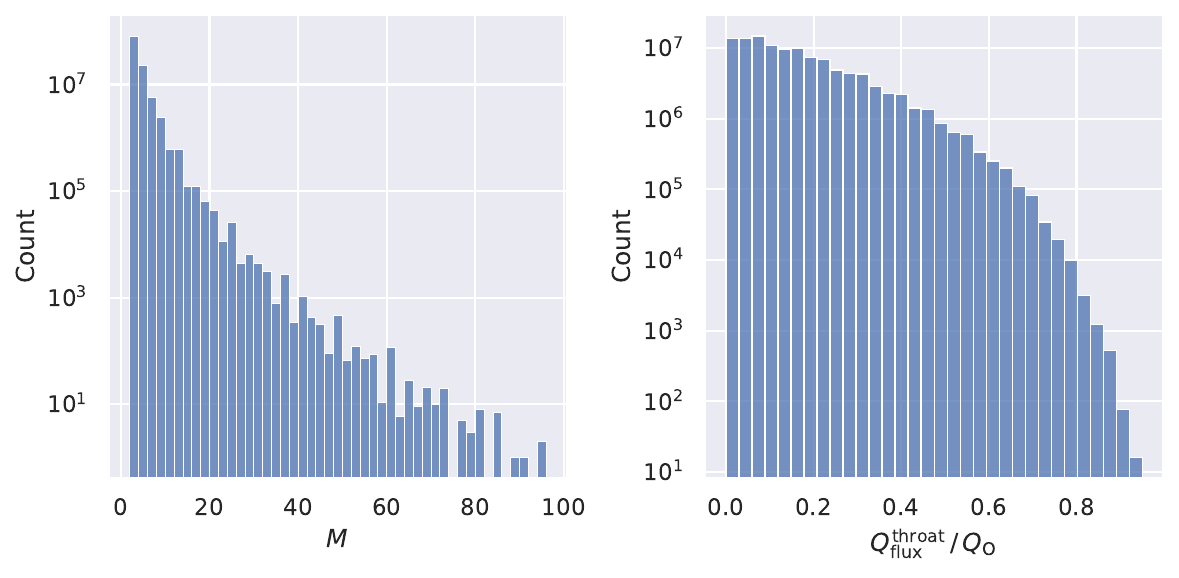}
\caption{
\emph{Left:} Log-scale histogram of the flux quantum $M$ on the conifold A-cycle, defined in \eqref{eq:coniMdef}. 
\emph{Right:} Log-scale histogram of D3-brane charge in the throat region, $Q_{\text{flux}}^{\text{throat}}$, as a fraction of the total charge $Q_{\text{O}}$.  See \eqref{eq:D3-charges} and \eqref{eq:D3_charge_in_throats} for definitions. 
We see that $M\gg 1$ occurs in our ensemble, as do
Klebanov-Strassler throats containing almost the entire D3-brane charge of the compactification, but both are exponentially rare.}
\label{fig:2histograms}
\end{figure}

\begin{figure}[!t]
\centering
\includegraphics[width=\linewidth]{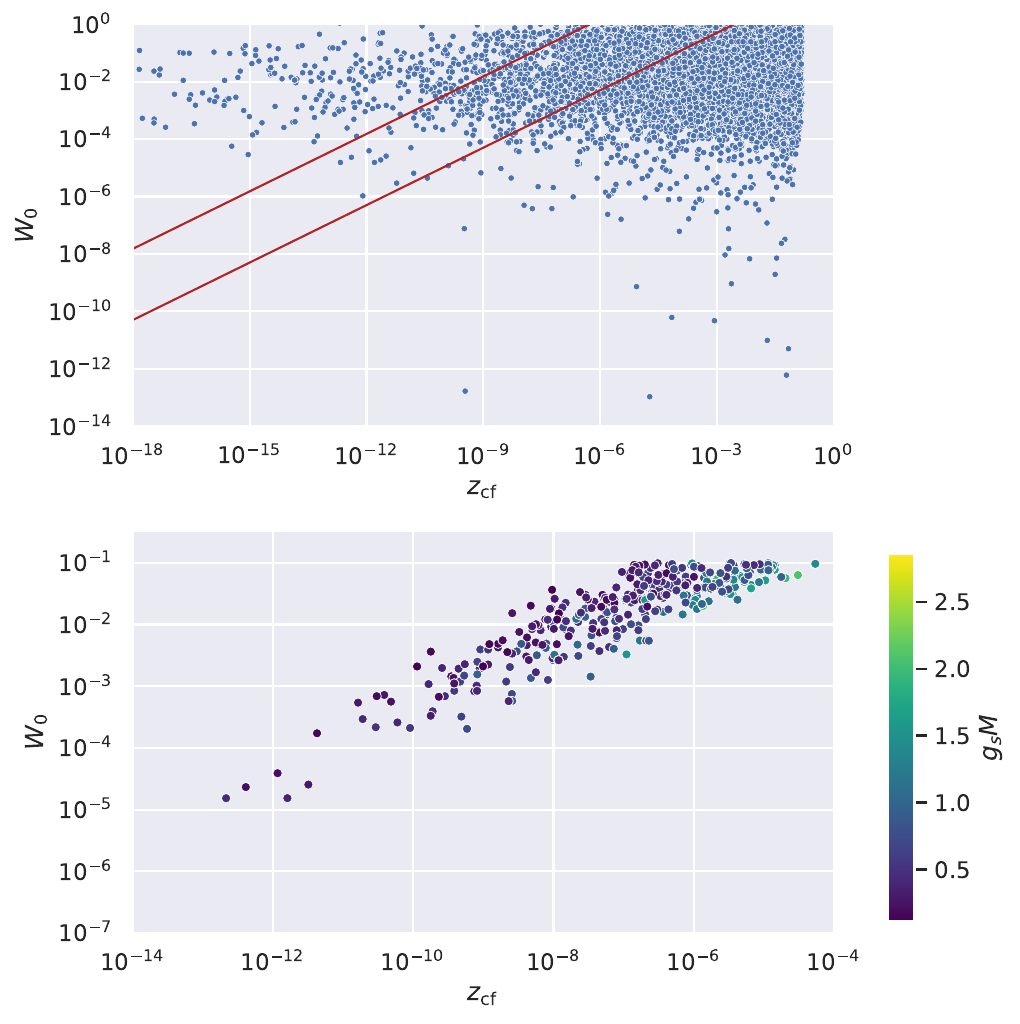}
\caption{Scatter plot of $W_0$ vs.~conifold modulus $z_{\text{cf}}$ for 
$33{,}371$
anti-D3-brane PFVs that we found.
The lines in the upper panel show the alignment bounds $0.1 \le \Xi \le 10$, with $\Xi$ defined in \eqref{eq:throat_tuning_maintext}.  
The lower panel zooms in on the 396 points that lie between these bounds and that also obey $g_s<0.4$ and $W_0<0.1$. 
}  
\label{fig:W0_zcf_comb}
\end{figure}

By employing the above algorithm, in  416 Calabi-Yau orientifolds we have enumerated 240{,}480{,}253
flux choices that solve \eqref{eq:flux_ellipsoid2} with $N_{\overline{\text{D3}}}=1$ and $N_{\text{D3}}=0$ with $k_{\text{max}}=500$.
In these we identified 141{,}594{,}222
two-term racetracks, and finally imposed \eqref{eq:flux_ellipsoid} and $M>12$. 
We call a conifold racetrack PFV that obeys these additional constraints an \emph{anti-D3-brane PFV}. 
We have found 33{,}371
anti-D3-brane PFVs: see Figure \ref{fig:W0_zcf_comb}.

\clearpage
 
\subsection{Kähler moduli stabilization and uplift}\label{ss:modstab}

Thus far we have shown how to construct a
Calabi-Yau orientifold $X/\calI$ equipped with
a conifold curve $\calC\subset\Xt$ and a choice of flux vectors $\vec{f},\vec{h}\in H^{3}(X,\bbZ)$ such that a single anti-D3-brane is required by Gauss's law.  

The stabilization of the complex structure moduli and axiodilaton by fluxes was already explained in \S\ref{sss:cpfv}:
we first search for a supersymmetric minimum 
$\tau = \langle \tau \rangle_{\mathrm{PFV}}$, $z^\alpha=p^\alpha \langle \tau \rangle_{\mathrm{PFV}}$ of the PFV effective theory defined in \S\ref{sss:cpfv}, and then use this minimum as a starting point for numerical root-finding.
When successful, this process yields vevs \eqref{eq:promotedvevs} of the complex structure moduli and axiodilaton at which the F-terms of these fields vanish.  Provided that $W_0$, $g_s$ and $\langle z_{\mathrm{cf}}\rangle_{\text{F}}$ are small, the solution is at weak coupling and is exponentially near a conifold.

At this stage, we have supersymmetrically stabilized the complex structure moduli, \emph{without considering the K\"ahler moduli or the anti-D3-brane.}
We now turn to stabilizing the remaining moduli in these compactifications. 
 
The stabilization procedure that we employ is intricate, and so following \cite{KKLT} we begin with an oversimplified model as a guide:
\begin{enumerate}
    \item Omitting the anti-D3-brane potential, find a supersymmetric AdS vacuum at a point $T_{\text{AdS}}$ in the K\"ahler moduli space.
    \item Including the anti-D3-brane potential, use $T_{\text{AdS}}$ from step (1) as a starting point for a numerical search for a de Sitter minimum $T_{\text{dS}}$ of the full potential \eqref{eq:Vfull}.
\end{enumerate}
In practice, in step (2) the complex structure moduli and axiodilaton vevs shift as a result of the uplift, requiring careful treatment that we explain below.

First we stabilize the K\"ahler moduli in a supersymmetric AdS vacuum, omitting the effect of the anti-D3-brane at this stage. Specifically, the expectation values of the moduli in the AdS vacuum\footnote{A common  misconception  about the KKLT scenario is that one first finds a fully consistent, physical, supersymmetric AdS$_4$ vacuum, and then adds (only) an anti-D3-brane to uplift to de Sitter. As stated, this is impossible: an anti-D3-brane carries $-1$ unit of D3-brane charge, so  the D3-brane tadpole differs by one unit between the two configurations, and therefore cannot be exactly cancelled in both.  Nonetheless, the AdS solution, which we sometimes refer to as a \emph{precursor},
is a  useful fiction: in our examples the number of moduli is large and the structure of the potential is intricate, so the starting guesses obtained from the AdS vevs are indispensable.} 
will be found by neglecting the contribution of the anti-D3-brane to the scalar potential and to Gauss's law, and instead simply minimizing the F-term potential as a function of the K\"ahler moduli.

Evaluated on the expectation values of the complex structure moduli and axiodilaton,
the full superpotential \eqref{eq:wlo3}
takes the form 
\begin{equation}\label{eq:wlo3evaluated}
    W = W_0 + \sum_D \sqrt{\frac{2}{\pi}}\frac{1}{4\pi^2}\, e^{-2\pi T_D/c_D}\, .
\end{equation}
We then seek to solve $D_i W = \partial_i W + \calK_i W =0$ 
for all K\"ahler moduli $T_i$, where for $\calK$ 
we use the leading-order expression $\calK_{\text{l.o.}}$ given in \eqref{eq:detailedform2}. 
We proceed in two steps:
\begin{enumerate}[label=\roman*.)]
\item First, following \cite{smallCCs}, we find a point $T_0$ in K\"ahler moduli space where each of the $T_i$ obeys
\begin{equation}
    \label{eq:T0def} \Re{T_i} = \Re{T_i^0} \coloneqq  \frac{c_i}{2\pi}\log{W_0^{-1}}\, .
\end{equation}
A point $t^i\in \calK_X$ satisfying \eqref{eq:T0def} is called a \textit{KKLT point}, and an algorithm for finding KKLT points (when they exist) was given in \cite{smallCCs}. 
While at a KKLT point the F-terms $D_iW$ do not vanish exactly, KKLT points are convenient starting points for numerical searches for genuine $D_i W=0$ points.
\item Starting from $T^0_i$, we look for a vev $T_i = \langle T_i \rangle_{\mathrm{F}}$ in K\"ahler moduli space where $D_iW=0$.
This is done numerically, using Newton's method.
\end{enumerate}

In a configuration without anti-D3-branes, a point
$\langle T_i \rangle_{\mathrm{F}}$ where $D_iW=0$ corresponds to a supersymmetric AdS vacuum with all moduli stabilized, as in \cite{smallCCs}.\footnote{The examples  presented in \cite{smallCCs} did not have Klebanov-Strassler throats, while the examples in \cite{Alvarez-Garcia:2020pxd} and \cite{coniLCS} did not have all K\"ahler moduli stabilized. In the course of the present work, we have obtained the first examples of fully-stabilized AdS vacua \textit{with} Klebanov-Strassler throats: see Appendix \ref{sec:KSInKS}.} 

So far we have omitted the effects of anti-D3-branes, and have found expectation values for the moduli in a supersymmetric AdS vacuum, with vacuum energy $V_F<0$.  The final step is to reintroduce the anti-D3-brane potential $V_{\overline{D3}}$, and
to locate critical points 
of the full scalar potential $V$ in \eqref{eq:Vfull} by solving
\begin{equation}
    \d_i V = 0
\end{equation}
for the K\"ahler moduli, complex structure moduli, and axiodilaton.

The scalar potential is a complicated function of $2h^{1,1}(X)+2h^{2,1}(X)+2$ variables, and so the numerical search requires some guidance.
We use the vevs $\langle T_i \rangle_{\mathrm{F}}$, $\langle \tau \rangle_{\mathrm{F}}$, $\langle z^\alpha \rangle_{\mathrm{F}}$ and $\langle z_{\mathrm{cf}} \rangle_{\mathrm{F}}$ that we obtained previously by solving all the F-terms --- neglecting the contribution of the anti-D3-brane potential ---  as yet another starting point for a search using Newton's method. Specifically, we introduce the anti-D3-brane potential, multiplied by some parameter $\epsilon$ that we adiabatically dial from zero to one. 
For sufficiently small step size, at each step all moduli --- both complex structure and Kahler moduli --- adjust by small amounts and the new minimum can be found. If the Hessian eigenvalues remain positive all the way to $\epsilon=1$ and the vacuum energy turns positive, we have succeeded in finding a candidate de Sitter minimum. We refer to the vevs in the de Sitter vacuum as $\langle T_i\rangle_{\mathrm{dS}}$, etc.

\newpage

\section{Candidate KKLT de Sitter Vacua}\label{sec:examples}

In \S\ref{sec:EFTs}-\S\ref{sec:methods} we defined the leading-order EFT for our class of compactifications, and we explained a procedure for finding de Sitter vacua of this theory through a large-scale computational search.

We now present five
compactifications that contain a total of 30 such vacua.
Key control parameters, for one example from each compactification, are summarized in Table \ref{tab:summary}. All of the data presented in this section  
is publicly available in a dedicated \href{https://github.com/AndreasSchachner/kklt_de_sitter_vacua}{{\color{MidnightBlue}GitHub repository}}, along with a demo notebook that shows how to interpret the data using \texttt{CYTools} \cite{Demirtas:2022hqf}.

\vspace{0.5cm}
\begin{table}[h]
\begin{centering}
\begin{tabular}{|c|c|c|c|c|c|c|c|c|c|}\hline
& & & & & & & & &\\[-1.3em]
ID & $h^{2,1}$ & $h^{1,1}$ & $M$ & $K'$ & $g_s$ & $W_0$ & $g_sM$ & $|z_{\mathrm{cf}}|$  & $V_0$ \\[0.1em]
\hline
\hline
& & & & & & & & &\\[-1.3em]
1 & 8 & 150 & 16 & $\frac{26}{5}$ & 0.0657 & 0.0115 & 1.051 &  2.822$\times10^{-8}$ &  +1.937$\times10^{-19}$ \\[0.1em]\hline 
2 & 8 & 150 & 16 & $\frac{93}{19}$ & 0.0571 & 0.00490 & 0.913 &  7.934$\times10^{-9}$ &  +1.692$\times10^{-20}$ \\[0.1em]\hline
3 & 8 & 150 & 18 & $\frac{40}{11}$ & 0.0442 & 0.0222 & 0.796 &  8.730$\times10^{-8}$ &  +4.983$\times10^{-19}$ \\[0.1em]\hline
4 & 5 & 93 & 20 & $\frac{17}{5}$ & 0.0404 & 0.0539 & 0.808 & 1.965$\times10^{-6}$ &  +2.341$\times10^{-15}$ \\[0.1em]\hline 
5 & 5 & 93 & 16 & $\frac{29}{10}$ & 0.0466 & 0.0304 & 0.746 & 8.703$\times10^{-7}$ &  +2.113$\times10^{-15}$ \\[0.1em]\hline 
\end{tabular}
\caption{Five candidate de Sitter vacua, in the order of their presentation in \S\ref{sec:examples}, and their control parameters. 
}
\label{tab:summary}
\end{centering}
\end{table}

\subsection{Example 1: $h^{1,1}=150$, $h^{2,1}=8$} \label{sec:manwe} 
 
As our first example, we consider the polytope $\De$ whose vertices are given by the columns of the matrix
\begin{equation}\label{eq:manwe_poly}
    \left(\begin{array}{ccccccccc}
    \phantom{-}1& -1& -1& -1& -1& -1& -1& -1& -1\\
    -1&  \phantom{-}2& -1& -1& -1&  \phantom{-}0& -1&  \phantom{-}0&  \phantom{-}0\\ 
    -1&  \phantom{-}1&  \phantom{-}0&  \phantom{-}2&  \phantom{-}2&  \phantom{-}0&  \phantom{-}0&  \phantom{-}0&  \phantom{-}1\\  
    -1&  \phantom{-}1&  \phantom{-}0&  \phantom{-}0&  \phantom{-}1&  \phantom{-}2&  \phantom{-}2&  \phantom{-}1&  \phantom{-}0\\
    \end{array}\right)\,.
\end{equation}
An FRST of $\De$ defines a toric variety whose generic anticanonical hypersurface is a smooth Calabi-Yau threefold $\Xt$ with Hodge numbers $h^{1,1}(\Xt)=8$ and  $h^{2,1}(\Xt)=150$, and similarly an FRST of the polar dual polytope  $\De^\circ$ defines a Calabi-Yau threefold $X$ with the mirror Hodge  numbers $h^{1,1}(X)=150$ and $h^{2,1}(X)=8$. The pair ($\De,\Dec$) is $\Delta$-favorable, $\Dec$-favorable, and trilayer, and in any phase $X$ obtained from an FRST of $\Dec$, there exists a sign-flip orientifold with $h^{1,1}_-(X/\mathcal{I})=h^{2,1}_+(X/\mathcal{I})=0$ \cite{orientifolds}.  
Thus, $\Dec$ meets  
our polytope criteria, and we will compactify on an orientifold of $X$.

We select fluxes in $H^{3}(X,\bbZ)$ that furnish a conifold PFV, as defined in \S\ref{sec:pfv}. There exists a particular FRST $\mathscr{T}$ of $\De$ that
yields a Calabi-Yau threefold $\Xt$ whose Mori cone has as a generator a conifold curve with  $n_{\text{cf}}=2$. Here and in subsequent examples, we choose a basis for $H_2(\Xt,\mathbb{Z})$ in which the conifold curve is represented by $\mathcal{C}_{\text{cf}}=(1,0,\ldots,0)$.
In this basis, the second Chern class of $\Xt$ is
\begin{equation}
     c_2(\Xt) = 
     \begin{pmatrix}  
        -2& -184& -112&  -10&  -10&  -26&   -2&    6
     \end{pmatrix} ^\top\, .
\end{equation}
One readily verifies that the flux vectors
\begin{align}
    \vec{M} &= \begin{pmatrix} 16& 10& -26& 8& 32& 30& 18& 28  \end{pmatrix}^\top\, , \\[0.4em]
    \vec{K}&=\begin{pmatrix}-6& -1& 0& 1& -3& 2& 0& -1\end{pmatrix}^\top\,,
\end{align}
furnish a conifold PFV as defined in \S\ref{sss:cpfv}, with 
\begin{equation}
\vec{p} = \frac{1}{40}\begin{pmatrix}0& -8&  0& -2&  4&  5&  5&  4\end{pmatrix}^\top\,.   
\end{equation}
The conifold flux quanta defined in \eqref{eq:coniMdef}
and \eqref{eq:Kprime_def} are 
\begin{equation}
    M = 16\,  \quad  \text{and }\quad  K' = \frac{26}{5}\,.
\end{equation}
Moreover, we have that
\begin{equation}
    -\vec{M}\cdot\vec{K} = 162 = 4+h^{1,1}(X)+h^{2,1}(X)\,,
\end{equation}
and so  
adding a single anti-D3-brane is necessary to fulfill Gauss's law \eqref{eq:gaussLaw}.

The leading terms in the bulk flux superpotential along the PFV locus are
\begin{equation}\label{eq:manwew}
    W_{\text{PFV}} = \frac{1}{\sqrt{8\pi^5}}\biggl(14\,e^{2\pi i \t \cdot \frac{1}{40}} - 80\,e^{2\pi i\t \cdot \frac{2}{40}} +
    118\,e^{2\pi i\t \cdot \frac{3}{40}} +\ldots
    \biggr)\,.
\end{equation} 
Using the location of the PFV minimum resulting from the first two terms in \eqref{eq:manwew} as the starting point for numerical minimization of $D_{z^a}W_{\text{flux}}$ and $D_{\tau}W_{\text{flux}}$, we find a true F-flat minimum, i.e.,~a solution to $DW_{\text{flux}}=0$ for the full flux superpotential 
\eqref{eq:flux_superpotential},
with parameters  
\begin{equation}
    g_s = 0.0732\, , \ z_{\text{cf}} = 1.390\times10^{-7}\, , \ W_0 = 0.0103\, , \ \text{and}\ g_sM = 1.171\,.
\end{equation}
The precision quoted  
here is sufficient for the remainder of our analysis, but if desired one can obtain extremely precise results for these quantities by keeping GV invariants up to high degree.

Having stabilized the complex structure moduli, we now stabilize the K\"ahler moduli, by considering nonperturbative effects on the 152 rigid prime toric divisors:
\begin{itemize}
    \item 35 rigid prime toric divisors host O7-planes that support $\mathfrak{so}(8)$ stacks, and so contribute gaugino condensate superpotential terms, as in \eqref{eq:wlo3}, with  $c_D=6$.
    \item 117 rigid prime toric divisors host Euclidean D3-branes, and so contribute to the superpotential with  $c_D=1$.  
\end{itemize}
One prime toric divisor hosting a Euclidean D3-brane intersects the conifold, and so as discussed in \S\ref{sec:conisec} its contribution is negligible. Removing this divisor, there are $h^{1,1}(X)+1=151$ nonperturbative superpotential terms, all of which are pure rigid in all phases.\footnote{The polytope $\Delta^{\circ}$ contains two two-faces of genus one, so the computation of $h^{2,1}(\widehat{D})$ in 
\eqref{eq:dpurerigid} is not completely trivial, but using
\cite{Kim:2021lwo} we have verified that all rigid divisors are pure rigid in all phases.} We adopt the Pfaffian normalization \eqref{eq:pfaffvalue} for all 151 contributions.

First we search for a candidate AdS minimum $T_{\text{AdS}}$ for the K\"ahler moduli. We do this using the K\"ahler potential in \eqref{eq:detailedform2}, including both the $\a'^3$ term \eqref{eq:Valphap3} and the worldsheet instanton corrections \eqref{eq:VWSI} to the Calabi-Yau volume \eqref{eq:detailedform3}; we also incorporate the corresponding terms 
in the holomorphic K\"ahler coordinates \eqref{eq:detailedform4}.

We find a single KKLT point inside the torically extended K\"ahler cone of $\Dec$. Using this point as the starting point for a numerical search, we find a unique AdS precursor geometry in a Calabi-Yau $X_{\text{AdS}}$ obtained as an FRST of $\Dec$. The complete $\a'$-corrected string-frame volume $\calV$ at $T_{\text{AdS}}$, defined as the sum of the uncorrected string frame volume of $X_{\text{AdS}}$, defined in \eqref{eq:Vst}, and the shifts due to the $\alpha'^3$ and worldsheet instanton corrections   defined in \eqref{eq:Valphap3} and \eqref{eq:VWSI}, is\footnote{Recall that we use units with $\ell_s=1$ throughout this work.}
\begin{equation}
     \calV = \calV^{(0)} + \de\calV_{\alpha'^3} + \de\calV_{\text{WSI}} = 665.447 - 0.344  -0.578 = 664.525\,.
\end{equation}
The Einstein-frame volume is  
\begin{equation}
    \calV_{\text{E}} = g_s^{-3/2}\calV = 3.304 \times 10^4\,.  
\end{equation}

At this stage we have supersymmetrically stabilized all of the moduli, and can proceed to incorporate supersymmetry breaking. As discussed in \S\ref{ss:modstab}, we iteratively uplift both the complex structure moduli
and the K\"ahler moduli until the two uplifts converge. 

In this example, we find a candidate de Sitter vacuum, with supersymmetry broken in both the complex structure and K\"ahler sectors, with 
vacuum energy 
\begin{equation}
    V_{\text{dS}} = +1.937\times10^{-19} M_\text{pl}^4\,.
\end{equation}
In the de Sitter vacuum, the complex structure parameters are given by  
\begin{subequations}
\begin{align}
    g_s &= 0.0657\,, \label{eq:param1uplift}\\ 
    W_0 &= 0.0115\,, \label{eq:param2uplift}\\ 
    z_{\text{cf}} &= 2.822\times10^{-8}\,, \label{eq:param3uplift}\\
    g_sM &= 1.051 \,. \label{eq:param4uplift}
\end{align}
\end{subequations}

\begin{figure}[!t]
\centering 
\includegraphics[width=.9\linewidth]{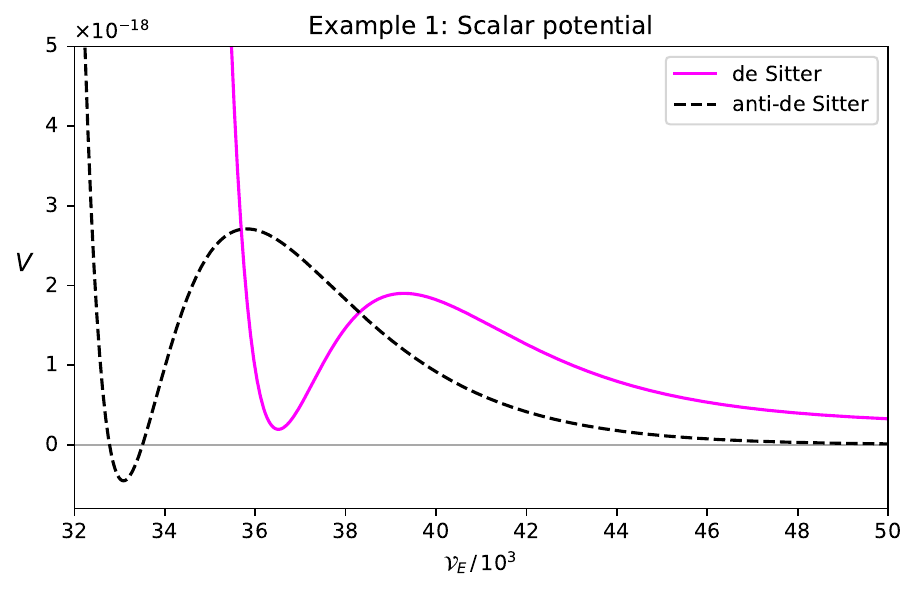}
\caption{Potential for the K\"ahler moduli before and after uplift for the example in \S\ref{sec:manwe}.}
\label{fig:manwe_uplift}
\end{figure} 

At the de Sitter point $T_{\text{dS}}$ in K\"ahler moduli space, the $\a'$-corrected string-frame volume  is
\begin{equation}
     \calV  = \calV^{(0)} + \de\calV_{\alpha'^3} + \de\calV_{\text{WSI}} = 614.834 - 0.344  -0.579 = 613.911\,,
\end{equation}
and the Einstein-frame volume is  
\begin{equation}
    \calV_{\text{E}} = g_s^{-3/2}\calV \approx 3.646 \times 10^4\,. 
\end{equation}
The effect of uplift on the potential for the K\"ahler moduli is shown in Figure \ref{fig:manwe_uplift}.\footnote{The existence of a  de Sitter  critical point in the  AdS potential is surprising from the point of view of the single-modulus toy example in \cite{KKLT}; nevertheless, it appears to be a generic feature of F-term potentials at large $h^{1,1}$. Such features
can be seen in a  single-modulus toy example in which the 
nonperturbative superpotential has two terms with opposite-sign
coefficients.}

\begin{figure}[!t]
\centering
\includegraphics[width=0.9\linewidth]{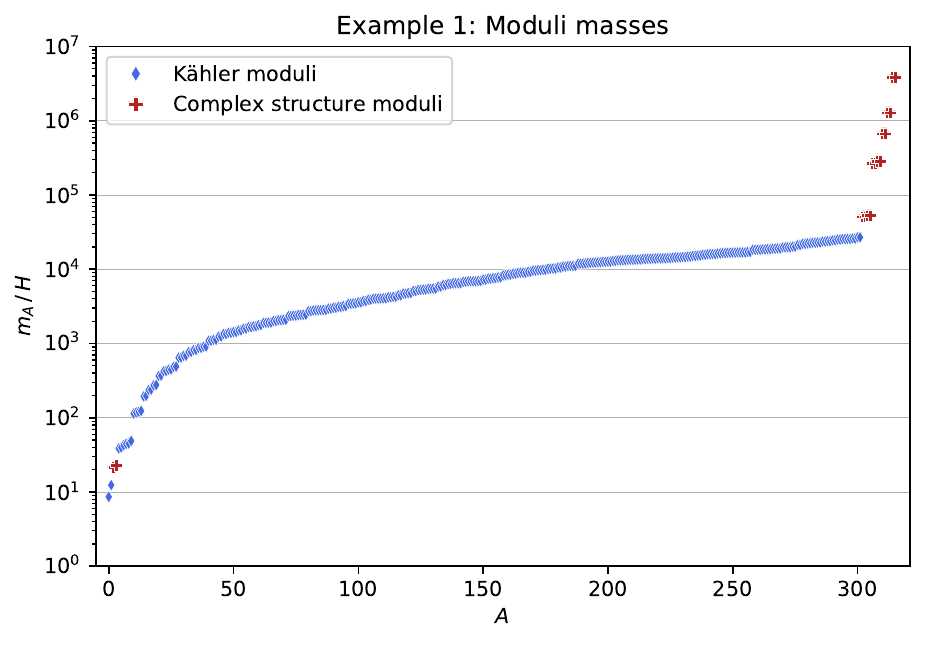}
\caption{The masses of the K\"ahler moduli, bulk complex structure moduli, and axiodilaton in the de Sitter vacuum of \S\ref{sec:manwe}.
The warped Kaluza-Klein scale \eqref{eq:mwkk} is $m_{wKK} \approx 10^4 H_{\text{dS}}$.} 
\label{fig:manwe_masses}
\end{figure}

\begin{figure}[!t]
\centering
\includegraphics[width=\linewidth]{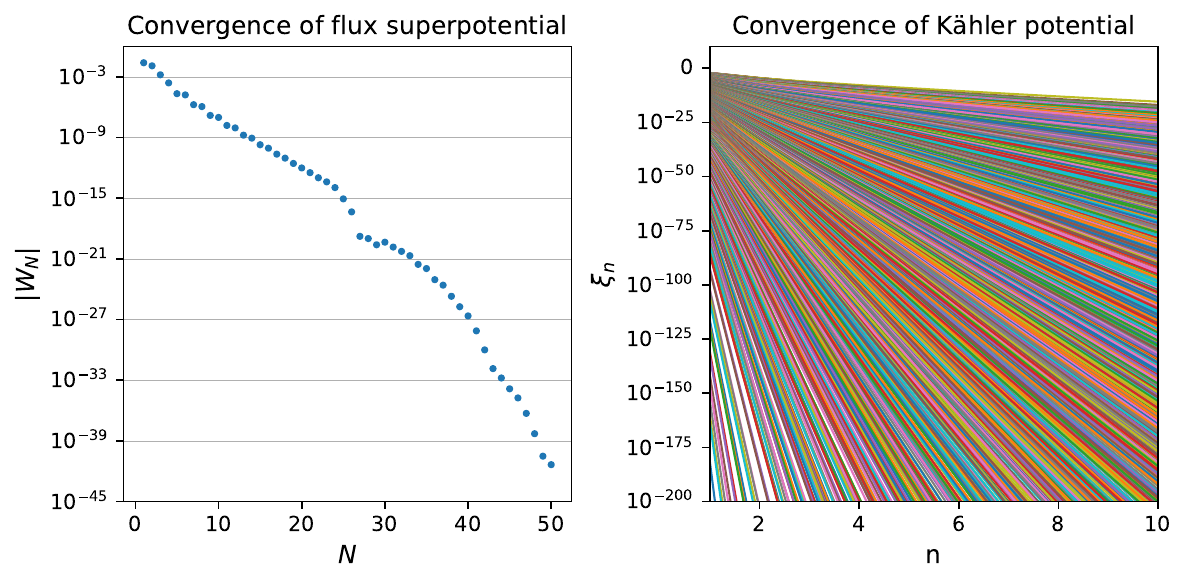}
\caption{Convergence test for the example in \S\ref{sec:manwe}. \emph{Left:} The convergence of the superpotential \eqref{eq:Widef} for the bulk complex structure moduli in the de Sitter vacuum. 
\emph{Right:} The convergence of the worldsheet instantons associated to $2{,}643$ potent rays spanning a $144$-dimensional subcone of $\mathcal{M}(X)$.}\label{fig:manwe_W_rainbow}
\end{figure}

All scalar modes have positive masses in this vacuum: the lightest mode, a K\"ahler modulus, has mass
\begin{equation}
    m_{\text{min}} = 8.616 H_{\text{dS}}\,.
\end{equation} where $ H_{\text{dS}}$ is the Hubble scale,
\begin{equation}
    H_{\text{dS}} = \sqrt{\tfrac{1}{3}V_{\text{dS}}} = 2.5 \times 10^{-10} M_\text{pl}\,.
\end{equation}
The mass spectra of the K\"ahler moduli, bulk\footnote{The conifold modulus $z_{\text{cf}}$ is not shown, as it gets replaced by the redshifted tower of modes localized near the infrared end of the throat, with masses of order the warped Kaluza-Klein scale $m_{wKK}$ in \eqref{eq:mwkk}.} complex structure moduli, and axiodilaton are shown in Figure \ref{fig:manwe_masses}.

\vspace{2mm}
We now check the consistency of the truncations made in finding the above 
solution.  

First, the flux superpotential involves a series of corrections from type IIA worldsheet instantons on the mirror $\widetilde{X}$: see
\eqref{eq:Widef}.  Convergence of  
this series 
is demonstrated on the left-hand side of Figure \ref{fig:manwe_W_rainbow}.

Second, despite the fact that the overall volume is quite large at $T_{\text{dS}}$, the string-frame volumes of various two-cycles are not. As a consequence, convergence of worldsheet instanton corrections to the K\"ahler potential and to the holomorphic coordinates, cf. \eqref{eq:detailedform3} and \eqref{eq:detailedform4}, is not guaranteed \emph{a priori}. 

In order to test whether the $\alpha'$ expansion is under good control in this example, we have sampled $2{,}643$ random potent rays --- i.e.,~rays hosting infinite sequences of nonzero GV invariants, cf.~\cite{smallCCs} --- inside low-dimensional faces of $\mathcal{M}(X)$.  
The rays we obtained turn out to span a $144$-dimensional subcone. 
For every potent ray $\{n\mathcal{C}\,|\, n\in\mathbb{Z}_{+}\}$ generated by a curve $\mathcal{C} \in \mathcal{M}(X)$ in some class $\mathbf{q}\in \mathcal{M}(X)\cap H_2(X,\mathbb{Z})$, we compute
\begin{equation}\label{eq:xin}
    \xi_n(\mathbf{t},\mathcal{C}) := \bigl |\mathscr{N}_{n\mathbf{q}}\, \mathrm{e}^{-2\pi n\, \mathbf{q}\cdot\mathbf{t}}\bigl |
\end{equation}
to high degree $n$, and plot the result on a log scale in the right panel of Figure~\ref{fig:manwe_W_rainbow}. Importantly, the corrections \eqref{eq:xin} decay exponentially with $n$.

The smallest potent curve $\mathcal{C}_{\text{min}}$ obtained in this way has
\begin{equation}
    \mathrm{Vol}_s(\mathcal{C}_{\text{min}}) \approx 0.971\,,\; \mathscr{N}_{\mathcal{C}_{\text{min}}} = 3\,,\; \mathscr{N}_{\mathcal{C}_{\text{min}}}\dfrac{\mathrm{Li}_2\bigl (\mathrm{e}^{-2\pi\mathrm{Vol}_s(\mathcal{C}_{\text{min}})}\bigl )}{(2\pi)^2} \approx 1.705\times 10^{-4} \, ,
\end{equation} 
where the final expression  
sets the size of the contribution of $\mathcal{C}_{\text{min}}$ to the holomorphic coordinates \eqref{eq:detailedform4} (see \S\ref{sec:rainbow}).
We conclude that the worldsheet instanton corrections associated with potent rays are indeed negligible in this vacuum.  On the other hand, some corrections from curves that shrink at finite distance in moduli space lead to non-negligible corrections to \eqref{eq:detailedform4}, but these effects are already fully incorporated in our expressions (at string tree level, cf.~the discussion in \S\ref{sec:known}).
 
Finally, we check that string loop corrections and Euclidean D(-1)-brane
and Euclidean D1-brane corrections that are inherited from the $\mathcal{N}=2$ parent model \cite{Robles-Llana:2006hby,Robles-Llana:2007bbv}, and contribute to $\delta\mathcal{K}_{\mathcal{N}=2}^{(g_s)}$
defined in \eqref{eq:deltaKN2}, are small. The leading correction comes from Euclidean D1-branes wrapping the smallest curves.  We find that this contribution is suppressed in comparison to the corresponding worldsheet instanton correction that we have incorporated in 
$\delta\mathcal{K}_{\mathcal{N}=2}^{\text{tree}}$  
by a factor
$\epsilon_{g_s}^{\mathcal{N}=2}\approx 0.005$: see \S\ref{sec:stringloop}.
 
In summary, we have found, in the effective theory defined by the approximations given in \S\ref{sec:summarysusyeft} and \S\ref{sec:KS}, a de Sitter vacuum in an explicit flux compactification on a Calabi-Yau orientifold. 

\newpage

\subsection{Example 2: $h^{1,1}=150$, $h^{2,1}=8$} \label{sec:lorien}

We again consider the polytope $\De$ defined in \eqref{eq:manwe_poly} and its dual $\Dec$, but now we consider an alternative FRST thereof, specified in the \href{https://github.com/AndreasSchachner/kklt_de_sitter_vacua}{GitHub repository}, and the associated Calabi-Yau $\widetilde{X}$.
The logic of \cite{Gendler:2023ujl} suffices to prove that the Calabi-Yau $\widetilde{X}$ considered here is distinct from that of \S\ref{sec:manwe}.

We find a conifold PFV furnished by the vectors 
\begin{align}
    \vec{M} &= \begin{pmatrix} 16& 9& -5& -32& -42& -4& 1& -9  \end{pmatrix}^\top\, , \\[0.4em]
    \vec{K} &=\begin{pmatrix}-6& -3& -2& 1& 1& 0& -2& -3\end{pmatrix}^\top\,,\\[0.4em]
    \vec{p} &= \begin{pmatrix}0& -5&  3& -14&  -13&  -2&  -3&  -4\end{pmatrix}^\top\,\times \frac{1}{38}\,,
\end{align}
with  
\begin{equation}
    M = 16\, \quad \text{and } \quad K' = \frac{93}{19}\,.
\end{equation} 
As before, we have that 
\begin{equation}
    -\vec{M}\cdot\vec{K}=162=4+h^{1,1}(X)+h^{2,1}(X),
\end{equation}
and so a single anti-D3-brane must be added to satisfy Gauss's law.

With these flux quanta, the superpotential for the bulk complex structure moduli along the PFV locus is 
\begin{equation}
    W_{\text{PFV}} = \frac{1}{\sqrt{8\pi^5}}\biggl(10\,e^{2\pi i \t \cdot \frac{1}{38}} - 104\,e^{2\pi i\t \cdot \frac{2}{38}} +4 e^{2\pi i\t \cdot \frac{3}{38}} + \ldots \Bigr)\,.
\end{equation} 
Using this superpotential to furnish a starting guess, we find a supersymmetric minimum for the complex structure moduli with parameters 
\begin{equation}
    g_s = 0.0595\, , \ z_{\text{cf}} = 1.591\times10^{-8}\, , \ W_0 = 0.00465\, , \ \text{and}\ g_sM = 0.953\,.
\end{equation}
As in \S\ref{sec:manwe}, there are 151 rigid divisors that do not intersect the conifold, all of which are pure rigid in every phase.
We find a single KKLT point, which gives rise to an AdS precursor with corrected Einstein frame volume
\begin{equation}
    \calV_E = 4.256\times 10^4\,.  
\end{equation}

\begin{figure}[!t]
\centering
\includegraphics[width=\linewidth]{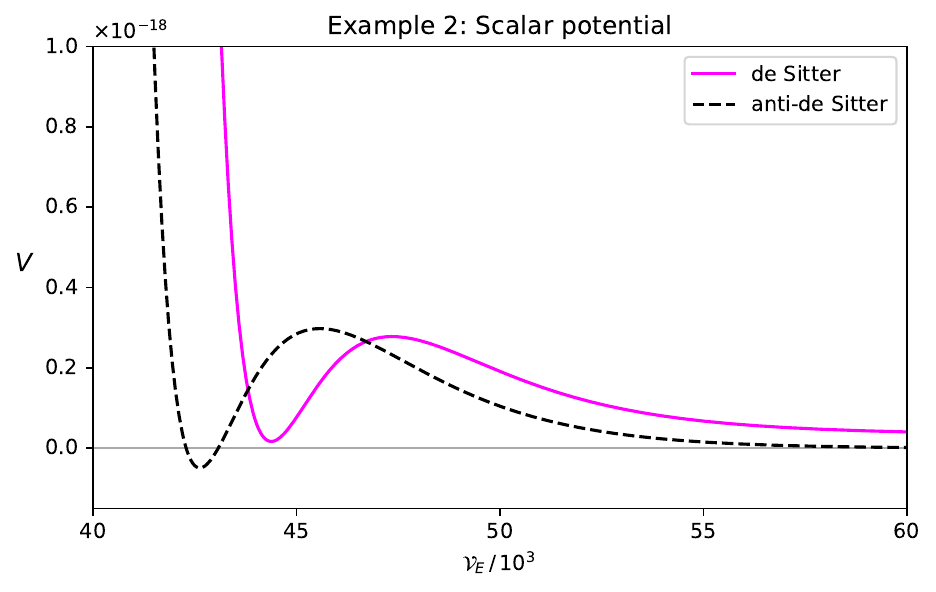}
\caption{Potential for the K\"ahler moduli before and after uplift for the example in \S\ref{sec:lorien}.}\label{fig:lorien_uplift}
\end{figure}

Incorporating the uplift potential of the anti-D3-brane, we find a
candidate de Sitter vacuum, with supersymmetry broken in both the complex structure and K\"ahler sections, and
with vacuum energy 
\begin{equation}
    V_{\text{dS}} = +1.692\times10^{-20}\, M_{\text{pl}}^4\,.
\end{equation}
The complex structure parameters in the de Sitter vacuum are
\begin{equation}
    g_s = 0.0571\, , \ z_{\text{cf}} = 7.934\times10^{-9}\, , \ W_0 = 0.00490\, , \ \text{and } \ g_sM = 0.913\,.
\end{equation} 
The fully-corrected Einstein frame volume in the de Sitter vacuum is  
\begin{equation}
    \calV_E = 4.431 \times 10^4\,. 
\end{equation}
The effect of uplift on the scalar potential for the K\"ahler moduli is shown in Figure \ref{fig:lorien_uplift}.

\begin{figure}[!t]
\centering
\includegraphics[width=.98\linewidth]{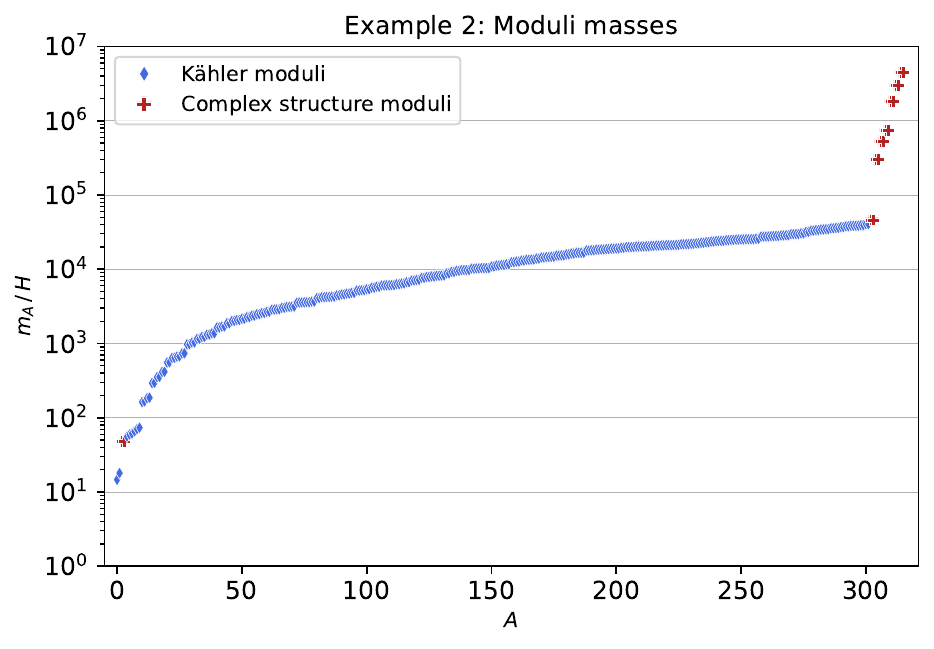}
\caption{The masses of the K\"ahler moduli, bulk complex structure moduli, and axiodilaton in the de Sitter vacuum discussed in  \S\ref{sec:lorien}.}\label{fig:lorien_masses}
\end{figure}

\begin{figure}[!t]
\centering
\includegraphics[width=\linewidth]{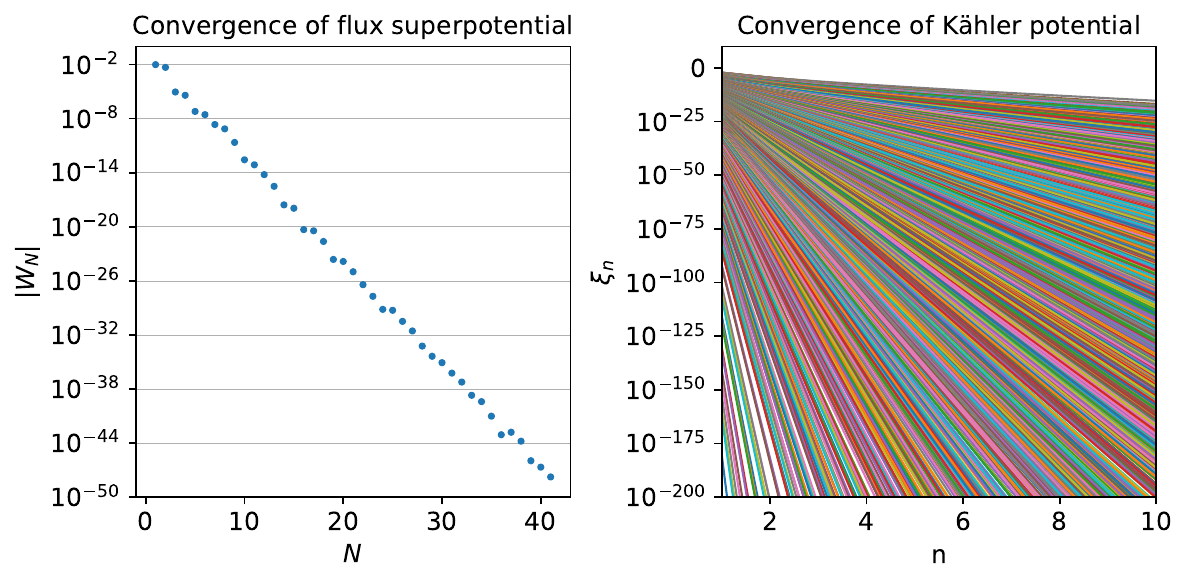}
\caption{Convergence test for the example in \S\ref{sec:lorien}. \emph{Left:} The convergence of the superpotential for the bulk complex structure moduli in the de Sitter vacuum. \emph{Right:} The convergence of the worldsheet instantons associated to 2{,}643 potent rays spanning a $144$-dimensional subcone of $\mathcal{M}(X)$.}\label{fig:lorien_rainbow}
\end{figure}

The masses of all moduli are positive, and the lightest mass is  
\begin{equation}
    m_{\text{min}} = 14.658\, H_{\text{dS}}\,.
\end{equation}
The mass spectra of the K\"ahler moduli and complex structure moduli are shown in Figure~\ref{fig:lorien_masses}.

Finally, we check for convergence of the worldsheet instanton series, both on $X$ and $\widetilde{X}$.
The terms in the flux superpotential \eqref{eq:Widef} are shown in the left panel of Figure~\ref{fig:lorien_rainbow}.
Turning to the K\"ahler sector, we have found 2{,}643 rays of potent curves, whose charges generate a lattice of rank 144. The convergence of the worldsheet instantons associated to these rays is demonstrated in Figure \ref{fig:lorien_rainbow}. The smallest potent curve $\mathcal{C}_{\text{min}}$ has
\begin{equation}
    \mathrm{Vol}_s(\mathcal{C}_{\text{min}}) \approx 0.966\,, \; \mathscr{N}_{\mathcal{C}_{\text{min}}} = 3\,,\; \mathscr{N}_{\mathcal{C}_{\text{min}}}\dfrac{\mathrm{Li}_2\bigl (\mathrm{e}^{-2\pi\mathrm{Vol}_s(\mathcal{C}_{\text{min}})}\bigl )}{(2\pi)^2} \approx 1.759\times 10^{-4} \, .
\end{equation}

\FloatBarrier

\newpage
\subsection{Example 3: $h^{1,1}=150$, $h^{2,1}=8$} \label{sec:tulkas}

We again consider the polytope $\De$ defined in \eqref{eq:manwe_poly} and its dual $\Dec$, but now we consider yet another alternative FRST thereof, specified in the \href{https://github.com/AndreasSchachner/kklt_de_sitter_vacua}{GitHub repository}, and the associated Calabi-Yau $\widetilde{X}$.
The logic of \cite{Gendler:2023ujl} once again suffices to prove that the Calabi-Yau $\widetilde{X}$ considered here is distinct from those of \S\ref{sec:manwe} and \S\ref{sec:lorien}.

We find a conifold PFV furnished by the vectors 
\begin{align}
    \vec{M} &= \begin{pmatrix} 18& 11& -19& -12& -2& -1-& 13& 5  \end{pmatrix}^\top\, , \\[0.4em]
    \vec{K} &=\begin{pmatrix}-5& -3& -1& 1& -1& 3& -1& -1\end{pmatrix}^\top\,,\\[0.4em]
    \vec{p} &= \begin{pmatrix}0& -11&  8& 2&  14&  -1&  2&  -8\end{pmatrix}^\top\,\times \frac{1}{44}\,,
\end{align}  
with  
\begin{equation}
    M = 18\, \quad \text{and } \quad K' = \frac{40}{11}\,.
\end{equation} 
As before, we have that 
\begin{equation}
    -\vec{M}\cdot\vec{K}=162=4+h^{1,1}(X)+h^{2,1}(X),
\end{equation}
and so a single anti-D3-brane must be added to satisfy Gauss's law.

With these flux quanta, the superpotential for the bulk complex structure moduli along the PFV locus is 
\begin{equation}
    W_{\text{PFV}} = \frac{1}{\sqrt{8\pi^5}}\biggl(-60\,e^{2\pi i \t \cdot \frac{1}{38}} + 736\,e^{2\pi i\t \cdot \frac{2}{38}} +2180 e^{2\pi i\t \cdot \frac{3}{38}} + \ldots \Bigr)\,.
\end{equation} 
Using this superpotential to furnish a starting guess, we find a supersymmetric minimum for the complex structure moduli with parameters 
before uplift
\begin{equation}
    g_s = 0.0450\, , \ z_{\text{cf}} = 1.121\times10^{-7}\, , \ W_0 = 0.0217\, , \ \text{and}\ g_sM = 0.811\,.
\end{equation}
As in \S\ref{sec:manwe}, there are 151 rigid divisors that do not intersect the conifold, all of which are pure rigid in every phase.
We find a single KKLT point, which gives rise to an AdS precursor with corrected Einstein frame volume
\begin{equation}
    \calV_E = 5.016\times 10^4\,.  
\end{equation}

\begin{figure}[!t]
\centering
\includegraphics[width=\linewidth]{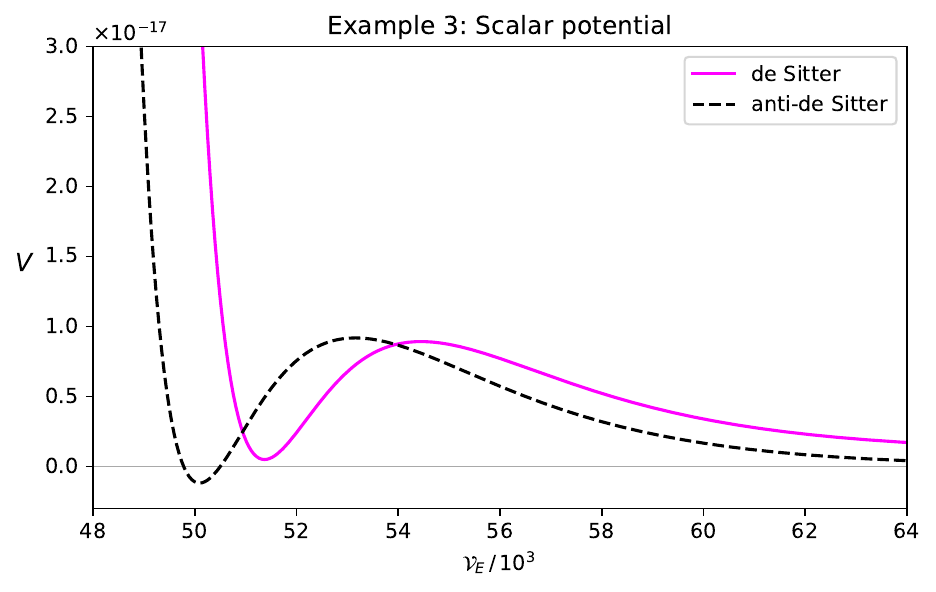}
\caption{Potential for the K\"ahler moduli before and after uplift for the example in \S\ref{sec:tulkas}.}\label{fig:tulkas_uplift}
\end{figure}

Incorporating the uplift potential of the anti-D3-brane, we find a
candidate de Sitter vacuum, with supersymmetry broken in both the complex structure and K\"ahler sections, and
with vacuum energy 
\begin{equation}
    V_{\text{dS}} = +4.983\times10^{-19}\, M_{\text{pl}}^4\,.
\end{equation}
The complex structure parameters in the de Sitter vacuum are
\begin{equation}
    g_s = 0.0442\, , \ z_{\text{cf}} = 8.730\times10^{-8}\, , \ W_0 = 0.0222\, , \ \text{and } \ g_sM = 0.796\,.
\end{equation}
The fully-corrected Einstein frame volume in the de Sitter vacuum is  
\begin{equation}
    \calV_E = 5.127 \times 10^4\,. 
\end{equation}
The effect of uplift on the scalar potential for the K\"ahler moduli is shown in Figure \ref{fig:tulkas_uplift}.

\begin{figure}[!t]
\centering
\includegraphics[width=.98\linewidth]{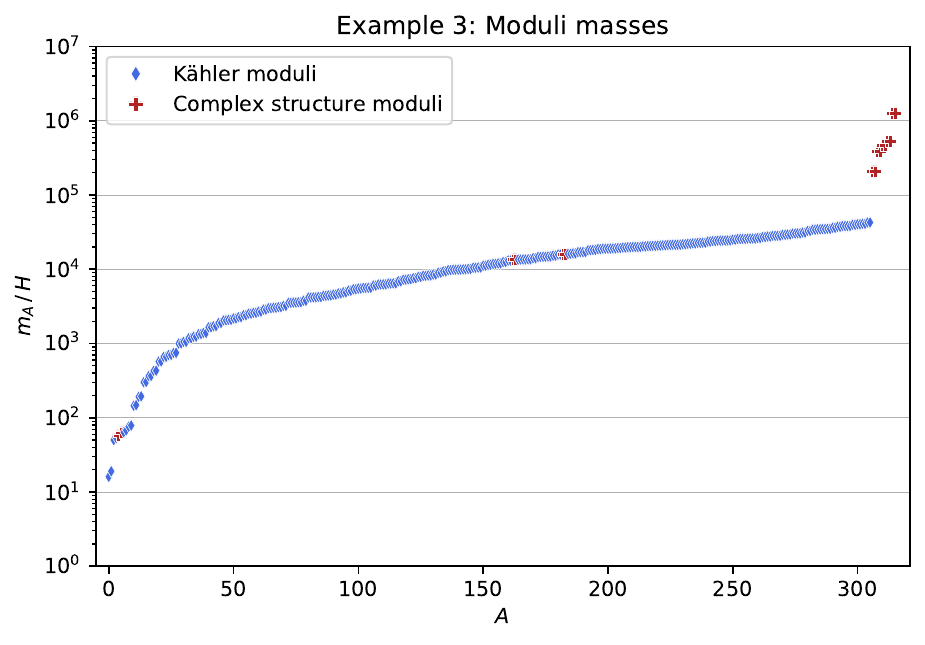}
\caption{The masses of the K\"ahler moduli, bulk complex structure moduli, and axiodilaton in the de Sitter vacuum discussed in  \S\ref{sec:tulkas}.}\label{fig:tulkas_masses}
\end{figure}

\begin{figure}[!t]
\centering
\includegraphics[width=\linewidth]{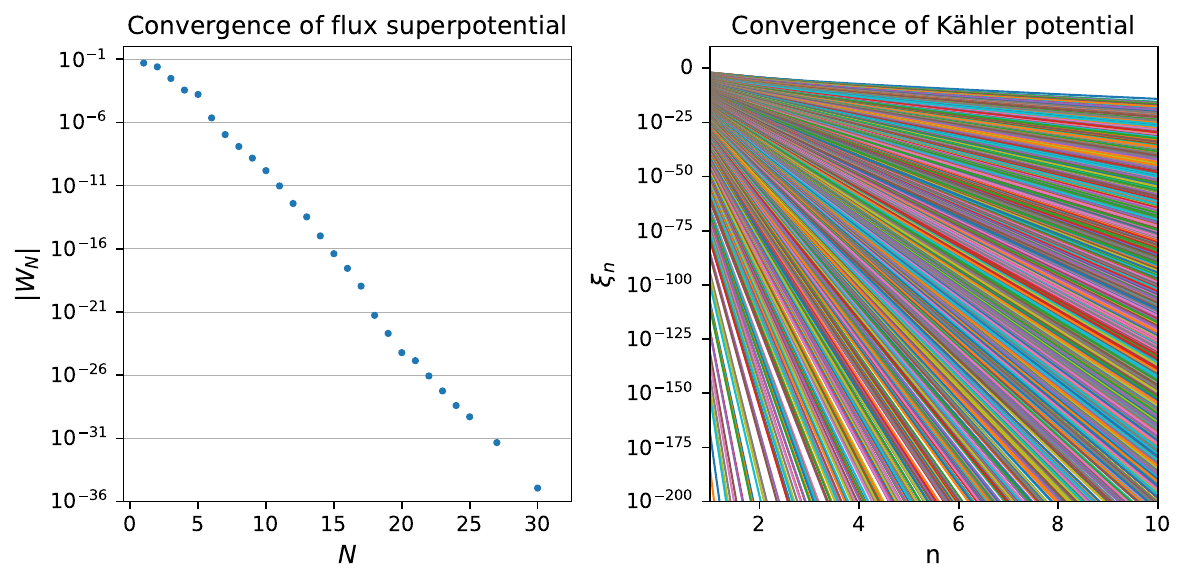}
\caption{Convergence test for the example in \S\ref{sec:tulkas}. \emph{Left:} The convergence of the superpotential for the bulk complex structure moduli in the de Sitter vacuum. \emph{Right:} The convergence of the worldsheet instantons associated to 2{,}643 potent rays spanning a $144$-dimensional subcone of $\mathcal{M}(X)$.}\label{fig:tulkas_rainbow}
\end{figure}

The masses of all moduli are positive, and the lightest mass is  
\begin{equation}
    m_{\text{min}} = 16.030\, H_{\text{dS}}\,.
\end{equation}
The mass spectra of the K\"ahler moduli and complex structure moduli are shown in Figure~\ref{fig:tulkas_masses}.

Finally, we check for convergence of the worldsheet instanton series, both on $X$ and $\widetilde{X}$.
The terms in the flux superpotential \eqref{eq:Widef} are shown in the left panel of Figure~\ref{fig:tulkas_rainbow}.
Turning to the K\"ahler sector, we have found 2{,}643 rays of potent curves, whose charges generate a lattice of rank 144. The convergence of the worldsheet instantons associated to these rays is demonstrated in Figure \ref{fig:tulkas_rainbow}. The smallest potent curve $\mathcal{C}_{\text{min}}$ has
\begin{equation}
    \mathrm{Vol}_s(\mathcal{C}_{\text{min}}) \approx 0.922\,,\;  \mathscr{N}_{\mathcal{C}_{\text{min}}} = 3\,,\;  \mathscr{N}_{\mathcal{C}_{\text{min}}}\dfrac{\mathrm{Li}_2\bigl (\mathrm{e}^{-2\pi\mathrm{Vol}_s(\mathcal{C}_{\text{min}})}\bigl )}{(2\pi)^2} \approx 2.318\times 10^{-4} \, .
\end{equation}

\FloatBarrier

\newpage
\subsection{Example 4: $h^{1,1}=93$, $h^{2,1}=5$} \label{sec:aule}

We consider the polytope $\De$ whose vertices are given by the columns of the matrix
\begin{equation}\label{eq:aule_poly}
    \left(\begin{array}{ccccccccc}
    \phantom{-}1& -1& -1& -1& -1& -1& -1& -1& -1\\
    -1&  \phantom{-}1& \phantom{-}0& \phantom{-}0& \phantom{-}0&  \phantom{-}1& \phantom{-}1&  \phantom{-}2&  \phantom{-}1\\ 
    -1&  \phantom{-}1&  \phantom{-}0&  \phantom{-}0&  \phantom{-}1&  \phantom{-}0&  \phantom{-}2&  \phantom{-}1&  \phantom{-}1\\  
    -1&  \phantom{-}0&  \phantom{-}0&  \phantom{-}1&  \phantom{-}0&  \phantom{-}1&  \phantom{-}1&  \phantom{-}2&  \phantom{-}2\\
    \end{array}\right)\,.
\end{equation}
A fine regular star triangulation of $\De$ defines a toric variety whose generic anticanonical hypersurface is a smooth Calabi-Yau threefold $\Xt$ with Hodge numbers $h^{1,1}(\Xt)=5$ and  $h^{2,1}(\Xt)=93$, and similarly an FRST of the polar dual polytope  $\De^\circ$ defines a Calabi-Yau threefold $X$ with the mirror Hodge  numbers $h^{1,1}(X)=93$ and $h^{2,1}(X)=5$. The pair ($\De,\Dec$) is $\Delta$-favorable, $\Dec$-favorable,
and trilayer, and in any phase $X$ obtained from an FRST of $\Dec$,
there exists a sign-flip orientifold with $h^{1,1}_-(X/\mathcal{I})=h^{2,1}_+(X/\mathcal{I})=0$ \cite{orientifolds}.  Of the $h^{1,1}(X)+4=97$ prime toric divisors of $\Dec$,
$h^{1,1}(X)+3=96$ are rigid. 
Thus, $\Dec$ meets all of our polytope criteria, and we will compactify on an orientifold of $X$.

There exists a particular FRST $\mathscr{T}$ of $\De$, whose data is specified in the \href{https://github.com/AndreasSchachner/kklt_de_sitter_vacua}{GitHub repository}, that
yields a Calabi-Yau threefold $\Xt$ whose Mori cone has as a generator a conifold curve with  $n_{\text{cf}}=2$. 
We find a conifold PFV furnished by the vectors 
\begin{align}
    \vec{M} &= \begin{pmatrix}  20&   4&   8& -18& -20  \end{pmatrix}^\top\, , \\[0.4em]
    \vec{K} &=\begin{pmatrix}-5& -1&  0&  1& -1\end{pmatrix}^\top\,,\\[0.4em]
    \vec{p} &= \begin{pmatrix}0& 1 & 2 & 1 &0\end{pmatrix}^\top\,\times \frac{1}{48}\, ,
\end{align}
with 
\begin{equation}
    M = 20\, \quad \ \text{and }\ \quad K' = \frac{17}{5}\,.
\end{equation}
Moreover, we have that
\begin{equation}
    -\vec{M}\cdot\vec{K} = 102 = 4+h^{1,1}(X)+h^{2,1}(X)\,,
\end{equation}
and so  
adding a single anti-D3-brane is necessary to fulfill Gauss's law \eqref{eq:gaussLaw}. 

With these flux quanta, the superpotential for the bulk complex structure moduli along the PFV locus is 
\begin{equation}\label{eq:aule_w} 
    W_{\text{PFV}} = \frac{1}{\sqrt{8\pi^5}}\biggl(108\,e^{2\pi i \t \cdot \frac{1}{48}} - 1120\,e^{2\pi i\t \cdot \frac{2}{48}} +
    60\,e^{2\pi i\t \cdot \frac{3}{48}} +\ldots
    \biggr)\,.
\end{equation} 
Using this superpotential to furnish a starting guess, we find a supersymmetric minimum for the complex structure moduli with parameters 
\begin{equation}
    g_s = 0.0410\, , \ z_{\text{cf}} = 2.369\times10^{-6}\, , \ W_0 = 0.0525\, , \ \text{and}\ g_sM = 0.821\,.
\end{equation}

There are 95 rigid divisors that do not intersect the conifold, all of which are pure rigid in all phases.
Unlike the examples in \S\ref{sec:manwe} and \S\ref{sec:lorien}, where we found a single KKLT point and correspondingly a single AdS precursor, in this geometry we find 36 KKLT points, 
which
give rise to 29 AdS precursors. 

We will for now focus on a single AdS precursor, which has fully corrected Einstein frame volume 
\begin{equation}
    \calV_E = 1.310 \times 10^4\,.   
\end{equation}

\begin{figure}[!t]
\centering
\includegraphics[width=\linewidth]{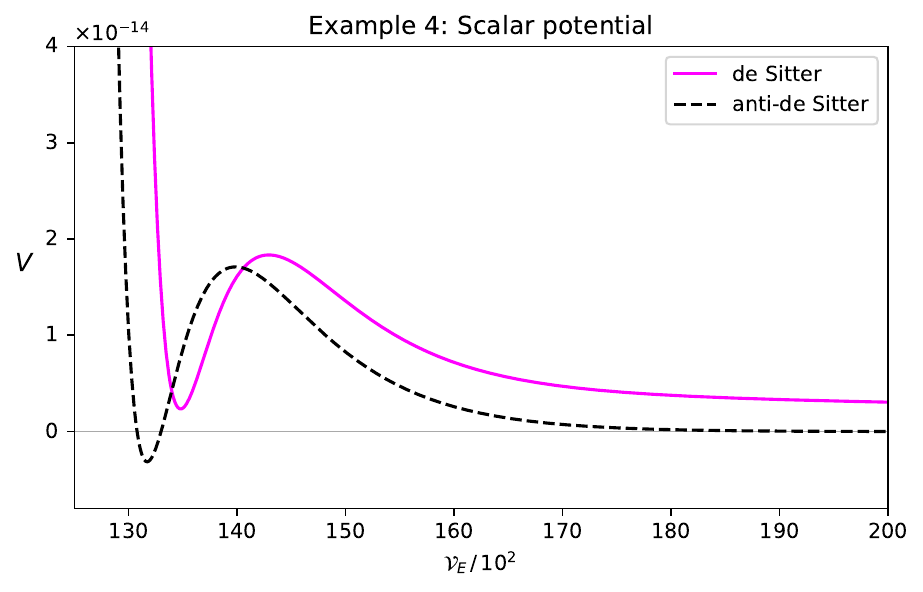}
\caption{Potential for the K\"ahler moduli before and after uplift for the example in \S\ref{sec:aule}.}\label{fig:aule_uplift}
\end{figure}

\begin{figure}[!t]
\centering
\includegraphics[width=\linewidth]{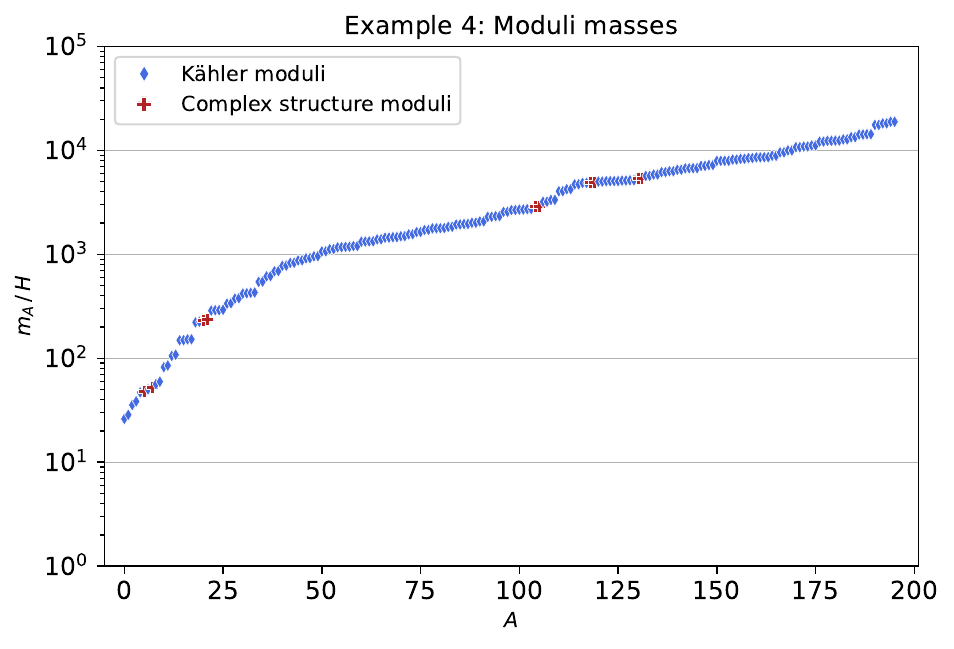}
\caption{The masses of the K\"ahler moduli, bulk complex structure moduli, and axiodilaton in the de Sitter vacuum discussed in \S\ref{sec:aule}.}\label{fig:aule_masses}
\end{figure}

Incorporating supersymmetry breaking from the anti-D3-brane, we find a nonsupersymmetric minimum for the complex structure moduli with parameters 
\begin{equation}
    g_s = 0.0404\, , \ z_{\text{cf}} = 1.965\times10^{-6}\, , \ W_0 = 0.0539\, , \ \text{and } \ g_sM = 0.808\,.
\end{equation}
We similarly find a nonsupersymmetric minimum $T_{\text{dS}}$ for the K\"ahler moduli.
After uplift, we end up with a de Sitter vacuum with vacuum energy 
\begin{equation}
    V_{\text{dS}} = +2.341\times10^{-15}\, M_{\text{pl}}^4\,.
\end{equation}
The fully corrected Einstein frame volume at $T_{\text{dS}}$ is
\begin{equation}
    \calV_E = 1.340 \times 10^4\,. 
\end{equation}
The effect of uplift on the scalar potential for the K\"ahler moduli is shown in Figure \ref{fig:aule_uplift}.
The vacuum is tachyon-free, and the smallest mass is
\begin{equation}
    m_{\text{min}} = 26.157\, H_{\text{dS}}\,.
\end{equation}
The full spectra of masses for the K\"ahler moduli and complex structure moduli are shown in Figure~\ref{fig:aule_masses}.

\begin{figure}[!t]
\centering
\includegraphics[width=\linewidth]{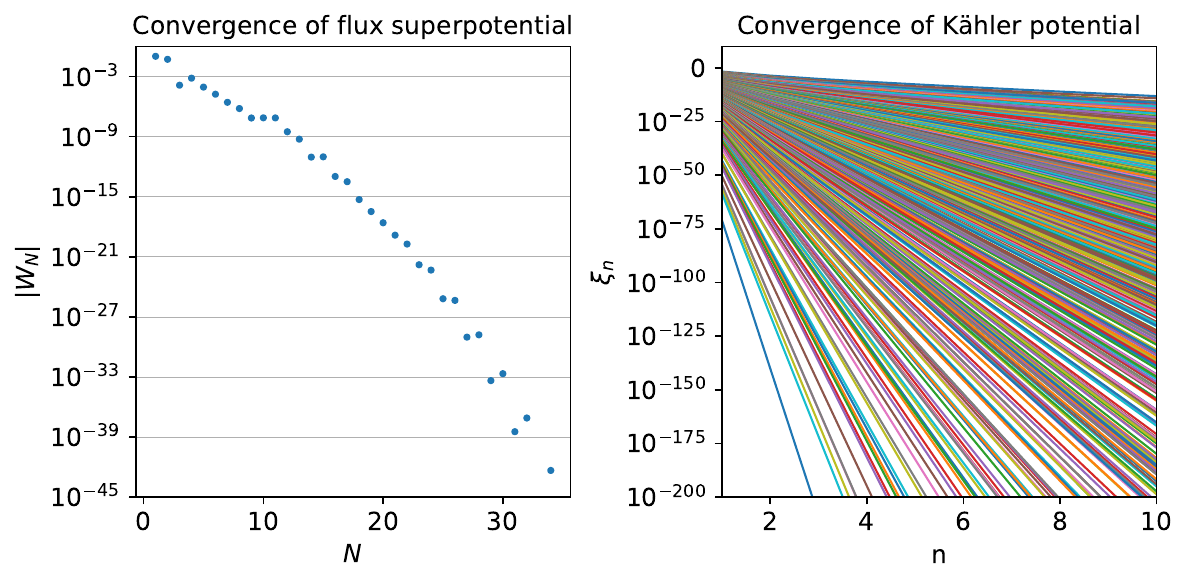}
\caption{Convergence test for the example in \S\ref{sec:aule}. \emph{Left:} The convergence of the superpotential for the bulk complex structure moduli in the de Sitter vacuum. \emph{Right:} The convergence of the worldsheet instantons associated to 1,201 potent rays spanning a $93$-dimensional subcone of $\mathcal{M}(X)$.}\label{fig:aule_rainbow}
\end{figure}

Finally, we check for convergence of the worldsheet instanton series, both on $X$ and $\widetilde{X}$.
The terms in the flux superpotential \eqref{eq:Widef} are shown in the left panel of Figure~\ref{fig:aule_rainbow}.
Turning to the K\"ahler sector, we have found 1,201 rays of potent curves, whose charges generate a lattice of rank 93. The convergence of the worldsheet instantons associated to these rays is demonstrated in Figure \ref{fig:aule_rainbow}.
The smallest potent curve $\mathcal{C}_{\text{min}}$ has
\begin{equation}
    \mathrm{Vol}_s(\mathcal{C}_{\text{min}}) \approx 0.795\,,\;  \mathscr{N}_{\mathcal{C}_{\text{min}}} = 3\,,\; \mathscr{N}_{\mathcal{C}_{\text{min}}}\dfrac{\mathrm{Li}_2\bigl (\mathrm{e}^{-2\pi\mathrm{Vol}_s(\mathcal{C}_{\text{min}})}\bigl )}{(2\pi)^2} \approx 5.168\times 10^{-4} \, .
\end{equation}

\begin{figure}[!t]
\centering
\includegraphics[width=\linewidth]{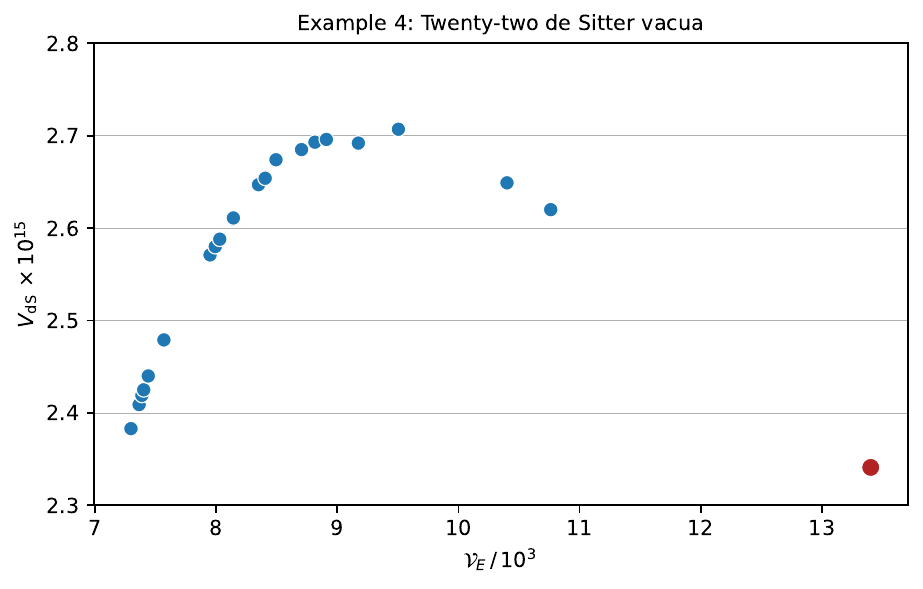}
\caption{Twenty-one other de Sitter vacua (blue dots) obtained in the same moduli space as, and from the same choice of flux quanta as, the de Sitter vacuum presented in full detail in \S\ref{sec:aule} (red dot).} \label{fig:aule_others}
\end{figure}

So far we have focused on the uplift that results from a single AdS precursor, but as mentioned above there are 28 others. From these 28 precursors, we find 21 additional de Sitter vacua, which are shown in Figure \ref{fig:aule_others}.

\newpage

\FloatBarrier
\subsection{Example 5: $h^{1,1}=93$, $h^{2,1}=5$} \label{sec:orome}

We consider the polytope $\De$ whose vertices are given by the columns of the matrix
\begin{equation}\label{eq:orome_poly}
    \left(\begin{array}{ccccccccc}
    \phantom{-}1& -1& -1& -1& -1& -1& -1& -1& -1\\
    -1&  \phantom{-}0& \phantom{-}0& \phantom{-}0& \phantom{-}0&  \phantom{-}0& \phantom{-}1&  \phantom{-}2&  \phantom{-}2\\ 
    -1&  \phantom{-}0&  \phantom{-}0&  \phantom{-}1&  \phantom{-}1&  \phantom{-}0&  \phantom{-}1&  \phantom{-}1&  \phantom{-}2\\  
    -1&  \phantom{-}0&  \phantom{-}1&  \phantom{-}0&  \phantom{-}1&  \phantom{-}1&  \phantom{-}0&  \phantom{-}2&  \phantom{-}1\\
    \end{array}\right)\,.
\end{equation}
A fine regular star triangulation of $\De$ defines a toric variety whose generic anticanonical hypersurface is a smooth Calabi-Yau threefold $\Xt$ with Hodge numbers $h^{1,1}(\Xt)=5$ and  $h^{2,1}(\Xt)=93$, and similarly an FRST of the polar dual polytope  $\De^\circ$ defines a Calabi-Yau threefold $X$ with the mirror Hodge  numbers $h^{1,1}(X)=93$ and $h^{2,1}(X)=5$. The pair ($\De,\Dec$) is $\Delta$-favorable, $\Dec$-favorable,
and trilayer, and in any phase $X$ obtained from an FRST of $\Dec$,
there exists a sign-flip orientifold with $h^{1,1}_-(X/\mathcal{I})=h^{2,1}_+(X/\mathcal{I})=0$ \cite{orientifolds}.  Of the $h^{1,1}(X)+4=97$ prime toric divisors of $\Dec$,
$h^{1,1}(X)+3=96$ are rigid. 
Thus, $\Dec$ meets all of our polytope criteria, and we will compactify on an orientifold of $X$.

There exists a particular FRST $\mathscr{T}$ of $\De$, whose data is specified in the \href{https://github.com/AndreasSchachner/kklt_de_sitter_vacua}{GitHub repository}, that
yields a Calabi-Yau threefold $\Xt$ whose Mori cone has as a generator a conifold curve with  $n_{\text{cf}}=2$. 
We find a conifold PFV furnished by the vectors 
\begin{align}
    \vec{M} &= \begin{pmatrix}  16&   12&   -10& -30& -14  \end{pmatrix}^\top\, , \\[0.4em]
    \vec{K} &= \begin{pmatrix}-4& 4&  1&  3& -1\end{pmatrix}^\top\,, \\[0.4em]
    \vec{p} &= \begin{pmatrix}0& 1 & 1 & 0 &5\end{pmatrix}^\top\, \times \frac{1}{40}\, , 
\end{align}
with 
\begin{equation}
    M = 16\, \quad \ \text{and }\ \quad  K' = \frac{29}{10}\,.
\end{equation}
Moreover, we have that
\begin{equation}
    -\vec{M}\cdot\vec{K} = 102 = 4+h^{1,1}(X)+h^{2,1}(X)\,,
\end{equation}
and so  
adding a single anti-D3-brane is necessary to fulfill Gauss's law \eqref{eq:gaussLaw}. 

With these flux quanta, the superpotential for the bulk complex structure moduli along the PFV locus is 
\begin{equation}\label{eq:orome_w} 
    W_{\text{PFV}} = \frac{1}{\sqrt{8\pi^5}}\biggl(-88\,e^{2\pi i \t \cdot \frac{1}{40}} + 1024\,e^{2\pi i\t \cdot \frac{2}{40}} +
    1368\,e^{2\pi i\t \cdot \frac{3}{40}} +\ldots
    \biggr)\,.
\end{equation} 
Using this superpotential to furnish a starting guess, we find a supersymmetric minimum for the complex structure moduli, with parameters 
\begin{equation}
    g_s = 0.0486\, , \ z_{\text{cf}} = 1.468\times10^{-6}\, , \ W_0 = 0.0291\, , \ \text{and}\ g_sM = 0.778\,.
\end{equation}

There are 95 rigid divisors that do not intersect the conifold, all of which are pure rigid in all phases.  In this geometry we find 59 KKLT points. 
We will for now focus on one of these, which leads to a single AdS precursor with fully corrected Einstein frame volume 
\begin{equation}
    \calV_E = 7.804 \times 10^3\,.  
\end{equation}

\begin{figure}[!t]
\centering
\includegraphics[width=\linewidth]{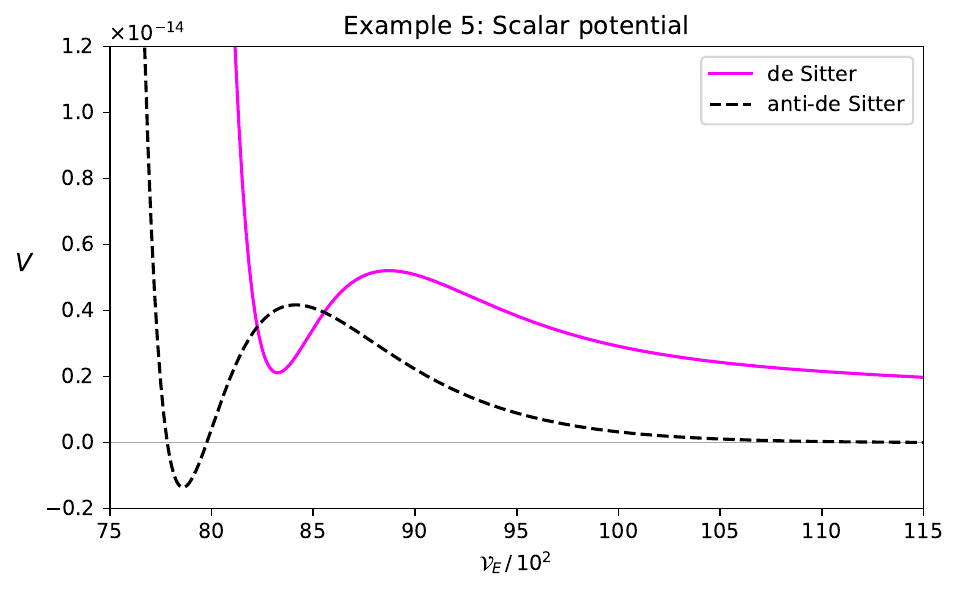}
\caption{Potential for the K\"ahler moduli before and after uplift for the example in \S\ref{sec:orome}.}\label{fig:orome_uplift}
\end{figure}

Incorporating supersymmetry breaking from the anti-D3-brane, we find a nonsupersymmetric minimum for the complex structure moduli with parameters 
\begin{equation}
    g_s = 0.0466\, , \ z_{\text{cf}} = 8.703\times10^{-7}\, , \ W_0 = 0.0304\, , \ \text{and } \ g_sM = 0.746\,.
\end{equation}
We similarly find a nonsupersymmetric minimum $T_{\text{dS}}$ for the K\"ahler moduli.
The fully corrected Einstein frame volume at $T_{\text{dS}}$ is
\begin{equation}
    \calV_E = 8.266 \times 10^3\,.  
\end{equation}
The effect of uplift on the scalar potential for the K\"ahler moduli is shown in Figure \ref{fig:orome_uplift}.

\begin{figure}[!t]
\centering
\includegraphics[width=\linewidth]{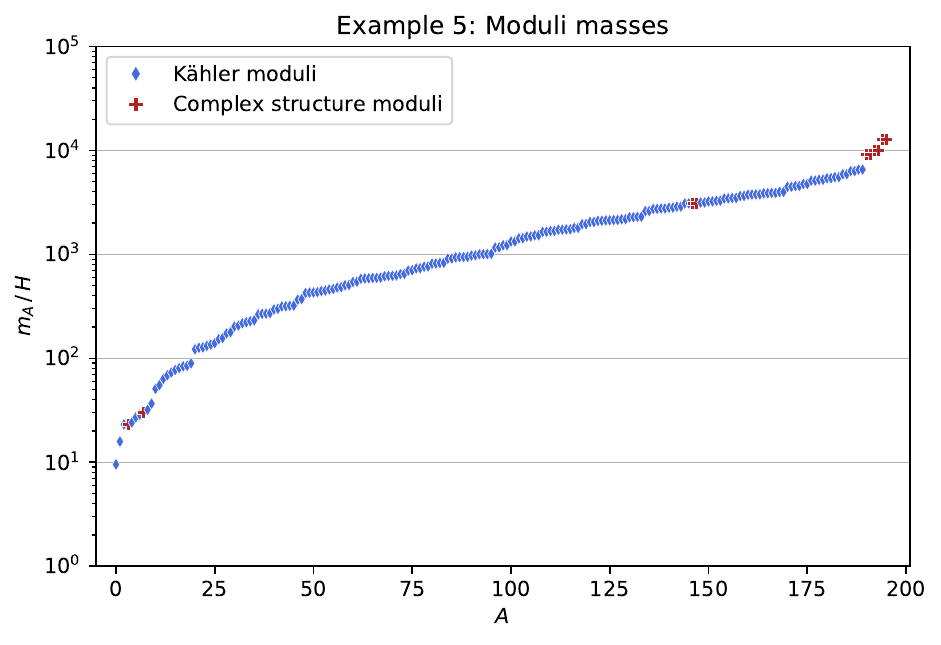}
\caption{The masses of the K\"ahler moduli, bulk complex structure moduli, and axiodilaton in the de Sitter vacuum discussed in  \S\ref{sec:orome}.}\label{fig:orome_masses}
\end{figure}

After uplift, we end up with a de Sitter vacuum with vacuum energy 
\begin{equation}
    V_{\text{dS}} = +2.113\times10^{-15}\, M_{\text{pl}}^4\,.
\end{equation}
The vacuum is tachyon-free, and the smallest mass is given in Hubble units by 
\begin{equation}
    m_{\text{min}} = 9.539\, H_{\text{dS}}\,.
\end{equation}
The full spectra of masses for the K\"ahler moduli and complex structure moduli are shown in Figure~\ref{fig:orome_masses}.

\begin{figure}[!t]
\centering
\includegraphics[width=\linewidth]{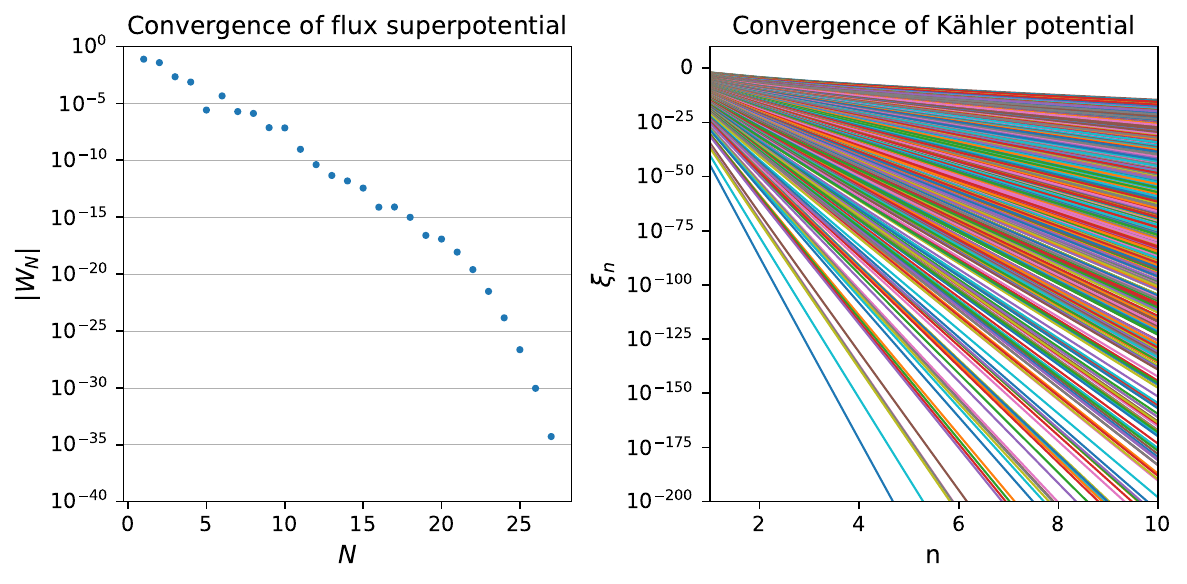}
\caption{Convergence test for the example in \S\ref{sec:orome}. \emph{Left:} The convergence of the superpotential for the bulk complex structure moduli in the de Sitter vacuum. \emph{Right:} The convergence of the worldsheet instantons associated to 1{,}181 potent rays spanning a $93$-dimensional subcone of $\mathcal{M}(X)$.}\label{fig:orome_rainbow}
\end{figure}

Finally, we check for convergence of the worldsheet instanton series, both on $X$ and $\widetilde{X}$.
The terms in the flux superpotential \eqref{eq:Widef} are shown in the left panel of Figure~\ref{fig:orome_rainbow}.
Turning to the K\"ahler sector, we have found 1{,}181 rays of potent curves, whose charges generate a lattice of rank 93. The convergence of the worldsheet instantons associated to these rays is demonstrated in Figure \ref{fig:orome_rainbow}.
The smallest potent curve $\mathcal{C}_{\text{min}}$ has
\begin{equation}
    \mathrm{Vol}_s(\mathcal{C}_{\text{min}}) \approx 0.891\,,\;  \mathscr{N}_{\mathcal{C}_{\text{min}}} = 3\,,\; \mathscr{N}_{\mathcal{C}_{\text{min}}}\dfrac{\mathrm{Li}_2\bigl (\mathrm{e}^{-2\pi\mathrm{Vol}_s(\mathcal{C}_{\text{min}})}\bigl )}{(2\pi)^2} \approx 2.809\times 10^{-4} \, .
\end{equation}

\begin{figure}[!t]
\centering
\includegraphics[width=0.95\linewidth]{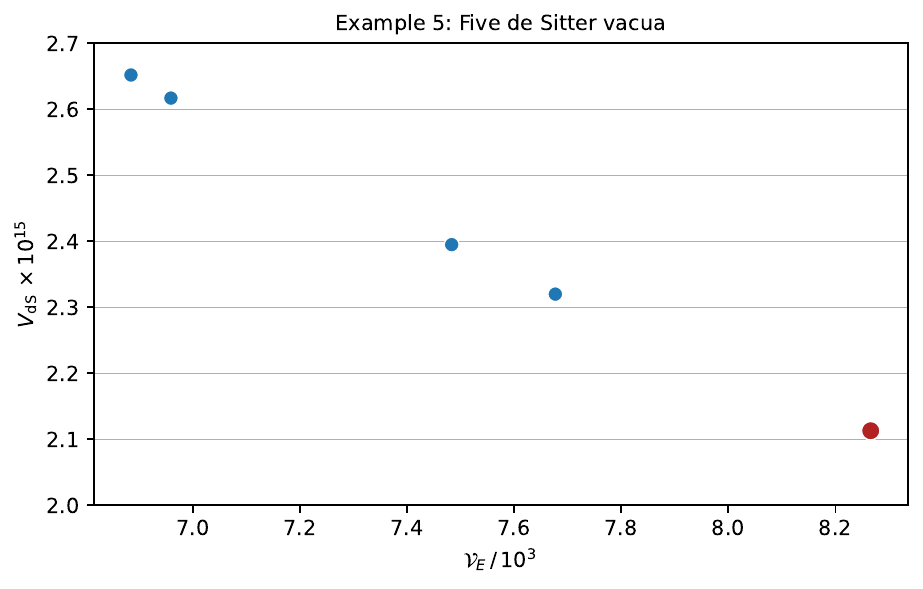}
\caption{Four other de Sitter vacua (blue dots) obtained in the same moduli space as, and from the same choice of flux quanta as, the de Sitter vacuum presented in full detail in \S\ref{sec:orome} (red dot).}\label{fig:orome_others} \ 
\end{figure}

So far we have focused on the uplift that results from a single AdS precursor. 
Upon considering the other precursors in this geometry, we find four other de Sitter vacua, which are shown in Figure \ref{fig:orome_others}. 
 
\section{Outlook}\label{sec:outlook} 

This work has prepared the foundations for a systematic study of KKLT de Sitter vacua.

Despite considerable effort, the candidate vacua constructed here have some limitations. We have neglected the effects of certain unknown corrections, including string loop corrections to the K\"ahler potential, $\a'$ corrections to the potential of an anti-D3-brane in a Klebanov-Strassler throat, and perturbations to the throat sourced by seven-branes in the bulk.
None of these corrections to the effective theory are parametrically large (see Appendix \ref{app:corr}), but neither are they small enough to be immediately negligible.

In the course of this work we studied in great detail, and on a large scale, how the vacuum structure in our solutions is modified by including \emph{known} corrections, particularly $\alpha'$ corrections to the bulk theory (cf.~\S\ref{sec:EFTs}).  While including such effects 
in a given example often led to destabilization, in an ensemble of solutions some were unaffected, and others were rendered more stable.
We anticipate a similar result from including the effects that we have enumerated above, but cannot yet compute: some candidate vacua will surely be spoiled, but some may survive as genuine de Sitter vacua of string theory. Put another way,  
while any particular vacuum might be destabilized 
by the unknown corrections, in a large enough ensemble of candidates we expect at least some to survive.

A further issue concerns flux quantization.
In certain toroidal orientifolds, odd integer flux quanta are consistent only in the presence of exotic O-planes \cite{Frey:2002hf}.
Whether the flux quantization conditions in Calabi-Yau orientifolds similarly restrict 
odd flux quanta is not known, and
resolving this question is a important target for future research. 
In any event, the construction of candidate de Sitter vacua with purely even flux quanta and control parameters comparable to those in \S\ref{sec:examples} appears to be merely a computational task, albeit a formidable one, and we see no principled obstacles: see Appendix \ref{sec:KSInKS}.

A number of technical advances in understanding compactifications of string theory would put our solutions on more secure footing. These include computing $\alpha'$ corrections to the potential of an anti-D3-brane in a Klebanov-Strassler throat, as initiated in \cite{Junghans:2022exo,Junghans:2022kxg,Hebecker:2022zme,Schreyer:2022len,Schreyer:2024pml}; determining the Pfaffian normalization $n_{D}$ in \eqref{eq:K0} by evaluating the sum over the string spectrum that was set up in \cite{Alexandrov:2022mmy}; computing string loop corrections in Calabi-Yau orientifolds; obtaining the Calabi-Yau metric numerically and propagating perturbations due to gaugino condensation across the bulk of the compactification; and characterizing the quantization conditions for three-form flux in Calabi-Yau orientifolds.  If all of these could be achieved, 
our methods could yield fully controlled KKLT de Sitter vacua, rather than what we have termed candidate vacua.

Even without fundamental developments in string theory, one could make some amount of progress through large-scale computation.
The restriction of our search to $h^{2,1} \le 8$, which we found necessary to limit the computational cost of this initial survey, removes the overwhelming majority of the search space for flux vacua in toric hypersurface Calabi-Yau threefolds.  Thus, the vacua we find here barely scratch the surface of what may exist in the Kreuzer-Skarke landscape, and there is much to discover in the future. In particular, there exist polytopes in the Kreuzer-Skarke list with much larger $Q_{\text{O}}$ than those we analyzed here; we expect these to be particularly rich targets for future study.

Even so, the computational effort required for this work was already nontrivial.  After adapting a host of efficient algorithms developed in \cite{smallCCs,Demirtas:2022hqf,CMS}, and inventing many new methods,
approximately 50 core-years of computation time were required to construct more than 100 million flux vacua. 
Deeper searches at larger $h^{2,1}$ would require novel algorithms, as well as computational resources beyond the scale of small clusters.

Moreover, some limitations are intrinsic to the construction, and cannot be remedied by statistics.    
We explained in \S\ref{sec:KS} that the minimum attainable $W_0$ in our setting 
is bounded in terms of the other parameters by \eqref{eq:cookedness},
which at strong 't Hooft coupling is a severe and inflexible constraint. 
But so far we have not come close to saturating the bound  \eqref{eq:cookedness}:
for the vacua constructed here it requires 
$W_0 \gtrsim 10^{-9}$, and we have reached only
$W_0 \sim 10^{-2}$. 
The reason for this gap is that most of the conifold PFVs we construct are far from saturating
$Q_{\text{flux}}^{\text{throat}} \approx Q_{\text{O}}$, as demonstrated by Figure \ref{fig:2histograms}.
 
In any event, the modest values of $W_0$ found in this work are not, in themselves, an insuperable obstacle to theoretical control.  Indeed, in the example of \S\ref{sec:aule} we found $W_0 \approx 0.05$, and yet this example passed all available tests of control of the $\alpha'$ expansion for the bulk supersymmetric theory, and moreover the string coupling $g_s \approx 0.04$ is comfortably small.  In contrast, $\alpha'$ corrections in the throat region, which are controlled by $1/(g_s M)$, are a risk in all our examples. 

With this limitation in mind, let us speculate on what could be found in future searches.
We expect that the Kreuzer-Skarke landscape contains large numbers of candidate de Sitter vacua akin to those we have constructed.  Some of these may well have $g_s M \gtrsim 5$, but  not $\gtrsim 15$.  Thus, $\alpha'$ corrections to the anti-D3-brane potential are a particularly acute risk.  Computing these corrections in full detail is therefore a crucial task for the future.
Even more valuable would be an understanding of dynamical supersymmetry breaking in the Klebanov-Strassler theory at weak 't Hooft coupling, or indeed in any other sector that can arise in these compactifications 
(see e.g.~\cite{Aganagic:2007py}).   
In our search we discovered many flux vacua with \emph{exponentially} small $W_0$ in which
$g_s M \ll 1$, but  
in such cases we have no method for supersymmetry breaking in hand (see Appendix \ref{sec:KSInKS}).

\section{Conclusions}\label{sec:conclusion}

To search for de Sitter vacua of string theory,
we constructed more than 100 million flux compactifications of type IIB string theory on Calabi-Yau orientifolds.
We found 33{,}371 
cases that contain an anti-D3-brane and a Klebanov-Strassler throat in which the anti-D3-brane is metastable 
according to the criterion of
\cite{KPV}.
Finally, we identified five compactifications in which
the potential for the complex structure moduli, axiodilaton, and K\"ahler moduli has a local minimum with positive vacuum energy, as foreseen in \cite{KKLT}.

The minima we have found are de Sitter vacua of the effective theory in which we performed our analysis.  They are \emph{candidate} de Sitter vacua of string theory because the effective theory makes approximations: in the closed string sector we worked at string tree level but to all orders in $\alpha'$, while for the anti-D3-brane potential we worked at leading order in both expansions. 

The vacuum structure in our models is controlled by instantons: Euclidean D3-brane contributions to the superpotential, worldsheet instanton corrections to the K\"ahler potential, and worldsheet instantons of type IIA on the mirror threefold.
We computed each of these series explicitly and then performed extensive tests of the validity of the truncations we made (\S\ref{sec:inlo}).  
The only undetermined parameters in the leading-order effective theory were Pfaffian numbers, which we marginalized over.

To definitively establish whether the solutions we have presented are indeed de Sitter vacua of string theory,
advances in several areas will be essential.  
One needs a computation of string loop corrections to the K\"ahler potential and K\"ahler coordinates in Calabi-Yau orientifolds \cite{Berg:2005ja,Kim:2023sfs,Kim:2023eut}, and of the warped metric in this setting.
The most urgent need is for a more complete understanding of $\alpha'$ corrections to the potential of an anti-D3-brane in a warped throat \cite{Junghans:2022exo,Junghans:2022kxg,Hebecker:2022zme,Schreyer:2022len,Schreyer:2024pml}. 
The string field theory framework for flux compactifications developed by Cho and Kim 
\cite{Cho:2023mhw} is a promising path for this computation.
 
We offer the compactifications we have constructed 
as a testing ground for addressing the many questions that remain, and hope they will serve as a foundation for 
better understanding the vacuum structure in cosmological solutions of string theory.

\section*{Acknowledgements}
 
We thank Naomi Gendler, Arthur Hebecker, Shamit Kachru, Miguel Montero, Natalie Paquette, Simon Schreyer, Eva Silverstein,  Irene Valenzuela, and Gerben Venken for helpful discussions, and we thank Naomi Gendler, Shamit Kachru, and Manki Kim
 for comments on a draft.
We thank   Mehmet Demirtas, Nate MacFadden, and Andres Rios-Tascon for collaboration in related works. We are especially grateful to Manki Kim for performing a calculation that gave independent evidence supporting our treatment of pure rigid divisors,
and to Nate MacFadden
for sharing the code developed in \cite{MacFadden:2023cyf}.
The work of L.M.~and A.S.~was supported in part by NSF grant PHY-2309456.
RAN is supported by a Klarman Fellowship at Cornell University, and thanks the NYU CCPP for their hospitality during the final stages of this work.   

\begin{appendix}
\addtocontents{toc}{\protect\setcounter{tocdepth}{1}}

\section{Control Analysis}
\label{app:corr}

In this Appendix we carefully analyze the candidate de Sitter vacua reported in \S\ref{sec:examples}, enumerating the possible sources of error and explaining what we have done to limit these errors.

Our definition of the leading-order EFT involves truncating infinite series of corrections, 
and we first present strong evidence that the truncations we have chosen are consistent,  
both in the nonperturbative superpotential (\S\ref{sec:autoch}) and in the K\"ahler potential (\S\ref{sec:rainbow}).
We also marginalize over the Pfaffian numbers $\mathcal{A}_D$, finding that our vacua survive in a very wide range (\S\ref{sec:pfaff}). 
Next we turn to corrections \emph{beyond} the leading-order EFT  (\S\ref{app:sscorr}).
By evaluating known corrections inherited from $\mathcal{N}=2$ supersymmetry, as well those obtained from a conservative model of $\mathcal{N}=1$ effects,
we argue that corrections in the string loop expansion  are plausibly well-controlled  (\S\ref{sec:stringloop}).
We then estimate the effects of the leading perturbations of the Klebanov-Strassler throat
(\S\ref{sec:bulkeffects}), 
as well as the effects of warping (\S\ref{ss:warping}).

\subsection{Control in the leading-order EFT}\label{sec:inlo}

\subsubsection{Truncation of the nonperturbative superpotential}\label{sec:autoch}

The nonperturbative superpotential given in \eqref{eq:Wnpgeneral},
\begin{equation}
    W_{\text{np}}(z,\t,T) = \sum_D \calA_D(z,\t)\, e^{-\frac{2\pi}{c_D}T_D}\,,
    \label{eq:wnpagain} 
\end{equation}
involves a generally infinite sum over effective divisors $D$.  
The importance of a nonperturbative correction to \eqref{eq:wnpagain} is set by the  action 
\begin{equation}\label{eq:sdef}
    S(D) \coloneqq  \frac{2\pi}{c_D}\, T_D\,,
\end{equation}
with smaller $S(D)$ indicating more important contributions. 
 
As a practical computational strategy, in each example  we 
found a
candidate minimum $T_{\text{dS}}$ for the K\"ahler moduli by first 
truncating the spectrum of divisors
to include only\footnote{We also omit any divisors that intersect the conifold, as explained in \S\ref{sec:conifoldselection}.} pure rigid prime toric divisors 
in \eqref{eq:wnpagain}.
We denote by 
$S_{\text{retain}}$
the largest pure rigid prime toric divisor action that we retained in this truncation.
Having found a vacuum through the above truncation, we then established its consistency \emph{a posteriori} 
by verifying that at $T_{\text{dS}}$ all other effective divisors that we omitted are in fact negligible.
That is, we computed the \emph{smallest omitted action}
\begin{equation}\label{eq:somit}
S_{\text{omit}}\coloneqq  \text{min}_{D_{\text{eff}}}\,S(D_{\text{eff}})\,,
\end{equation} with the minimum running over all (known) effective divisors that 
we did not retain,
and verified 
that $S_{\text{omit}} \gg S_{\text{retain}}$.

Let us now explain how this consistency check was performed.
The $h^{1,1}+4$ prime toric divisors $D_I$ defined in \eqref{eq:primetoric} generate $H_4(X,\mathbb{Z})$ over the integers, and in particular every effective divisor $D$ 
can be written as
\begin{equation}
D = \sum_I c^I D_I\,, \qquad c^I \in \mathbb{Z}\,. 
\end{equation}
If $c^I \in \mathbb{Z}_{\ge 0}$, then $D$ is inherited from an effective divisor $\mathscr{D} \subset V$ as $D = \mathscr{D} \cap X$, and we call $D$ an inherited effective divisor.  If $D$ is effective but not inherited, i.e.,~if one or more of the $c^I$ is negative, we call $D$ autochthonous. 
 
Whether a prime toric divisor is rigid is determined purely by polytope data, but whether it is pure rigid can in principle depend on a choice of triangulation of the polytope, i.e.~on the particular Calabi-Yau phase.  We refer to divisors that are pure rigid for some triangulations but not for others as phase-dependent pure rigid divisors, and to those that are pure rigid for all triangulations as phase-independent pure rigid.

In summary, effective divisors may be classified as
\begin{itemize}
\item inherited or autochthonous,
\item if inherited, then either prime ($D_I$) or non-prime $\bigl(D = \sum_I c^I D_I\,,~c^I \ge 0\bigr)$,
\item rigid or non-rigid,
\item if rigid, then either pure rigid ($\calA_D=\text{const}$) or non-pure rigid ($\calA_D=\calA_D(z,\t)$),
\item if pure rigid, then either phase-dependent or phase-independent.
\end{itemize}
In all of the examples in \S\ref{sec:examples}, 
every rigid divisor is pure rigid in every phase.

Because there are at most four non-rigid prime toric divisors  
in our examples, it is easy to check that these are negligible,
with  
$S_{\text{min,omit}}^{\text{prime}}/S_{\text{retain}}  \gtrsim 30$.
  
Next, inherited effective divisors that are not prime toric are positive sums of prime toric divisors.
Sorting the pure rigid actions as 
\begin{equation}
    S_{1} \le S_{2} \le \cdots \le S_{\text{retain}}\,,
\end{equation}
we find in our examples that, in particular,
$2S_{1} \sim S_{1}+S_{2} \lesssim S_{\text{retain}}$.  Denoting by $D_i$ the divisor producing the term with action $S_i$, at first sight it seems that Euclidean D3-branes wrapping $2 D_1$ or $D_1+D_2$ could yield significant contributions.
However, as we explained in \S\ref{sec:wnp}, sums of rigid divisors have too many zero modes to contribute to $W$, and so are immaterial for this work, regardless of their volumes. 

Finally, we examine contributions from autochthonous divisors.
No general algorithm for finding all such divisors is available (though see \cite{Gendler:2022ztv} for progress on this question).
However, a special class of autochthonous divisors known as min-face divisors can be computed from polytope data (see \cite{smallCCs,Gendler:2022ztv}). In examples where the effective cone can be computed exactly \cite{Gendler:2022qof}, min-face divisors have been found to be the smallest autochthonous divisors.  In the vacua studied here, we find that the smallest min-face divisors have $S^{\text{aut}}_{\text{min}} \gtrsim 10^3$, so we are extremely well-justified in neglecting the contribution of this class of autochthonous divisors to the superpotential. 
 
In summary, in all the examples presented in \S\ref{sec:examples}, we find 
$S^{\text{aut}}_{\text{min}}  >  S^{\text{prime}}_{\text{min,omit}} = S_{\text{omit}}  \gg S_{\text{retain}}$, and so the 
superpotential contributions of all identifiable effective divisors that we have \emph{not} included are exponentially subleading relative to the terms that we have kept: see Table \ref{tab:pot_rays_auto}.

\begin{table}[H]
\centering
\begin{tabular}{|c|c|c|c|c|c|c|c|c|c|}
\hline 
& & &  &  &  &  & &  & \\[-1.3em]
ID & $S_{\text{min}}^{\text{ED3}}$ & $S_{\text{min}}^{\lambda\lambda}$ & $S_{\text{retain}}$ & $S_{\text{min,omit}}^{\text{prime}}$ & $S_{\text{min}}^{\text{aut}}$ & \# pot.~rays & rank & $\mathrm{Vol}_s(\mathcal{C}_{\text{min}})$ & $\lambda$ \\ [0.2em] 
\hline
\hline
 &  &  &  &  &  & &  & &\\[-1.3em] 
1 & 8.264 & 6.419 & 16.641 & 676 & 5363 &  2643 & 144 & 0.971  & 1.776 \\ [0.2em]  
\hline 
 &  &  & & &  & &  & &\\[-1.3em]
2 & 9.197 & 7.351 & 17.631 & 814 & 6018 &  2643 & 144 & 0.966 &  1.766\\ [0.2em] 
\hline 
 &  &  & & &  & &  & &\\[-1.3em]
3 & 7.768 & 5.922 & 16.324 & 931 & 5013  & 2643 & 144 & 0.922  & 1.686\\ [0.2em] 
\hline 
 &  &  &  & &  & &  & &\\[-1.3em]
4 & 6.050 & 4.122 & 14.278  &  492  &  3875  & 1201 & 93  & 0.795   &  1.660 \\ [0.2em] 
\hline 
 &  &  &  & &  & &  & &\\[-1.3em]
5 & 6.739 & 4.906 & 11.470  & 734 & 1943  &  1181 & 93 &  0.891  &  1.640  \\ [0.2em] 
\hline
\end{tabular} 
\caption{Tests of truncation of the nonperturbative superpotential and the K\"ahler potential.  
We show  
the minimum and maximum actions from pure rigid divisors that we retained,
the smallest action
$S_{\text{min,omit}}^{\text{prime}}$
from prime toric divisors that were not included,
the smallest action $S_{\text{min}}^{\text{aut}}$ of an autochthonous divisor that we found, 
the number and rank of rays of potent curves, the smallest volume of a potent curve, and the parameter $\lambda$ defined in \eqref{eq:lambdadef}.}\label{tab:pot_rays_auto}
\end{table}

\subsubsection{Truncation of the K\"ahler potential}\label{sec:rainbow}

Just as we truncated the sum over divisors $D$ in \eqref{eq:wnpagain}, we must also truncate the sum over curves when evaluating corrections to the K\"ahler potential and the holomorphic K\"ahler coordinates, as defined in \eqref{eq:detailedform2}-\eqref{eq:detailedform4}. 
 
The leading contribution to the effective theory from a curve ${\mathcal{C}}$ is a correction to the K\"ahler coordinates 
\eqref{eq:detailedform4},
\begin{equation}\label{eq:delc}
    \Delta_{\mathcal{C}} \mathcal T_i =
\frac{1}{(2\pi)^2} \, 
 \mathcal{C}_i \mathscr{N}_{\mathcal{C}} \,\text{Li}_2\Bigl((-1)^{\mathbf{\gamma}\cdot \mathbf{\mathcal{C}}}e^{-2\pi \mathbf{\mathcal{C}}\cdot \mathbf{t}}\Bigr)\,.
\end{equation}
The main limitation in including such effects is computational, rather than technical: using the algorithm in \cite{CMS}, in each geometry we can easily obtain the GV invariants of hundreds of thousands of curves, and thus obtain
the resulting corrections \eqref{eq:delc}.
However, evaluating the polylogarithms 
associated to such a large set of curves \emph{at each step} in our lengthy root-finding process is computationally intractable.  
Moreover, most of the above curves are quite large, and so their contributions to the K\"ahler potential and K\"ahler coordinates are negligible.  

As in \cite{smallCCs}, our approach is to 
first directly incorporate the effects of curves that can be realized as the intersection of toric divisors; these 
curves and their GV invariants are described in \S6 of \cite{CMS}. In addition, we include complete intersection curves obtained from a set of birationally-equivalent toric ambient varieties, constructed from neighboring triangulations as well as next-to-nearest neighbors. 
For this we use the methods developed by MacFadden in \cite{MacFadden:2023cyf}.

Then, once we have found a minimum by incorporating the effects of such toric curves in \eqref{eq:detailedform2}-\eqref{eq:detailedform4}, 
we retroactively verify that we were justified in neglecting all other curves by searching for additional small curves using \cite{CMS}, computing the corresponding corrections  \eqref{eq:delc},  and verifying that these are negligible.

Specifically, for each example 
we sampled  
inside low-dimensional faces of $\mathcal{M}(X)$,  and also sampled directly in the full-dimensional cone up to a cutoff volume.  In this way we found nilpotent curves that had not been identified as complete intersection curves in a nearby toric phase, and we then incorporated
the corresponding corrections to the K\"ahler potential and K\"ahler coordinates.

Moreover, from the above search we found potent curves $\mathcal{C}$, which generate potent rays $n\mathcal{C}$, $n\in\mathbb{N}$.
We denote the collection of generating curves in our sample by $\mathcal{S}_{\text{pot}}$.
We record in
Table \ref{tab:pot_rays_auto}  the largest correction
$\Delta_{\mathcal{C}} \mathcal T_i$ from any potent curve $\mathcal{C} \in \mathcal{S}_{\text{pot}}$.

A further diagnostic of control of the $\alpha'$ expansion involves potent rays.
Given a potent ray generated by a curve $\mathcal{C} \in \mathcal{S}_{\text{pot}}$, there exists a real constant $\lambda_{\mathcal{C}}\in \mathbb{R}_+$ such that the contribution from worldsheet instantons along the ray $n\mathcal{C}$ diverges when evaluated for the rescaled Kähler parameters
\begin{equation}
\mathbf{t}_{\mathcal{C}}\coloneqq \dfrac{1}{\lambda_{\mathcal{C}}}\, \mathbf{t}_{\text{dS}}\; ,\, \lambda_{\mathcal{C}}>0\, .
\end{equation}
Here $\mathbf{t}_{\text{dS}}$ denotes the K\"ahler parameters at the de Sitter minimum. 
In Table~\ref{tab:pot_rays_auto}, we report the minimal value of $\lambda_{\mathcal{C}}$ for all potent curves in our sample $\mathcal{S}_{\text{pot}}$, i.e.,
\begin{equation}\label{eq:lambdadef}
    \lambda\coloneqq \min_{\mathcal{C} \in \mathcal{S}_{\text{pot}}} \lambda_{\mathcal{C}}\,,
\end{equation}
which can be interpreted as 
the (multiplicative) safety margin of the K\"ahler parameters in the de Sitter vacuum with respect to the worldsheet instanton series on the rays of potent curves that we found.

\begin{figure}[!t]
\centering
\includegraphics[width=\linewidth]{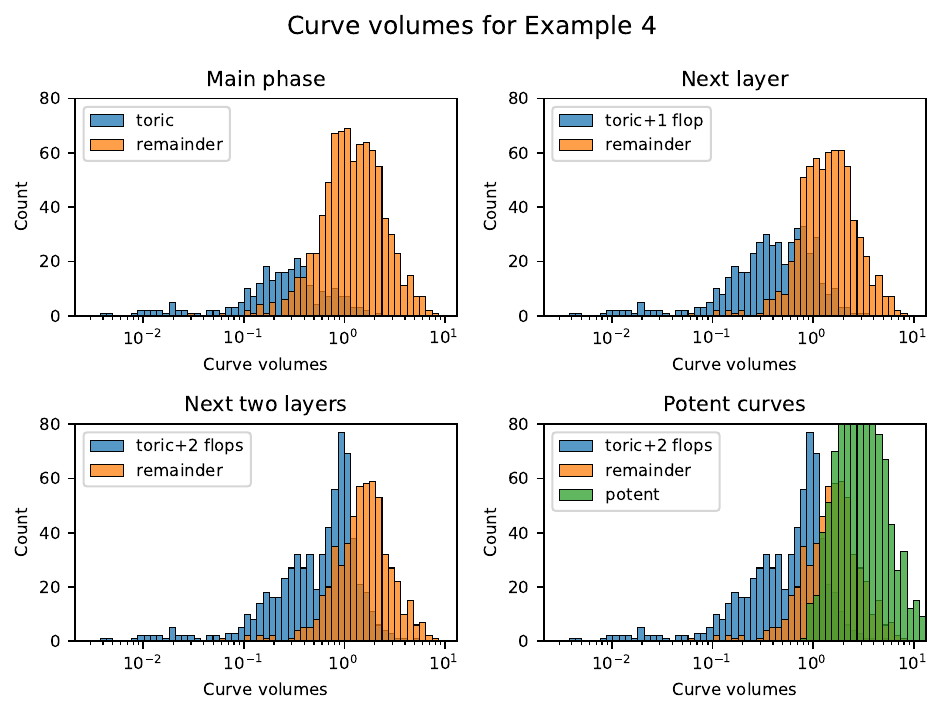}
\caption{Minimum-volume curves found in the example of \S\ref{sec:aule}, according to the categories explained in the text.}\label{fig:onions}
\end{figure}

Figure \ref{fig:onions} shows the volumes of the nilpotent toric curves found in the triangulation corresponding to the de Sitter vacuum, in the neighboring triangulations, and in the next-to-nearest neighbors, as well as nilpotent and potent curves found by direct computation as in  \cite{CMS}.  As explained above, for all the nilpotent curves shown, we explicitly incorporated  the corresponding corrections to the K\"ahler potential and K\"ahler coordinates.  For the potent curves, we checked that the largest contribution was small.

\subsubsection{Effects of Pfaffian numbers}\label{sec:pfaff}

Thus far we have taken the Pfaffians $\calA_{D}$ in the nonperturbative superpotential to be given by setting $n_D=1$ in \eqref{eq:K0}, 
\begin{equation}
\calA_{D} \equiv \sqrt{\frac{2}{\pi}}\frac{n_D}{4\pi^2} \rightarrow \sqrt{\frac{2}{\pi}}\frac{1}{4\pi^2}\,.
\end{equation}  
However, the values of the numerical constants $n_D$ are presently unknown,
and so we have also repeated our analysis for a wide range of values of $n_D$.
In each of the examples of \S\ref{sec:examples}, we have verified a successful 
uplift to a de Sitter vacuum over the range
\begin{equation}\label{eq:ndrange}
    10^{-3} \le n_D \le 10^4\,.
\end{equation}
These bounds are the smallest and largest values we attempted, not values at which uplift failed.

We remark that in each case the minimal value $n_D^{\text{min}}$ is such that 
\begin{equation}
\mathcal{A}^{\text{min}}= \sqrt{\frac{2}{\pi}}\frac{n_D^{\text{min}}}{4\pi^2} < \frac{1}{16\pi^2}\,,
\end{equation}
Thus, a reader who expects that 
$\mathcal{A}=\frac{k}{16\pi^2}$ with $k \in \mathbb{Z}$ should expect our examples to uplift to de Sitter vacua.

Increasing $n_D$ increases the Calabi-Yau volume $\calV_E$ at the point $T_{\text{AdS}}$
in K\"ahler moduli space where the supersymmetric AdS precursor exists.
Comparing to the alignment condition \eqref{eq:throat_tuning_maintext}, we conclude that
increasing $n_D$ has the effect of increasing  $\Xi$.
Thus, as $n_D$ increases, examples that uplifted to non-supersymmetric AdS vacua at $n_D=1$ will transition into de Sitter vacua,
and examples that uplifted to de Sitter vacua at $n_D=1$ will eventually transition into runaways.
Conversely, as $n_D$ becomes very small, examples that were runaways at $n_D=1$ may eventually become de Sitter vacua.
Thus, even if the correct values of $n_D$ are outside the range where the 
examples presented in \S\ref{sec:examples} uplift to de Sitter, we expect that, for any reasonable value of $n_D$, at least some examples of the large dataset 
shown in Figure \ref{fig:W0_zcf_comb} 
will successfully uplift.

So far we have considered uniform rescalings of the Pfaffians, i.e. we have taken the Pfaffians of all contributing divisors to have the same value of $n_D$.  We have also studied unequal Pfaffians, 
randomly drawing each of the $n_D$ from a distribution around a central value $\bar{n}$.
For modest variations around a central value in the range \eqref{eq:ndrange}, we find that the vacua typically survive.

\subsection{Corrections beyond leading order}\label{app:sscorr}

Having addressed the practical limitations related to finite computations within the leading-order EFT, we now turn to corrections from beyond leading order.

\subsubsection{String loop corrections}\label{sec:stringloop}

The K\"ahler coordinates $T_i$ on K\"ahler moduli space can be written,
as in \eqref{eq:firstdefT},
\begin{align}\label{eq:firstdefTapp}
       T_i  &= T^{\text{tree}}_i + 
          T_i^{(\alpha')^2} + T_i^{\text{WSI}} 
          +\de T^{\calN=2,(g_s)}_i+ \de T^{\calN=1}_i \,.
\end{align} 
Thus far we have worked with the leading-order 
K\"ahler coordinates $T_i^{\text{l.o.}}$, 
\begin{equation}\label{eq:tloapp}
     T_i^{\text{l.o.}}  = T^{\text{tree}}_i + 
          T_i^{(\alpha')^2} + T_i^{\text{WSI}} 
\end{equation}
as given explicitly
in \eqref{eq:detailedform4}.
Likewise, for the K\"ahler potential\footnote{
The $\alpha'$ corrections   
to the K\"ahler potential
are small in all our examples, as shown in \S\ref{sec:examples}.
Even so, they 
have a non-negligible effect through the implicit dependence of $\calV$ on the holomorphic coordinates $T_i$,
and cannot be neglected when computing the vacuum point in moduli space.}
we have used  $\calK^{\text{l.o.}}$
as given in \eqref{eq:Kahlerpotential}.
The expression \eqref{eq:tloapp} incorporates all $\alpha'$ corrections, perturbative and nonperturbative, at string tree level.

At higher orders in the string loop expansion, one expects corrections proportional to powers of $g_s \ll 1$, encoded in the terms 
$\de T^{\calN=2,(g_s)}_i$ and $\de T^{\calN=1}_i$ in \eqref{eq:firstdefTapp}.
In this Appendix we directly evaluate $\de T^{\calN=2,(g_s)}_i$, as computed by \cite{Robles-Llana:2006hby,Robles-Llana:2007bbv}, and show that this term is negligible in our examples.  We then estimate corrections from $\de T^{\calN=1}_i$, which has not been computed in Calabi-Yau compactifications, but has been modeled in \cite{Berg:2007wt} following the toroidal orientifold computation of \cite{Berg:2005ja}.

We begin with Euclidean D1-brane corrections inherited from the $\mathcal{N}=2$ parent model.
Specifically, we have evaluated the 
corrections computed in \cite{Robles-Llana:2006hby,Robles-Llana:2007bbv} for the smallest curves that arise in our models. 
For each such curve, with string-frame volume $t$ and B-field $b$, the full $SL(2,\mathbb{Z})$-invariant expressions for the K\"ahler potential and K\"ahler coordinates receive corrections of the form
\begin{equation}
    I = \sum_{(0,0)\neq (m,n)\in \mathbb{Z}^2 }\frac{1}{|m\tau +n|} e^{2\pi i z_{m,n}}\, ,\quad z_{m,n}:=nb+i|m\tau+n|t\, .
\end{equation}
The $m=0$ part of this sum corresponds to the effect of worldsheet instantons,
\begin{equation}
    I_{m=0}=2\,\text{Li}_1\Bigl((-1)^{2b} e^{-2\pi t}\Bigr)\, ,
\end{equation}
while the remainder can be Poisson-resummed to yield \cite{Robles-Llana:2006hby}
\begin{equation}
    I_{m\neq 0}=-\frac{g_s}{\pi}\sum_{m\neq 0}\sum_{n\in \mathbb{Z}}\left|\frac{b+n+it}{mt}\right|e^{-2\pi i m C_0(b+n)}K_1\left(\frac{2\pi}{g_s}|m|\cdot |b+n+it|\right)\, ,
\end{equation}
with $K_1$ the modified Bessel function of the second kind, and amounts to the series of Euclidean D1-brane corrections (including the sum over worldvolume flux). In our case, $b\in \{0,1/2\}$ and the leading term in the two cases is
\begin{equation}
    I_{m\neq 0}\simeq \begin{cases}
        -\frac{2g_s}{\pi}K_1\left(\frac{2\pi}{g_s}t\right) & b=0\\
        -\frac{4g_s}{\pi}\sqrt{1+\frac{1}{4t^2}}\cos(\pi C_0)K_1\left(\frac{2\pi}{g_s}\sqrt{t^2+1/4}\right) & b=\frac{1}{2}\, .
    \end{cases}
\end{equation}
We define the control parameter
\begin{equation}\label{eq:epsN2def}
\epsilon_{g_s}^{\mathcal{N}=2}:=\max_{\textbf{curves}}\left(\frac{I_{m\neq 0}}{I_{m=0}} \right)\, ,
\end{equation}
which represents the size of the leading contribution from
$\de T^{\calN=2,(g_s)}_i$, compared to the worldsheet instanton terms that we have already included. 
The value of $\epsilon_{g_s}^{\mathcal{N}=2}$ in each of our models is given in Table \ref{tab:N2loopSummary}.
We conclude that the effects of $\de T^{\calN=2,(g_s)}_i$ are safely negligible in our constructions.

\begin{table}[t!]
\begin{centering}
\begin{tabular}{|c|c|c|}\hline
ID & $\epsilon_{g_s}^{\mathcal{N}=2}$   \\\hline
1 & $5.3 \times 10^{-3}$    \\\hline   
2 & $3.9 \times 10^{-3}$   \\\hline   
3 & $2.5 \times 10^{-3}$   \\\hline
4 & $5.6 \times 10^{-4}$ \\\hline  
5 & $3.1 \times 10^{-8}$  \\\hline   
\end{tabular}
\caption{
Relative correction $\epsilon_{g_s}^{\mathcal{N}=2}$, defined in \eqref{eq:epsN2def}, from the string loop corrections
$\de T^{\calN=2,(g_s)}_i$ to the K\"ahler coordinates.}
\label{tab:N2loopSummary}
\end{centering}
\end{table}

We now turn to $\de T^{\calN=1}_i$,
which we will express in terms of
\begin{equation}
\calT_i^{\text{l.o.}} := \text{Re}\, T_i^{\text{l.o.}}\,,
\end{equation} with $T_i^{\text{l.o.}}$ 
the string tree level K\"ahler coordinate including all $\alpha'$ corrections, as in \eqref{eq:detailedform4} and \eqref{eq:tloapp}.  The $\calT_i^{\text{l.o.}}$ are Einstein frame volumes, so corrections can be written as powers of $1/\calT_i^{\text{l.o.}} = g_s/\calT_{i,s}^{\text{l.o.}}$, with $\calT_{i,s}^{\text{l.o.}}$ the corresponding string frame volumes.
 
We will write 
\begin{equation}\label{eq:ModelForCorrectionsToTscheme}
   \de T^{\calN=1}_i =  \de T^{\calN=1}_{i,\text{divisor}} + \de T^{\calN=1}_{i,\text{curve}} \, ,
\end{equation}
and first consider the corrections involving divisor volumes.
Na\"ive dimensional analysis suggests a series of corrections in powers of 
\begin{equation}
\frac{g_i^2 N}{16\pi^2} = \frac{c_{D_i}}{4\pi\calT_i^{\text{l.o.}}}\,,
\end{equation}
where the left-hand side is the expectation for a gauge theory with coupling $g_i$ and $N$ colors, and the right-hand side is the corresponding expression in our notation.
Thus, we write\footnote{An additional possibility
is a term proportional to $\log(\mathcal{V}_E)$
\cite{Conlon:2010ji}. The variation of such a term with respect to the K\"ahler parameters is of order $1/\mathcal{V}_E$. Therefore, to leading order in $1/\mathcal{V}_E$
such a term can be absorbed with an appropriate rescaling of the Pfaffians
$\mathcal{A}_D$.  As we marginalize over the $\mathcal{A}_D$ in a wide range (\S\ref{sec:pfaff}),
we do not include a log term in \eqref{eq:ModelForCorrectionsToTdiv}.} 
\begin{equation}\label{eq:ModelForCorrectionsToTdiv}
   \de T^{\calN=1}_{i,\text{divisor}} = \frac{c_{D_i}}{4\pi}\sum_{n=1}^\infty k^{i}_{\text{self},n} \,\Biggl(\frac{c_{D_i}}{4\pi} \cdot \frac{g_s}{\calT_{i,s}^{\text{l.o.}}}\Biggr)^n+\ldots  \,,
\end{equation}
with some unknown coefficients $k^i_{\text{self}}$.
By comparing with \cite{Berg:2004ek},\footnote{See also
\cite{Berg:2007wt,Conlon:2010ji,Gao:2022uop}.} which explicitly computed corrections to the K\"ahler coordinates in $\mathcal{N}=1$ supersymmetric toroidal orientifolds,
we expect that the $k^i_{\text{self}}$ are  $\mathcal{O}(1)$.   Directly computing these coefficients in a Calabi-Yau orientifold is an important task for the future.

The one-loop term in \eqref{eq:ModelForCorrectionsToTdiv}, $n=0$, defines the Pfaffian $\mathcal{A}_D$, which we are treating separately (\S\ref{sec:pfaff}), and so the first non-trivial correction could arise at $\mathcal{O}(g_s)$, i.e., $n=1$.

Turning to the corrections $\de T^{\calN=1}_{i,\text{curve}}$ from curves, we expect the most dangerous effects to come from small-volume cycles.
It was argued in \cite{Berg:2005ja,Berg:2007wt}
that the dominant effect in this regime involves winding strings that wrap the curve of intersections of two stacks of D7-branes.  In our compactifications, the D7-brane stacks do not intersect at all, but one might speculate that the curve of intersection of a D7-brane stack with a Euclidean D3-brane could support modes that source a correction.
We are thus led to a conservative model of $\mathcal{N}=1$ string loop corrections from small-volume curves, of the form 
\cite{Berg:2005ja,Berg:2007wt,Conlon:2010ji} (see also \cite{Gao:2022uop})
\begin{equation}\label{eq:ModelForCorrectionsToT}
   \de T^{\calN=1}_{i,\text{curve}} = 
    \sum_{\mathcal{C}_{ij}\subset D_i\cap D_j}
    \theta(g)\, k_{\mathcal{C}_{ij}}\,\frac{g_s}{\vol_s{\mathcal{C}}_{ij}} + \ldots \, ,
\end{equation}
where $\theta$ is the Heaviside step function, $g$ is the genus of $\mathcal{C}_{ij}$, and $k_{\mathcal{C}_{ij}}$ is an unknown constant.
That is, for coordinate $T_i$ we consider curves $\mathcal{C}_{ij}$ corresponding to the intersection of other divisors $D_j$ with $D_i$.
If $\mathcal{C}_{ij}$ has genus $g>0$, then winding modes along $\mathcal{C}_{ij}$ can correct $T_{i}$.

Now we turn to evaluating \eqref{eq:ModelForCorrectionsToT}.
In each of our examples, all seven-brane stacks, as well as all of the leading Euclidean D3-branes (see \S\ref{sec:autoch}), wrap prime toric divisors.
So suppose that $D_i$ and $D_j$ are toric divisors, and that $\mathcal{C}_{ij} \subset D_i \cap D_j$.
For $X$ a Calabi-Yau hypersurface obtained from a 
$\Delta^{\circ}$-favorable polytope 
whose dual
is $\Delta$-favorable 
(a condition that holds in all our examples: see \S\ref{sec:polytopeselection}),
$\mathcal{C}_{ij}$
is a curve of genus $>0$ if and only if 
$D_i$ and $D_j$ are each associated with 
vertices of a one-face of $\Delta^\circ$, without interior point, such that the dual two-face of $\Delta$ has at least one interior point.\footnote{See e.g.~\cite{Braun:2017nhi} for an accessible approach to related results on stratification.} 
In our models this situation does not arise, and so there are \emph{no contributions} to 
\eqref{eq:ModelForCorrectionsToT} from toric complete intersection curves $\mathcal{C}_{ij} \subset D_i \cap D_j$. 
That is, if we consider winding strings on D7-brane/Euclidean D3-brane intersection curves, following \cite{Berg:2005ja,Berg:2007wt}, then we find $\de T^{\calN=1}_{i,\text{curve}} = 0$.

In an abundance of caution, we suppose that there might be contributions 
\begin{equation}\label{eq:potloopest}
\de T^{\calN=1}_{i,\text{curve}} =  k_{\mathcal{C}}\,\frac{g_s}{\vol_s{\mathcal{C}}}\,, 
\end{equation}
for \emph{some} curve $\mathcal{C}$, even one that is not the intersection of divisors supporting D7-branes and Euclidean D3-branes (cf.~\cite{Gao:2022uop}).
We still require that $\mathcal{C}$ be potent, so that it might in principle have $g>0$ and so support winding modes.
Thus, we take the string frame volume $\vol_s{\mathcal{C}}$ of the smallest potent curve $\mathcal{C}$  
as a template for corrections to the holomorphic coordinates $\calT_{i}$. 
As we currently lack a method to determine whether higher genus Riemann surfaces exist in any given curve class, we simply assume that such a representative exists. For this reason, the corresponding correction is very likely an overestimate.

Table  \ref{tab:loopSummary} records our estimates of the largest divisor self-correction from \eqref{eq:ModelForCorrectionsToTdiv} and
the largest curve correction from \eqref{eq:potloopest} 
in each model, taking
$k_{\mathcal{C}}=k^i_{\text{self},1}=1$.

\begin{table}[H]
\begin{centering}
\begin{tabular}{|c|c|c|c|}\hline
ID     & Self-correction & Curve correction \\\hline
1  & 0.00609  & 0.0511 \\\hline
2  & 0.00463 & 0.0401   \\\hline
3  & 0.00713 & 0.0387  \\\hline    
4  & 0.0147 & 0.0543  \\\hline
5  & 0.0104  & 0.0481  \\\hline
\end{tabular}
\caption{Summary of largest relative corrections 
$\de T^{\calN=1}_i/T_i$ to 
\eqref{eq:ModelForCorrectionsToTdiv} and
\eqref{eq:potloopest}.}
   \label{tab:loopSummary}
\end{centering}
\end{table}

\subsubsection{Perturbations of the throat}\label{sec:bulkeffects} 
 
We now turn to ten-dimensional supergravity in order to examine 
the effects of relevant operators that perturb the infrared region of the throat. 

Gaugino condensation on a stack of seven-branes wrapped on a divisor $D$ sources three-form flux \cite{Koerber:2007xk,Baumann:2010sx}, breaking the no-scale symmetry of the Klebanov-Strassler solution.
The strength of the source is proportional to the nonperturbative superpotential term $W_{\text{np}}^{(D)}$, which is small.  However, if the flux excites a relevant perturbation of the throat in the UV, this mode will grow toward the IR, and can become significant if the hierarchy of scales is large.  We now give a quantitative estimate of this effect,
following \cite{Baumann:2010sx}.

We consider the leading mode in the throat that breaks no-scale: this is a mode\footnote{This mode was classified as Series I flux in \cite{Baumann:2010sx}: see Table 7 and \S7.2.2 of \cite{Baumann:2010sx}.}  of the three-form flux $G_3$, 
    corresponding to an operator in the dual field theory with dimension $\Delta=\frac{5}{2}$  
    \cite{Baumann:2010sx}.   
    Given a UV source $j_{\text{UV}}^{(D)}$ from a divisor $D$, this mode grows toward the IR as
    \begin{equation}
    \delta\phi(r)=\left(\frac{r_{\text{UV}}}{r}\right)^{\frac{3}{2}}j_{\text{UV}}^{(D)}\, .
    \end{equation}    
As the vev of the three-form $G_3$ at the tip of the throat is of order $M$, we estimate the relative size of the backreaction at the tip of the throat to be of order
\begin{equation}\label{eq:throat_control_parameter_setup}
\varepsilon^{(D)}:=\frac{\delta\phi(r_{\text{IR}})}{M}=\left(\frac{r_{\text{UV}}}{r_\text{IR}}\right)^{\frac{3}{2}}\frac{j_{\text{UV}}^{(D)}}{M}\,.
\end{equation}

We first compute the strength of the source.
Writing the full superpotential  \eqref{eq:wlo3evaluated} as
\begin{equation}\label{eq:wlo3evaluatedapp}
    W = W_0 + \sqrt{\frac{2}{\pi}}\frac{1}{4\pi^2}\sum_D \, e^{-2\pi T_D/c_D} = 
   W_0 + \sum_D W_{\text{np}}^{(D)} \,, \end{equation}
and using\footnote{The right-hand side is as in\cite{Kachru:2019dvo}, with their $e^{8u}$ replaced by our (equivalent) factor $\mathcal{V}_E^{\frac{4}{3}}$.} \cite{Kachru:2019dvo,Hamada:2018qef}   
\begin{equation}
c_D \frac{\langle \lambda\lambda\rangle^{(D)}}{32\pi^2}\simeq e^{\frac{K}{2}} W_{\text{np}}^{(D)}\times \mathcal{V}_E^{\frac{4}{3}}\,,
\end{equation}
we estimate 
    \begin{equation}
        j_{\text{UV}}^{(D)}\sim \frac{1}{\sqrt{g_s}}\frac{\langle \lambda\lambda\rangle^{(D)}}{32\pi^2} \sim   \frac{\mathcal{V}_E^{\frac{1}{3}}}{\sqrt{128\widetilde{\mathcal{V}}}}\frac{1}{c_D}W_{\text{np}}^{(D)}\, .
    \end{equation}

To evaluate $\varepsilon^{(D)}$, we compute the hierarchy of scales $r_{\text{IR}}/r_\text{UV}$. 
Uplift to a de Sitter vacuum requires the relation
\eqref{eq:throat_tuning} between $z_{\text{cf}}$ and $W_0$, as derived in Appendix \ref{sec:alignment}.  For $\Xi \sim 1$ 
the hierarchy of scales is given by \eqref{eq:throathier}, so we arrive at
\begin{equation}\label{eq:throat_control_parameter}
\varepsilon^{(D)} \sim 0.14\times \frac{\calV_E^{\frac{1}{3}}}{c_DM (g_s^3 \widetilde{\calV})^{\frac{1}{8}}}\times \frac{W_{\text{np}}^{(D)}}{W_0^{\frac{3}{4}}}\, ,
\end{equation}
 
Table \ref{tab:bulk_effects} shows that
$\varepsilon^{(D)} \ll 1$ for all divisors $D$ hosting nonperturbative effects,
for all of the examples of \S\ref{sec:examples}.
We therefore expect the infrared region of the warped throat to be unaffected by the dynamics in the bulk.

\begin{table}[t!]
\begin{centering}
\begin{tabular}{|c|c|c|}\hline
ID & $\operatorname{max}\varepsilon^{(D)}$   \\\hline
1 & $5.6 \times 10^{-5}$    \\\hline  
2 & $5.2 \times 10^{-5}$   \\\hline  
2 & $7.2 \times 10^{-5}$ \\\hline  
4 & $2.3 \times 10^{-4}$ \\\hline 
5 & $1.2 \times 10^{-4}$  \\\hline   
\end{tabular}
\caption{Sizes of perturbations to the Klebanov-Strassler throat, as estimated in \eqref{eq:throat_control_parameter}.}
\label{tab:bulk_effects}
\end{centering}
\end{table}

If one instead found $\varepsilon^{(D)} \gtrsim 1$ for some $D$, the impact would depend on the location of $D$ in relation to the throat.  A divisor $D_{\text{far}}$ sourcing fluxes that have to propagate across the compactification to reach the throat will have smaller effects on the infrared than a divisor $D_{\text{near}}$ that extends into the upper reaches of the throat.
However, quantifying the suppression that results from propagation would require knowing the Calabi-Yau metric in the bulk, and so is beyond the scope of this work.

\subsubsection{Warping Effects}\label{ss:warping}
Next we consider the $\calN=1$ effects due to warping, parameterized in (4.17) of 
\cite{smallCCs} as
\begin{equation}
    g_{\calN=1}^i = \frac{|N_{\text{D3}}(\omega_i)|}{\calT_{i,E}^{(0)}}\label{eq:smallCCs417}
\end{equation}
for divisors $D_i$, and
\begin{equation}
    g_{\calN=1}^X = \frac{Q_{\text{O}}}{\calV_E^{2/3}}\label{eq:smallCCs418}
\end{equation}
for the entire Calabi-Yau. The sizes of these corrections in each of the examples are summarized in Table \ref{tab:warpingSummary}. We see that the characteristic size of warping effects is $\calO(30\%)$ for divisors and $\calO(15\%)$ for the entire Calabi-Yau.

The quantities \eqref{eq:smallCCs417} and \eqref{eq:smallCCs418} provide estimates\footnote{
In principle one could use $Q_{\text{flux}}^{\text{throat}}$ rather than $Q_{\text{O}}$ in an estimate for the singular bulk problem, but $Q_{\text{flux}}^{\text{throat}}$ is not a great deal smaller than $Q_{\text{O}}$ in our examples.} of the margin of safety from the effects of backreaction of seven-branes and  O3-planes, respectively, on the warp factor, a phenomenon studied in \cite{Carta:2019rhx,Gao:2020xqh,Carta:2021lqg}
and referred to as the \emph{singular bulk problem} in \cite{Gao:2020xqh}.

\begin{table}[h]
\begin{centering}
\begin{tabular}{|c|c|c|}\hline
ID & $\operatorname{max}(g_{\calN=1}^i)$ & $g_{\calN=1}^X$ \\\hline
1 & 0.273 & 0.146 \\\hline
2 & 0.244 & 0.128 \\\hline
3 & 0.292 & 0.116 \\\hline
4 & 0.363 & 0.177 \\\hline
5 & 0.374 & 0.245 \\\hline
\end{tabular}
\caption{Summary of the characteristic size of warping effects, given in \eqref{eq:smallCCs417} and \eqref{eq:smallCCs418}, in the five de Sitter vacua, labeled as in \S\ref{sec:examples}.}
\label{tab:warpingSummary}
\end{centering}
\end{table}

\section{Normalization of the Anti-D3-brane Potential} \label{sec:alignment}

In this Appendix we compute the energy of a warped anti-D3-brane, in conventions that accord with our treatment of the supersymmetric theory.  
 
We write the K\"ahler potential \eqref{eq:detailedform2} as
\begin{equation}
		\mathcal{K} = -\log\bigl(16\mathcal{V}_{E}^2/g_s\bigr)-\log\bigl(||\Omega||^2\bigr)\, 
	\end{equation}
where $\mathcal{V}_{E} = \calV/g_s^{3/2}$ is the Einstein frame volume of $X$ in units of $\ell_s^2\equiv (2\pi)^2\alpha'$, including perturbative and nonperturbative corrections in $\alpha'$, cf.~\eqref{eq:detailedform3}, 
and $||\Omega||^2\coloneqq -i\int_X \Omega\wedge \overline{\Omega}$.

Denoting  $W_0 \coloneqq  |\langle W_{\text{flux}} \rangle |$ with 
$W_{\text{flux}}$ given in \eqref{eq:flux_superpotential},
the supersymmetric bulk minimum lies at 
\begin{equation}
		|V_F| \coloneqq \frac{3}{16}\cdot \frac{g_s}{\mathcal{V}_E^2}\cdot \frac{|W_0|^2}{||\Omega||^2}\,.
	\end{equation}
Next, we compute the contribution to the scalar potential from a warped anti-D3-brane. The warped ten-dimensional metric in Einstein frame reads (see e.g.~\cite{Giddings:2005ff,Frey:2008xw})
	\begin{equation}
		ds^2=\frac{\ell_s^2}{4\pi}\frac{e^{2A}}{t}dx^2+\ell_s^2 e^{-2A}ds^2_{CY}\, ,\quad e^{-4A}=e^{-4A_0}+t-t_0\, ,
	\end{equation}
	in terms of the overall volume modulus $t$, and a reference solution of the warp factor $e^{-4A_0}$, and with the Calabi-Yau metric $ds^2_{CY}$ normalized to unit volume, and finally with
	\begin{equation}
		t_0\coloneqq \int_X \sqrt{g_{CY}}e^{-4A_0}\, .
	\end{equation}
	In this system of units, the four dimensional reduced Planck mass measured with the four-dimensional Weyl-rescaled metric $dx^2$ is given by
$M_\text{pl}=1$.	
	We now assume that complex structure moduli are stabilized near a conifold point where a special Lagrangian three-sphere $A$ degenerates. Its volume measured with the (unit volume) metric $ds^2_{CY}$ is then given by \cite{Koerber:2010bx}  
	\begin{equation}\label{eq:cys3vol}
 2^{-3/2}\cdot \text{Vol}(A)_{CY}=\frac{\int_A \Omega}{||\Omega||}=: \frac{|z_{\text{cf}}|}{||\Omega||}=: \hat{z}_{\text{cf}}\, .
	\end{equation}
	In contrast, the physical volume measured with the full ten-dimensional Einstein frame metric is given by 
	\begin{equation}\label{eq:physs3vol}
			\text{Vol}(A)= e^{-3A_{IR}}	\text{Vol}(A)_{CY}\ell_s^3\, ,
	\end{equation}
	where $e^{A_{IR}}$ is the value of the warp factor at the tip of the throat.
	
	The Klebanov-Strassler solution \cite{Klebanov:2000hb} implies the relation
	\begin{equation}\label{eq:ksvol}
		\text{Vol}(A)_{\text{KS}}=\frac{\Sigma^{\frac{3}{4}}}{2\pi\cdot \sqrt{6} } \left(\sqrt{g_s}M\right)^{\frac{3}{2}} \ell_s^3 \, ,
	\end{equation}
	with
	\begin{align}\label{eq:ksint}
    \Sigma&\coloneqq 2^{\frac{2}{3}}\int_0^\infty dx \frac{x\, \coth(x)-1}{\sinh^2(x)}\bigl(\sinh(2x)-2x\bigr)^{\frac{1}{3}}\approx 1.13983\ldots \, .
	\end{align}
	Comparing \eqref{eq:cys3vol}, \eqref{eq:physs3vol}, and \eqref{eq:ksvol}, we find that
	\begin{equation}\label{eq:warpz}
		e^{3A_{IR}}\approx \gamma \cdot \frac{|\hat{z}_{\text{cf}}|}{(\sqrt{g_s}M)^{\frac{3}{2}}}\, , \quad \gamma\coloneqq \frac{8\pi\cdot \sqrt{3}}{\Sigma^{3/4}}\approx 39.4612\, .
	\end{equation}
	Therefore, the warped anti-D3-brane potential can be written
\begin{equation}
		V_{\overline{D3}} \approx 2T_{D3}\cdot \Bigl(\frac{\ell_s^2}{4\pi}\Bigr)^2\,\frac{e^{4A_{IR}}}{t^2}=\frac{1}{4\pi} \frac{e^{4A_{IR}}}{t^2}\equiv \frac{3\zeta}{32}\cdot\frac{g_s}{\mathcal{V}_E^{4/3}} \left( \frac{\mathcal{V}_E^{4/3}}{t^2} \cdot \frac{|\hat{z}_{\text{cf}}|^{\frac{4}{3}}}{(g_s M)^2}\right)\, ,
	\end{equation} 
in terms of
\begin{equation}\label{eq:zetadef}
\zeta \coloneqq  \frac{128}{\Sigma}\cdot\left(\frac{\pi}{3}\right)^{1/3}\approx 114.037\,.
\end{equation}
At large volume $t\gg 1$, we may identify
	\begin{equation}
		\mathcal{V}_E=\int_X d^6y\sqrt{g_{CY}}e^{-6A}\approx t^{\frac{3}{2}}\, ,
	\end{equation}
	up to corrections of order $N_{\text{D3}}/\mathcal{V}_E^{\frac{2}{3}}$ --- see \S\ref{ss:warping}.
 
The above expressions are only valid when warping is significant near the tip of the throat, i.e., when 
	\begin{equation}
		 e^{3A_{IR}}\ll e^{3A_{UV}}\, ,
	\end{equation}
	or equivalently when
 \begin{equation}\label{eq:strongly_warped_def}
    |z_{\text{cf}}| \ll  (g_s M^2)^{\frac{3}{4}}\,\frac{||\Omega||}{\gamma \sqrt{\mathcal{V}_E}}\, .
\end{equation}
Finally,  at large complex structure we have
	\begin{equation}
		||\Omega||^2\approx 8 \widetilde{\mathcal{V}}\, ,
	\end{equation}
	where $ \widetilde{\mathcal{V}}$ is the volume of the mirror threefold measured in type IIA string frame.
	
We can therefore write
\begin{equation}\label{eq:throat_tuning}
\Xi:= \dfrac{V_{\overline{D3}}}{|V_F|} = \frac{|z_{\text{cf}}|^{\frac{4}{3}}}{|W_0|^2}\frac{\mathcal{V}_E^{\frac{2}{3}}\widetilde{\mathcal{V}}^{\frac{1}{3}}}{(g_s M)^2}\cdot \zeta
\,,  
\end{equation}
where the numerical constant $\zeta\approx 114.037$ was given in \eqref{eq:zetadef}.

Configurations with $V_{\overline{D3}}/|V_F| \gg 1$ will quickly decompactify, while configurations with $V_{\overline{D3}}/|V_F| < 1$ will have negative vacuum energy.
The de Sitter vacua found in this work arise in cases where $1 \lesssim V_{\overline{D3}}/|V_F| \lesssim 3$.

For $\Xi=1$, 
the throat hierarchy of scales reads  
\begin{equation}\label{eq:throathier}
        \left(\frac{r_{\text{IR}}}{r_\text{UV}}\right)^3\coloneqq \frac{e^{3A_{\text{IR}}}}{e^{3A_{\text{UV}}}}\approx 
        \frac{(3\pi)^{3/4}}{2^{15/4}}
         \times W_0^{3/2}\biggl(\frac{g_s}{\widetilde{\mathcal{V}}}\biggl)^{\frac{3}{4}}  \, .
    \end{equation}
Finally, the warped Kaluza-Klein scale $m_{wKK}$ in four-dimensional Planck units is
\begin{equation}\label{eq:mwkk}
    m_{wKK}^2\approx \Sigma^{-\frac{1}{2}} \left(\frac{3}{8\pi}\right)^\frac{1}{3}\frac{|z_{\mathrm{cf}}|^\frac{2}{3}}{(\sqrt{g_s}M)^3\mathcal{V}_E^{\frac{2}{3}}\tilde{\mathcal{V}}^{\frac{1}{3}}}\, .
\end{equation}

\FloatBarrier
\section{A Landscape of Vacua}  
\label{sec:KSInKS}

In this Appendix we present a selection of anti-de Sitter vacua, both supersymmetric and non-supersymmetric, that we found in the course of our search for de Sitter vacua. We plan on presenting a more thorough analysis of this dataset in future work \cite{coni1.5}. 
This Appendix is merely intended to provide context for the ensemble from which our de Sitter candidates are drawn, as also illustrated in Figures \ref{fig:2histograms} and \ref{fig:W0_zcf_comb}.

\subsection{KS in KS}

Supersymmetric AdS vacua with all moduli stabilized, but without conifolds, were constructed in \cite{smallCCs}, and conversely flux compactifications with conifolds but without all moduli stabilized were constructed in 
\cite{Alvarez-Garcia:2020pxd,coniLCS}. In Table \ref{tab:KS_in_KS} we give the first examples of supersymmetric AdS vacua with conifolds, \emph{and} with all moduli stabilized.

\begin{table}[H] 
\centering
\resizebox{\columnwidth}{!}{%
\begin{tabular}{|c|c|c|c|c|c|c|c|c|c|c|c|c|} \hline
& & & & & & & & & & \\[-1.3em]
ID & $(h^{2,1},h^{1,1})$    & $M$ & $K'$ & $N_{\text{D3}}$ & $g_s$ & $W_0$ & $g_sM$ & $|z_{\mathrm{cf}}|$  & $-V_F$ & $\Xi$  \\[0.1em]
\hline
\hline
& & & & & & & & & &\\[-1.3em]
a & $(6,160)$   & 8 & $\frac{1}{15}$ & 2 & $3 \cdot 10^{-3}$ & $1.0 \cdot 10^{-35}$ & 0.021 & $6.0 \cdot 10^{-6}$ & $2.5 \cdot 10^{-90}$ & $10^{70}$    \\[0.1em]\hline
& & & & & & & & & &\\[-1.3em]
b & $(7, 155)$  & 8& 2 & 0 &0.18 & $7.4 \cdot 10^{-18}$& 1.46 & $2.1 \cdot 10^{-3}$ & $5.1 \cdot 10^{-50}$  & $10^{34}$   \\[0.1em]\hline
& & & & & & & & & &\\[-1.3em]
c & $(6,160)$   & 2 & $10$ & 0 & 0.015 & $1.6 \cdot 10^{-27}$ & 0.30 & $2.4 \cdot 10^{-47}$ & $5.8 \cdot 10^{-72}$  & 0.06    \\[0.1em]\hline
& & & &  & & & & & &\\[-1.3em]
d & $(6, 160)$  &2 &$\frac{33}{2}$ & 11 &0.27 &$3.2 \cdot 10^{-25}$ & 0.55 & $1.3 \cdot 10^{-42}$ & $2.3 \cdot 10^{-66}$  & 0.65   \\[0.1em]\hline
& & & & & & & & & &\\[-1.3em]
e & $(8, 150)$    & 14 & 4 & 0  & 0.075 & 0.032 & 1.05 & $9.1 \cdot 10^{-7}$&$1.8\cdot 10^{-17}$ & 3.38    \\[0.1em]\hline
\end{tabular}
}
\caption{The first examples of supersymmetric AdS flux vacua with Klebanov-Strassler throats and with all moduli stabilized.}\label{tab:KS_in_KS}
\end{table}

Examples (a)-(e) in Table \ref{tab:KS_in_KS} are all supersymmetric AdS vacua
with Klebanov-Strassler throats.
Each of these compactifications
can fulfill Gauss's law \eqref{eq:gaussLaw} with $N_{\overline{\text{D3}}}=0$,
and so has
no net anti-D3-branes.  
Net D3-branes are present in (a) and (d), but not in (b), (c), or (e).
We have stabilized all closed string moduli in each of these examples; in (a) and (d) the additional moduli associated to the D3-branes are presumably lifted as in \cite{Kachru:2003sx,DeWolfe:2007hd}, but are not studied here. 

Example (a) has the smallest value of $W_0$ in our ensemble, while example (b) has the largest $g_sM$ in a case with $W_0<1\times10^{-10}$. However, in both (a) and (b)
the alignment parameter $\Xi$
defined in \eqref{eq:throat_tuning_maintext} is exponentially large, so there is no prospect of controllable supersymmetry breaking from physics in the throat in either case.

Example (c) has the smallest $z_{\text{cf}}$ in a well-aligned vacuum. 
Example (d) has the largest $g_sM$ in a well-aligned, strongly-warped vacuum with very small $W_0$.

Finally, example (e) is a very special vacuum.
Gauss's law \eqref{eq:gaussLaw} is fulfilled with $N_{\overline{\text{D3}}}=0$, 
so there is a supersymmetric AdS vacuum with a strongly-warped Klebanov-Strassler throat.
If one now introduces an
anti-D3-brane at the tip of the throat \emph{and} a D3-brane elsewhere in the compactification, the tadpole is unchanged. One can check that because the vacuum is well-aligned, the brane-antibrane pair does not cause immediate decompactification, but instead uplifts to a configuration with positive energy. Moreover, because
$M>12$, the anti-D3-brane is metastable against brane-flux annihilation, at least at the level of the leading order in $\alpha'$ approximation made throughout this work.

Thus, compactification (e) provides a totally explicit setting for the inflationary scenario of \cite{Kachru:2003sx}.  
The closed string moduli are not initially destabilized by the energy of the brane-antibrane pair, and the D3-brane position can evolve under the influence of the nonperturbative superpotential.  We expect (but have not shown) that for suitable initial conditions the D3-brane will fall down the throat and annihilate against the anti-D3-brane.
We hasten to state that we have \emph{not} computed the full potential for D3-brane motion, only the potential for the closed string moduli in the presence of the anti-D3-brane energy.  Determining whether the D3-brane evolution is actually slow-roll inflation, rather than a rapid plunge followed by tachyon condensation, would at a minimum require applying the results of \cite{Baumann:2010sx}, but more realistically will require a numerical supergravity solution.  This is a natural target for future work.

\subsection{Vacua with even flux quanta}
\label{sec:even}

It was shown in \cite{Frey:2002hf} that, in flux compactifications on \emph{toroidal} orientifolds, odd flux quanta are only permitted in the presence of exotic O3-planes, which modify the definition of $Q_{\text{O}}$ away from the form given in \eqref{eq:D3-charges}. However, in the toroidal context, this constraint follows from the existence of ``half-cycles" $\sim T^3/\mathbb{Z}_2$  in the quotient that do not exist in the unorientifolded parent torus.
Corresponding cycles and constraints have not been examined in 
Calabi-Yau orientifolds, and therefore it is unclear whether the conclusions of \cite{Frey:2002hf} apply in the Calabi-Yau context. 
 
In any event, finding de Sitter vacua with purely even flux quanta is beyond the scope of this work.
This is because, 
for the orientifolds we considered, one always has $Q_{\text{O}}=2+h^{1,1}(X)+h^{2,1}(X)\equiv0\, \operatorname{mod}\, 4$. Thus, if the PFV flux vectors $\vec{M}$ and $\vec{K}$ are both even, then in light of \eqref{eq:gaussLaw} we must have an even number of anti-D3-branes, $p \in 2\mathbb{Z}$. As a result, to meet the leading-order metastability criterion \eqref{eq:KPV_constraint}, we would need to find configurations with \emph{two} antibranes, and correspondingly flux vectors with $M>24$. From Figure \ref{fig:2histograms}
it is clear that 
flux vacua with extremely large $M$ do exist, but at the same time 
the computational methods used here are poorly suited to finding them on a large scale.

\begin{figure}[H]
\centering
\includegraphics[width=\linewidth]{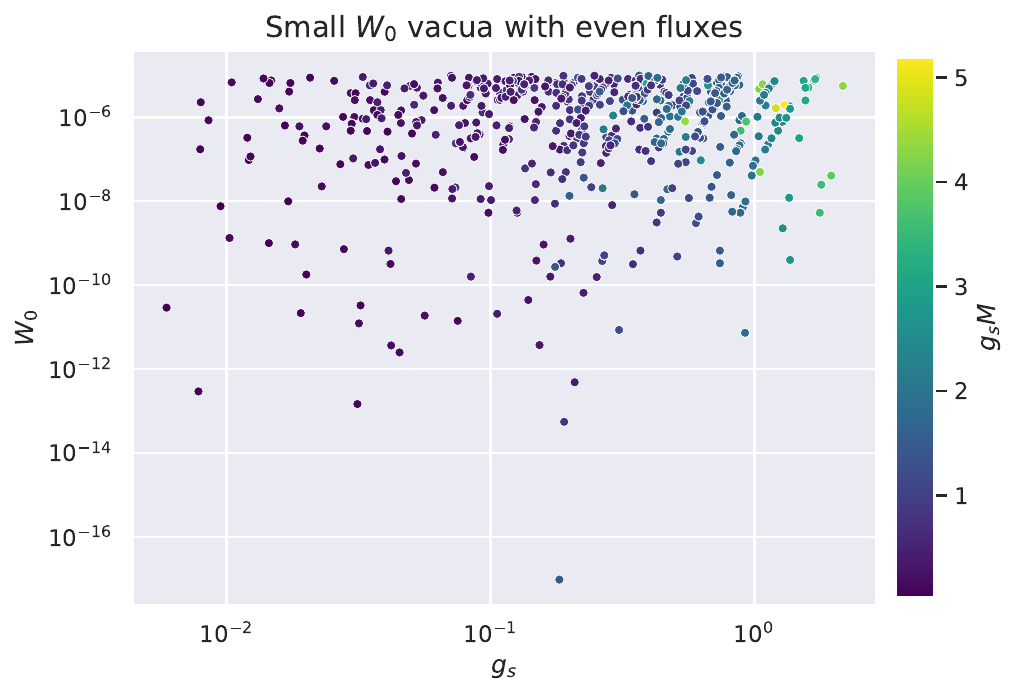}
\caption{Examples of 2{,}412 conifold PFVs with purely even flux quanta and small $W_0$.
}\label{fig:evenScatter}
\end{figure}

While the construction of de Sitter with even fluxes will require a dedicated search and improved methods, the construction of supersymmetric AdS vacua with even fluxes and Klebanov-Strassler throats 
is well within reach. We have thus far constructed 183,069
conifold PFVs with purely even flux quanta, among which those with small $W_0$ and $g_s$ are shown in Figure \ref{fig:evenScatter}. 
To illustrate some of the possible parameter values, we present three of these configurations in Table \ref{tab:even_examples}.   

\begin{table}[t!]
\centering
\resizebox{\columnwidth}{!}{%
\begin{tabular}{|c|c|c|c|c|c|c|c|c|c|c|c|}\hline
& & & & & & & & & &\\[-1.3em]
ID & $(h^{2,1},h^{1,1})$ & $M$ & $K'$ & $N_{\text{D3}}$ & $g_s$ & $W_0$ & $g_sM$ & $|z_{\mathrm{cf}}|$  & $V_F$ & $\Xi$  \\[0.1em]
\hline
\hline
& & & & & &  & & & &\\[-1.3em]
f & $(7,155)$  & 8 & 2 & 0 & 0.18 & $9.7 \cdot 10^{-18}$ & 1.46 & $2.1\cdot10^{-3}$ &  $ - 8.3 \cdot 10^{-50}$ & $10^{35}$     \\[0.1em]\hline 
& & & & & & & & & & \\[-1.3em]
g & $(8,150)$  & 8 & $\frac{54}{7}$ & 0  &  0.23 & $2.3 \cdot 10^{-2}$ & 1.86 & $3.1 \cdot 10^{-7}$ & $-1.6 \cdot 10^{-16}$ & 0.28    \\[0.1em]\hline
& & & & &  & & & & &\\[-1.3em]
h & $(6,160)$  & 4 & $\frac{7}{2}$ & $-2$ & 0.056 & $1.9 \cdot 10^{-11}$ & 0.23 & $1.0\cdot 10^{-22}$ & $-2.2 \cdot 10^{-38}$ & 0.22    \\[0.1em]\hline
\end{tabular}
}
\caption{A selection of vacua with even flux quanta and their control parameters. }
\label{tab:even_examples}
\end{table}

Examples (f) and (g) in Table \ref{tab:even_examples} have no net D3-branes or anti-D3-branes. In each case we have carried out stabilization of all moduli, arriving at a supersymmetric AdS vacuum with a Klebanov-Strassler throat.
Vacuum (f) has the smallest value of $W_0$ amongst even PFVs, but has $\Xi \gg 1$, whereas vacuum (g) is a well-aligned vacuum but has more moderate $W_0$.

Finally, we turn to example (h), which has the smallest $W_0$ occurring in a well-aligned PFV with even fluxes. Moreover, (h) contains $p=2$ anti-D3-branes.  Because $g_s M$ is quite small, it is far from clear that this anti-D3-brane configuration is metastable, and so we do not refer to this as a vacuum. Nevertheless, we present example (h) as a curiosity, and as evidence that aligned configurations with anti-D3-branes and small $W_0$ can be found in compactifications with purely even fluxes.

\subsection{Non-supersymmetric AdS vacua}

Thus far in this Appendix we have focused on supersymmetric AdS vacua. We now turn to nonsupersymmetric AdS vacua. In \S\ref{sec:examples} we considered nonsupersymmetric vacua in which the supersymmetry breaking energy of the anti-D3-brane was sufficient to uplift to de Sitter space, which
required $\Xi>1$ --- see \eqref{eq:throat_tuning_maintext}.
If instead we consider vacua with $\Xi<1$, the energy of the anti-D3-brane is insufficient to uplift the vacuum to positive energy, and we obtain a nonsupersymmetric AdS vacuum. 
 
Following the framework laid out in the main text, we have found three nonsupersymmetric AdS vacua, whose properties are summarized in Table \ref{tab:susy_broken_AdS}. We note that although in a de Sitter vacuum the masses of all moduli must be positive, in an AdS vacuum the Hessian of the potential is allowed to have negative eigenvalues as long as the resulting tachyonic mass is above the BF bound \cite{Breitenlohner:1982jf}. In particular, example (i) has two BF-allowed tachyons, whereas examples (j) and (k) have positive-definite masses.

We have not subjected the nonsupersymmetric AdS vacua presented here to the full battery of tests that we applied to the de Sitter vacua of \S\ref{sec:examples}; we defer a complete study of AdS vacua in our ensemble to future work \cite{coni1.5}.

\begin{table}[H]
\centering
\begin{tabular}{|c|c|c|c|c|c|c|c|c|c|}\hline
& & & & & & & & \\[-1.3em]
ID & $(h^{2,1},h^{1,1})$   & $M$ & $K'$ & $g_s$ & $W_0$ & $g_sM$ & $|z_{\mathrm{cf}}|$  & $V_0$ \\[0.1em] \hline
\hline
& & & & & & & & \\[-1.3em]
i & $(7,155)$  & 14 & $\frac{40}{7}$ & 0.15 & 0.29 & 2.05 & $2.0\cdot10^{-5}$ & $-1.7\cdot10^{-13}$\\[0.1em]\hline
& & & & & & & & \\[-1.3em]
j& $(7,147)$   & 14 &$\frac{183}{33}$ & 0.059  & $7.9\cdot10^{-4}$ & 0.83 & $1.1\cdot10^{-10}$ & $-1.2\cdot10^{-20}$\\[0.1em]\hline
& & & & & & & & \\[-1.3em]
k& $(6,160)$   & 14 &$\frac{52}{11}$ & 0.079  & $5.9\cdot10^{-2}$ & 1.10 & $2.5\cdot10^{-7}$ & $-8.8\cdot10^{-16}$\\[0.1em]\hline

\end{tabular}
\caption{Nonsupersymmetric AdS vacua.}
\label{tab:susy_broken_AdS}
\end{table}

\section{Formulas for the Potential and its Derivatives}
\label{sec:formulas}

In this section we derive formulas for evaluating the relevant terms in the scalar potential, and for computing first and second derivatives with respect to the moduli. We divide the problem into three parts: in \S\ref{sec:formulas_Kahler} we consider the K\"ahler moduli (for fixed $g_s$), in \S\ref{sec:formulas_CS} we consider the complex structure moduli and the dilaton, and in \S\ref{sec:formulas_uplift_potential} we consider the uplift potential \eqref{eq:anti-D3-potential0} and its derivatives.

In each case we used finite-difference methods and automatic differentiation to perform consistency checks of the expressions for the derivatives and Hessian.
 
\subsection{K\"ahler moduli sector}\label{sec:formulas_Kahler}

We consider the K\"ahler potential 
\begin{equation}
	K=-2\log(g_s^{-2}\Phi)\, ,
\end{equation}
with
\begin{align}
	\Phi=&\mathcal{V}+\delta\mathcal{V}_{\text{BBHL}}-\frac{1}{2}[\mathcal{F}-t^i\mathcal{F}_i]\, ,\\
	\text{Re}(T_i)=&g_s^{-1}\mathcal{T}_i\, ,\quad \mathcal{T}_i\coloneqq \tau_i-\frac{\chi_i}{24}+\mathcal{F}_i\, ,
\end{align}
in terms of $\delta\mathcal{V}_{\text{BBHL}}\coloneqq -\frac{\zeta(3)\chi}{4(2\pi)^3}$, and
\begin{equation}
	\mathcal{F}\coloneqq -\frac{1}{(2\pi)^3}\sum_q n_q^0 \text{Li}_3((-1)^{\gamma\cdot q}e^{-2\pi q\cdot t})\, ,\quad \mathcal{F}_i\coloneqq \del_{t^i}\mathcal{F}\, ,\quad \text{etc.}
\end{equation}
It will be useful to define the following quantities:
\begin{align}
	y_i&\coloneqq \mathcal{T}_{ij}t^j\equiv 2\tau_i+\mathcal{F}_{ij}t^j\, ,\quad y_{ij}\coloneqq \mathcal{T}_{ijk}t^k\equiv \mathcal{\kappa}_{ij}+\mathcal{F}_{ijk}t^k\, ,\\
	z_i&\coloneqq  \mathcal{T}_{ijk}t^jt^k\equiv 2\tau_i+\mathcal{F}_{ijk}t^jt^k\, ,\quad z_{ij}\coloneqq \mathcal{T}_{ijkl}t^kt^l\equiv \mathcal{F}_{ijkl}t^kt^l\, ,
\end{align}
where we define $\mathcal{T}_{ij}\coloneqq \del_{t^i}\mathcal{T}_j$, $\mathcal{T}_{ijk}\coloneqq \del_{t^i}\mathcal{T}_{jk}$ etc.

We have the following useful relation:
\begin{equation}
	\del_{t^i}\Phi=\frac{1}{2}y_i\, ,\quad \del_{t^i}y_j=y_{ij}+\mathcal{T}_{ij}\, .
\end{equation}
The inverse K\"ahler metric takes the form
\begin{equation}
	g_{i\bar{j}}=\frac{4\Phi}{g_s^2}\left[\epsilon y_i y_j-\mathcal{T}_{ij}\right]\, ,
\end{equation}
and the F-term potential can be conveniently reorganized as
\begin{equation}
	V=\frac{4g_s^2}{\Phi}\left[-Y+\epsilon \left(4X\overline{X}-4g_s\text{Re}[X\overline{W}]+g_s^2\eta|W|^2\right)\right]\, ,
\end{equation}
with $W^i\coloneqq \del_{T_i}W$, $W^{ij}\coloneqq \del_{T_i}W^j$ etc, and
\begin{align}
	X&\coloneqq \frac{1}{2}y_i W^i\, ,\\
	Y&\coloneqq \mathcal{T}_{ij}W^i\overline{W^j}\, ,\\
	\epsilon&\coloneqq  \left(\mathcal{T}_{ij}t^it^j-2\Phi\right)^{-1}\, ,\\
	\eta&\coloneqq  \frac{6\Phi-\mathcal{T}_{ij}t^it^j}{4\Phi}\equiv \frac{6\delta\mathcal{V}_{\text{BBHL}}-3\mathcal{F}+3t^i\mathcal{F}_i-\mathcal{F}_{ij}t^it^j}{4\Phi}\, .
\end{align}
We define
\begin{align}
	&\Omega_i\coloneqq \mathcal{T}_{ij}W^j\, ,\quad \Gamma^i\coloneqq W^{ij}\mathcal{T}_{jk}\overline{W^k}\, , \quad \xi^i\coloneqq W^{ij}y_j\, ,\quad \xi^{ij}\coloneqq W^{ijk}y_k\,  ,\nonumber\\
	&\mathcal{T}^{(W)}_{ij}\coloneqq \mathcal{T}_{ijk}W^k\, ,\quad \mathcal{T}^{(\Gamma)}_{ij}\coloneqq \mathcal{T}_{ijk}\Gamma^k\, ,\quad \mathcal{T}^{(\xi)}_{ij}\coloneqq \mathcal{T}_{ijk}\xi^k\, ,\\
	& \mathcal{T}^{(W,\overline{W})}_{ij}\coloneqq \mathcal{T}_{ijkl}W^k\overline{W^l}\, ,\quad \mathcal{T}^{(W,t)}_{ij}\coloneqq \mathcal{T}_{ijkl}W^k t^l\, ,\quad W^{ij}_{(\overline{\Omega})}\coloneqq W^{ijk}\overline{\Omega_k}\, .
\end{align}
Denoting $\phi_i\coloneqq \text{Im}(T_i)$, we find
\begin{align}
	X_i\coloneqq \del_{t^i}X=&\frac{1}{2}(y_{ij}+\mathcal{T}_{ij})W^j+\frac{1}{2}g_s^{-1}\mathcal{T}_{ij}\xi^j\, ,\\
	X^i\coloneqq \del_{\phi_i}X=&\frac{i}{2}\xi^{i}\, ,\\
	X_{ij}\coloneqq \del_{t^j}X_i=&\mathcal{T}_{ij}^{(W)}+\frac{1}{2}\mathcal{T}_{ij}^{(t,W)}+\frac{1}{2g_s}\mathcal{T}_{ij}^{(\xi)}+\frac{1}{2g_s}\left[(y_{ik}+\mathcal{T}_{ik})W^{kl}\mathcal{T}_{lj}+(i\leftrightarrow j)\right]+\frac{1}{2g_s^2}\mathcal{T}_{ik}\xi^{kl}\mathcal{T}_{lj}\, ,\nonumber\\
	{X^i}_j\coloneqq \del_{\phi_i}X_j=&\frac{i}{2g_s}\xi^{ik}\mathcal{T}_{kj}+\frac{i}{2}W^{ik}(y_{kj}+\mathcal{T}_{kj})\, ,\\
	X^{ij}\coloneqq \del_{\phi_i}X^j=&-\frac{1}{2}\xi^{ij}\, ,\\
	Y_i\coloneqq \del_{t^i}Y=&\mathcal{T}_{ij}^{(W)}\overline{W^j}+\frac{2}{g_s}\mathcal{T}_{ij}\text{Re}\left[\Gamma^j\right]\, ,\\
	Y^i\coloneqq \del_{\phi_i}Y=&-2\text{Im}(W^{ij}\overline{\Omega_j})\, ,\\
	Y_{ij}\coloneqq \del_{t^j}Y_i=&\mathcal{T}_{ij}^{(W,\overline{W})}+\frac{2}{g_s}\left[\mathcal{T}_{ik}\text{Re}\left[W^{kl}\overline{\mathcal{T}_{lj}^{(W)}}\right]+(i\leftrightarrow j)\right]+\frac{2}{g_s}\text{Re}\left[\mathcal{T}_{ij}^{(\Gamma)}\right]\\
	&+\frac{2}{g_s^2}\left[\mathcal{T}_{ik}W^{kl}_{(\overline{\Omega})}\mathcal{T}_{lj}+\mathcal{T}_{ik}W^{kl}\mathcal{T}_{lm}\overline{W^{mn}}\mathcal{T}_{nj}\right]\, ,\\
	{Y^i}_j\coloneqq \del_{t^j}Y^i=&-2\text{Im}\left(\frac{1}{g_s}W^{ik}_{(\overline{\Omega})}\mathcal{T}_{kj}+W^{ik}\overline{\mathcal{T}_{kj}^{(W)}}+\frac{1}{g_s}W^{ik}\mathcal{T}_{kl}\overline{W^{lm}}\mathcal{T}_{mj}\right)\, ,\\
	Y^{ij}\coloneqq \del_{\phi_i}Y^j=&-2\text{Re}\left[W^{ij}_{(\overline{\Omega})}-W^{ik}\mathcal{T}_{kl}\overline{W^{lj}}\right]\, .
\end{align}
Moreover,
\begin{align}
	\epsilon_i\coloneqq &\del_{t^i}\epsilon=-\epsilon^2\left(y_i+z_i\right)\, ,\\
	\epsilon_{ij}\coloneqq &\del_{t^j}\epsilon_i=2\epsilon^3 (y_i +z_i)(y_j+z_j)-\epsilon^2 \left(z_{ij}+3 y_{ij}+\mathcal{T}_{ij}\right)\, ,\\
	\eta_i\coloneqq &\del_{t^i}\eta=-\frac{y_i}{2\Phi}\eta+\frac{\mathcal{F}_{ij}t^j-\mathcal{F}_{ijk}t^jt^k}{4\Phi}\, ,\\
	\eta_{ij}\coloneqq &-\frac{y_{ij}+\mathcal{T}_{ij}}{2\Phi}\eta-\frac{y_i\eta_j+\eta_iy_j}{2\Phi}+\frac{\mathcal{F}_{ij}-\mathcal{F}_{ijk}t^k-\mathcal{F}_{ijkl}t^kt^l}{4\Phi}\, .
\end{align}
Using these expressions, one finds for the gradient of the F-term potential:
\begin{align}
	V_i\coloneqq \del_{t^i}V=&-\frac{y_i}{2\Phi}V-\frac{4g_s^2}{\Phi}\text{Re}\left[Y_i-\epsilon_i \left(4X\overline{X}-4g_s X\overline{W}+\eta g_s^2 |W|^2\right)\right.\nonumber\\
	&\left.\phantom{\frac{1}{1}}+\epsilon \left(4X_i(g_s \overline{W}-2\overline{X})+2\Omega_i(2\overline{X}-g_s\eta \overline{W})-g_s^2\eta_i|W|^2 \right) \right]\, ,\\
	V^i\coloneqq \del_{\phi_i}V=&-\frac{4g_s^2}{\Phi}\text{Re}\left[Y^i+\epsilon(4X^i (g_s\overline{W}-2\overline{X})+2ig_sW^i(2\overline{X}-g_s\eta \overline{W}))\right]\, .
\end{align}
For the Hessian $V^{ij}\coloneqq \del_{\phi_j}V^j$, ${V^i}_j\coloneqq \del_{t^j}V^i$ and $V_{ij}\coloneqq \del_{t^j}V_i$ we find
\begin{align}
	V^{ij}=&-\frac{4g_s^2}{\Phi}\text{Re}\left[Y^{ij}+\epsilon \left(4X^{ij}(g_s\overline{W}-2\overline{X})-8X^i\overline{X^j}-4g_s i (X^i\overline{W^j}+\overline{W^i}X^j)\right.\right.\nonumber\\
	&\quad\quad\quad\quad\quad\quad\quad\quad\quad\left.\phantom{\frac{1}{1}}\left.-2g_s W^{ij}(2\overline{X}-g_s\eta \overline{W})-2g_s^2\eta W^i \overline{W^j}\right)\right]\, ,\\
	{V_i}^j=&-\frac{y_i}{2\Phi}V^j-\frac{4g_s^2}{\Phi}\text{Re}\left[\phantom{\frac{1}{1}}{Y_i}^j-\epsilon_i\left(4X^j(2\overline{X}-g_s\overline{W})-2ig_sW^j(2 \overline{X}-\eta g_s\overline{W})\right)\right.\nonumber\\
	&\quad\quad\quad\quad\quad\quad\quad\quad +\epsilon\left(4{X_i}^j(g_s\overline{W}-2\overline{X})-8X_i\overline{X^j}-4i g_s X_i \overline{W^j}\right.\nonumber\\
	&\quad\quad\quad\quad\quad\quad\quad\quad\quad\quad  +2i \mathcal{T}_{ik}W^{kj}(2\overline{X}-g_s\eta \overline{W})+2\Omega_i (2\overline{X^j}+ig_s\eta \overline{W^j})\nonumber\\
	&\quad\quad\quad\quad\quad\quad\quad\quad\quad\quad\left.\left.-2i g_s^2\eta_i W^j \overline{W}\right)\phantom{\frac{1}{1}}\right]\, ,\\
	V_{ij}=&-\frac{y_{ij}+\mathcal{T}_{ij}}{2\Phi}V-\frac{y_iV_j+V_iy_j}{2\Phi}\nonumber\\
	&-\frac{4g_s^2}{\Phi}\text{Re}\left[\phantom{\frac{1}{1}}Y_{ij}-\epsilon_{ij}\left(4X\overline{X}-4g_sX\overline{W}+\eta g_s^2|W|^2\right)\right.\nonumber\\
	&\quad\quad\quad\quad+\epsilon_i\left(4X_j(g_s\overline{W}-2\overline{X})+2\Omega_j(2\overline{X}-g_s\eta \overline{W})-g_s^2\eta_j|W|^2\right)+\left(i\leftrightarrow j\right)\nonumber\\
	&\quad\quad\quad\quad+\epsilon\left(4X_{ij}(g_s\overline{W}-2\overline{X})-8X_i\overline{X_j}+4(X_i\overline{\Omega_j}+X_j\overline{\Omega_i})\right.\nonumber\\
	&\quad\quad\quad\quad\quad\quad\quad+2(\mathcal{T}_{ij}^{(W)}+\frac{1}{g_s}\mathcal{T}_{ik}W^{kl}\mathcal{T}_{lj})(2\overline{X}-g_s\eta \overline{W})\nonumber\\
	&\quad\quad\quad\quad\quad\quad\quad\left.\phantom{\frac{1}{1}} \left.-2g_s(\Omega_i\eta_j+\Omega_j\eta_i)\overline{W}-g_s^2\eta_{ij}|W|^2\right)\right] \, .
\end{align}
Importantly, we have written the above formulas in a way that requires numerical evaulation of tensors of rank at most equal to two. For this reason, we are able to efficiently evaluate our expressions numerically, even at the largest values of of $h^{1,1}=\mathcal{O}(100)$.

\subsection{Complex structure moduli sector}\label{sec:formulas_CS}
In the complex structure moduli  and axiodilaton sector, we have the following superpotential and K\"ahler potential:
\begin{align}
	\sqrt{\dfrac{\pi}{2}}W=&\frac{1}{2}\mathbb{N}_{ab}z^az^b-\tau \mathbb{K}_az^a-\frac{1}{(2\pi)^2}\sum_{\vec{q}}\hat{n}_{\vec{q}}^0 (\vec{M}\cdot \vec{q}) e^{2\pi i \vec{q}\cdot \vec{z}}\, ,\\
	K=&-\log\bigl(-i(\tau-\bar{\tau})\bigr)-\log\bigl(\widetilde{\mathcal{V}}\bigr)\, ,\\ \widetilde{\mathcal{V}}=&\frac{1}{6}\kappa_{abc}\,\text{Im}(z^a)\text{Im}(z^b)\text{Im}(z^c)+\frac{\zeta(3)\chi(X)}{4(2\pi)^3}\nonumber\\
		&+\frac{1}{2(2\pi)^3}\sum_{\vec{q}} n^0_{\vec{q}}\,\text{Re}\left[\text{Li}_3(e^{2\pi i \vec{q}\cdot \vec{z}})+2\pi \text{Im}(\vec{q}\cdot \vec{z})\text{Li}_2(e^{2\pi i \vec{q}\cdot \vec{z}})\right]\, .
\end{align}
It follows that
\begin{align}
	\mathcal{T}_a\coloneqq &2i\del_{a}\widetilde{\mathcal{V}}=\tau_a-\frac{1}{2(2\pi)^2}\sum_{\vec{q}} n_{\vec{q}}^0\, q_a \left[2\pi  \text{Im}(\vec{q}\cdot \vec{z})\text{Li}_1\left(e^{2\pi i \vec{q}\cdot \vec{z}}\right)   +     i\text{Im}(\text{Li}_2\left(e^{2\pi i \vec{q}\cdot \vec{z}}\right))   \right]\, ,\\
	\mathcal{K}_{a\bar{b}}\coloneqq &2i\del_{a}\overline{\mathcal{T}_b}=\kappa_{ab}-\frac{1}{2\pi}\sum_{\vec{q}}n_{\vec{q}}^0 \,q_a q_b \text{Re}(\text{Li}_1 \left(e^{2\pi i \vec{q}\cdot \vec{z}}\right))\, ,\\
	\widehat{\mathcal{K}}_{ab}\coloneqq &2i \del_a \mathcal{T}_b=\kappa_{ab}+\sum_{\vec{q}} n_{\vec{q}}^0\,q_aq_b \text{Im}(\vec{q}\cdot \vec{z})\text{Li}_0\left(e^{2\pi i \vec{q}\cdot \vec{z}}\right)\, ,\\
	\mathcal{K}_{ab\bar{c}}\coloneqq &2i \del_{a}\mathcal{K}_{b\bar{c}}=\kappa_{abc}+\sum_{\vec{q}} n^0_{\vec{q}}\,q_aq_bq_c\text{Li}_0\left(e^{2\pi i \vec{q}\cdot \vec{z}}\right)\, ,\\
	\widehat{\mathcal{K}}_{abc}\coloneqq &2i\del_{a} \widehat{\mathcal{K}}_{bc}=\mathcal{K}_{ab\bar{c}}-4\pi \sum_{\vec{q}}n_{\vec{q}}^0\, q_a q_b q_c \text{Im}(\vec{q}\cdot \vec{z}\,)\text{Li}_{-1}\left(e^{2\pi i \vec{q}\cdot \vec{z}}\right)\, ,\\
	\Kcal_{abc\bar{d}}\coloneqq &2i \del_a \Kcal_{bc\bar{d}}=-4\pi\sum_{\vec{q}} n^0_{\vec{q}}\, q_aq_bq_cq_d \text{Li}_{-1}\left(e^{2\pi i \vec{q}\cdot \vec{z}}\right)\, ,
\end{align}
where $\tau_a\coloneqq \frac{1}{2}\kappa_{abc}\text{Im}(z^b) \text{Im}(z^c)$, and $\kappa_{ab}\coloneqq \kappa_{abc}\text{Im}(z^c)$. It follows that $\frac{1}{2}\mathcal{K}_{ab}\text{Im}(z^b)=\text{Re}(\mathcal{T}_a)$.

The K\"ahler connection and metric read
\begin{align}
	K_a&=i\frac{\mathcal{T}_a}{2\widetilde{\mathcal{V}}}\, ,\\
	g_{a\bar{b}}&=\frac{1}{4\widetilde{\mathcal{V}}}\left(\frac{\mathcal{T}_a \overline{\mathcal{T}_b}}{\widetilde{\mathcal{V}}}-\mathcal{K}_{a\bar{b}}\right)\, ,
\end{align}
and we further define
\begin{equation}
	g^{hol}_{ab}\coloneqq \frac{1}{4\widetilde{\mathcal{V}}}\left(\frac{\mathcal{T}_a \mathcal{T}_b}{\widetilde{\mathcal{V}}}-\widehat{\mathcal{K}}_{a\bar{b}}\right)\, .
\end{equation}
Denoting $W_{a_1\ldots a_k}\coloneqq \del_{z^{a_1}}\ldots \del_{z^{a_k}}W$, the covariant derivatives of the superpotential are
\begin{align}
	D_{a}W=&W_a+i \frac{\Tcal_a}{2\Vtcal}W\, ,\\
	\mathcal{D}_a D_b W=&W_{ab}+\frac{i}{2\Vtcal}\left(\Tcal_a D_b W+\Tcal_b D_a W\right)-\Gamma_{ab}^c D_c W
	+\frac{\hat{\Kcal}_{ab}}{4\Vtcal}W\, ,\\
	\mathcal{D}_a\mathcal{D}_bD_cW=&W_{abc}+\frac{i}{2\Vtcal}\left(\Tcal_a \mathcal{D}_bD_c W+\Tcal_b \mathcal{D}_aD_c W+\Tcal_c \mathcal{D}_aD_b W\right)\nonumber\\
	&-\Gamma^d_{ab}\mathcal{D}_dD_cW-\Gamma^d_{ac}\mathcal{D}_dD_bW-\Gamma^d_{bc}\mathcal{D}_dD_aW\nonumber\\
	&+\frac{1}{4\Vtcal}\left(\hat{\Kcal}_{ab}D_cW+\hat{\Kcal}_{ac}D_bW+\hat{\Kcal}_{bc}D_aW\right)\nonumber\\
	&+\frac{i}{2\Vtcal}\left(\Tcal_a \Gamma^d_{bc}+\Tcal_b \Gamma^d_{ac}+\Tcal_c \Gamma^d_{ab}\right)D_dW\, ,\nonumber\\
	&-(\del_a \Gamma^d_{bc}+\Gamma_{bc}^m \Gamma_{am}^d)D_dW-\frac{i}{8\Vtcal}\hat{\Kcal}_{abc}W\, ,
\end{align}
with
\begin{align}
	\Gamma_{ab}^c&= g^{c\bar{d}}\del_a g_{b\bar{d}}=-\frac{i}{2}g^{c\bar{d}}\gamma_{ab\bar{d}}\, ,\\
	\gamma_{ab\bar{c}}&\coloneqq 2i\del_{a} g_{b\bar{c}}=-\frac{\Tcal_a\Tcal_b\overline{\Tcal_c}}{2\Vtcal^3}+\frac{\Tcal_a \Kcal_{b\bar{c}}+\Tcal_b \Kcal_{a\bar{c}}+\widehat{\mathcal{K}}_{ab}\overline{\Tcal_c}}{4\Vtcal^2}-\frac{\Kcal_{ab\bar{c}}}{4\Vtcal}\, ,
\end{align}
and
\begin{align}
	\del_a \Gamma^d_{bc}+\Gamma_{bc}^m\Gamma^d_{am}&=-\frac{1}{4}g^{d\bar{m}}\gamma_{abc\bar{m}}\, ,\\
	\gamma_{abc\bar{d}}\coloneqq 2i \del_a \gamma_{bc\bar{d}}&=-\frac{\Kcal_{abc\bar{d}}}{4\Vcal}+\frac{\Kcal_{ab\bar{d}}\Tcal_c+\Kcal_{ac\bar{d}}\Tcal_b+\Kcal_{bc\bar{d}}\Tcal_a}{4\Vcal^2}+\frac{\widehat{\mathcal{K}}_{abc}\overline{\mathcal{T}_{d}}}{4\Vcal^2}\nonumber\\
	&+\frac{\widehat{\mathcal{K}}_{ab}\Kcal_{c\bar{d}}+\widehat{\mathcal{K}}_{ac}\Kcal_{b\bar{d}}+\widehat{\mathcal{K}}_{bc}\Kcal_{a\bar{d}}}{4\Vcal^2}-\frac{(\widehat{\mathcal{K}}_{ab}\Tcal_c+\widehat{\mathcal{K}}_{ac}\Tcal_b+\widehat{\mathcal{K}}_{bc}\Tcal_a)\overline{\Tcal_d}}{2\Vcal^3}\nonumber\\
	&-\frac{\Kcal_{a\bar{d}}\Tcal_b\Tcal_c+\Kcal_{b\bar{d}}\Tcal_a\Tcal_c+\Kcal_{c\bar{d}}\Tcal_a\Tcal_b}{2\Vcal^3}+\frac{3\Tcal_a\Tcal_b\Tcal_c\overline{\Tcal_d}}{2\Vcal^4}\, .
\end{align}
The potential gradient is
\begin{equation}
	V_a\coloneqq \del_{z^a}V=e^{K}\left(\mathcal{D}_a D_b W g^{b\bar{c}}\overline{D_cW}-2D_aW \overline{W}\right)\, .
\end{equation}
Moreover, one finds
\begin{align}
	V_{ab}\coloneqq \del_a\del_b V=&e^{K}\left(\mathcal{D}_a \mathcal{D}_b D_c W g^{c\bar{d}}\overline{D_dW}-\mathcal{D}_a D_b W \overline{W}\right)+\Gamma^c_{ab}V_c\, ,\\
	V_{a\bar{b}}\coloneqq \del_{a}\del_{\bar{b}} V=&e^K\left(g_{a\bar{b}} (g^{c\bar{d}}D_c W \overline{D_dW}-2|W|^2)+ {{\mathcal{R}_{a\bar{b}}}^c}_d D_cW  g^{d\bar{m}}\overline{D_mW} \right.\nonumber\\
	&\left.\quad\quad+ g^{c\bar{d}}\mathcal{D}_a D_cW\overline{\mathcal{D}_b D_d W}-D_aW \overline{D_bW}\right)\, ,
\end{align}
with ${{\mathcal{R}_{a\bar{b}}}^c}_d=-\del_{\bar{b}}\Gamma_{ad}^c$.
Finally, we have   
\begin{align}
	\del_{\bar{b}} \Gamma_{ad}^c&=\frac{1}{4}\left(g^{c\bar{m}}\hat{\gamma}_{ad\bar{b}\bar{m}}-g^{c\bar{n}}g^{p\bar{m}}\overline{\gamma_{nb\bar{p}}}\gamma_{ad\bar{m}}\right)\, ,
\end{align}
in terms of
\begin{align}
	\hat{\gamma}_{ab\bar{c}\bar{d}}\coloneqq 2i\del_{a} \gamma_{b\bar{c}\bar{d}}=&
	\frac{\Kcal_{a\bar{c}\bar{d}}\Tcal_b+\Kcal_{b\bar{c}\bar{d}}\Tcal_a+\Kcal_{ab\bar{c}}\overline{\Tcal_d}+\Kcal_{ab\bar{d}}\overline{\Tcal_c}}{4\Vcal^2}\nonumber\\
	&+\frac{\mathcal{K}_{a\bar{d}}\Kcal_{b\bar{c}}+\mathcal{K}_{a\bar{c}}\Kcal_{b\bar{d}}+\overline{\widehat{\mathcal{K}}_{cd}}\widehat{\mathcal{K}}_{ab}}{4\Vcal^2}
	-\frac{\hat{\Kcal}_{ab}\overline{\Tcal_c}\overline{\Tcal_d}+\overline{\hat{\Kcal}_{cd}}\Tcal_a\Tcal_b}{2\Vcal^3}\nonumber\\
	&
	-\frac{\Kcal_{a\bar{c}}\Tcal_b\overline{\Tcal_d}+\Kcal_{a\bar{d}}\Tcal_b\overline{\Tcal_c}+
	\Kcal_{b\bar{c}}\Tcal_a\overline{\Tcal_d}+\Kcal_{b\bar{d}}\Tcal_a\overline{\Tcal_c}}{2\Vcal^3}+\frac{3\Tcal_a\Tcal_b\overline{\Tcal_c}\overline{\Tcal_d}}{2\Vcal^4}   \, .
\end{align}
Next, we consider the contribution from the axiodilaton:
\begin{align}
	K_{\tau}&=-\frac{1}{\tau-\bar{\tau}}\, ,\quad 
	g_{\tau\bar{\tau}}=-\frac{1}{(\tau-\bar{\tau})^2}\, ,\\ \Gamma^\tau_{\tau\tau}&=-\frac{2}{\tau-\bar{\tau}}\, ,\quad \del_{\tau}\Gamma^\tau_{\tau\tau}+(\Gamma^\tau_{\tau\tau})^2=\frac{6}{(\tau-\bar{\tau})^2}\, ,\quad {{\mathcal{R}_{\tau\bar{\tau}}}^\tau}_{\tau}=\frac{2}{(\tau-\bar{\tau})^2}\, .
\end{align}
Using this one finds
\begin{align}
	\sqrt{\dfrac{\pi}{2}} D_\tau W&=-\mathbb{K}_a z^a-\sqrt{\dfrac{\pi}{2}}\frac{W}{\tau-\bar{\tau}}\, ,\\
	D_{\tau}^2 W &\equiv \del_\tau^2W =0\, . 
\end{align}
As $[\mathcal{D}_a,\mathcal{D}_{\tau}]=0$, we conclude that
\begin{align}
	\sqrt{\dfrac{\pi}{2}}\mathcal{D}_a D_\tau W =&-\mathbb{K}_a-i\frac{\Tcal_a}{2\Vtcal}(\vec{\mathbb{K}}\cdot \vec{z})-\sqrt{\dfrac{\pi}{2}}\frac{D_a W}{\tau-\bar{\tau}}\, ,\\
	\sqrt{\dfrac{\pi}{2}}\mathcal{D}_a \mathcal{D}_b D_{\tau}W =& \frac{i}{2\Vtcal}\left(\Tcal_a \mathbb{K}_b+\Tcal_b \mathbb{K}_a\right)+\frac{\vec{\mathbb{K}}\cdot \vec{z}}{2\Vtcal^2}\Tcal_a\Tcal_b +\Gamma_{ab}^c\left(\mathbb{K}_c+\frac{i}{2\Vtcal}(\vec{\mathbb{K}}\cdot \vec{z})\Tcal_c\right)\nonumber\\
	&-\frac{\vec{\mathbb{K}}\cdot \vec{z}}{4\Vtcal}\hat{K}_{ab}-\sqrt{\dfrac{\pi}{2}}\frac{\mathcal{D}_aD_b W}{\tau-\bar{\tau}}\ , ,
\end{align}  
are the only remaining independent and non-vanishing superpotential derivatives. 

The resulting $\tau$-derivatives of the F-term potential are
\begin{align}
	V_{\tau}&\coloneqq \del_{\tau} V=e^K(\mathcal{D}_a D_\tau W g^{a\bar{b}}\overline{D_bW}-2\,D_\tau W \overline{W})\, ,\\
	V_{\tau\tau}&\coloneqq \del_\tau^2 V=e^K(-\del_\tau^2 W \overline{W})+\frac{i}{\text{Im}(\tau)}V_\tau=\frac{i}{\text{Im}(\tau)}V_\tau\, ,\\
	V_{\tau a}&\coloneqq \del_{\tau}V_a=e^K(\mathcal{D}_a \mathcal{D}_b D_\tau W g^{b\bar{c}}\overline{D_cW}-\mathcal{D}_a D_\tau W \,\overline{W})\, ,\\
	V_{\tau\overline{\tau}}&\coloneqq \del_{\tau}\del_{\bar{\tau}}V=e^K\left(\frac{g^{a\bar{b}}D_aW\overline{D_bW}-2|W|^2}{4\text{Im}(\tau)^2}+g^{a\bar{b}}\mathcal{D}_aD_{\tau}W\overline{\mathcal{D}_bD_{\tau}W}-2|D_\tau W|^2\right)\, ,\\
	V_{a\bar{\tau}}&\coloneqq \del_a \del_{\bar{\tau}}V=e^K(g^{b\bar{c}}\mathcal{D}_aD_bW \overline{\mathcal{D}_c D_{\tau}W}-D_a W \overline{D_\tau W})\, .
\end{align}
In practice, we evaluate the above formulas near a conifold limit, by integrating out the (heavy) conifold modulus explicitly, supersymmetrically, using the formula \eqref{eq:conifold_vev}, and considering only the light fields corresponding to fluctuations $\delta z^a$ such that $\delta z^a \mathbf{q}^{\mathrm{cf}}_a\equiv 0$. 
The enormous hierarchy between the masses of the light degrees of freedom and the conifold modulus makes a numerical treatment involving all fields impractical. Happily, corrections to our treatment are of order $|z_{\mathrm{cf}}|$, which is of order $10^{-6}-10^{-8}$ in our solutions.

\subsection{Uplift potential}\label{sec:formulas_uplift_potential}
We are interested in computing derivatives of the anti-D3-brane potential,
in an effective theory in which a conifold modulus as been integrated out supersymmetrically. Thus again, we use the formula \eqref{eq:conifold_vev}, i.e., 
\begin{equation}
	 |z_{\mathrm{cf}}| =\frac{1}{2\pi}\exp\left(-\frac{2\pi }{n^0_{\mathcal{C}}M^2}\delta \right)
\end{equation}
with
\begin{equation}
	\delta\coloneqq \left.\left(2Q_{\text{flux}}\text{Im}(\tau)+M^aM^b\mathcal{K}_{a\bar{b}}\right)\right|_{z_{\mathrm{cf}}=0}+\mathcal{O}(z_{\mathrm{cf}})\, ,.
\end{equation}
We record,
\begin{align}
	\delta_\tau\coloneqq \del_{\tau}\delta=-iQ_{\text{flux}}\, ,\quad \delta_{a}&\coloneqq \del_{z^a}\delta=-\frac{i}{2}\mathcal{K}_{ab\bar{c}}M^bM^c\, ,\\
	\delta_{ab}&\coloneqq \del_{z^a}\del_{z^b}\delta=-\frac{1}{4}\mathcal{K}_{abc\bar{d}}M^c M^d\, ,
\end{align}
and $\delta_{a\bar{b}}\equiv 0$.

Having integrated out the heavy conifold modulus, its vev still varies adiabatically as a function of the light moduli, specifically
\begin{align}
	\del_{\tau} |z_{\mathrm{cf}}|&=-\frac{2\pi}{n^0_{\mathcal{C}}M^2}\delta_\tau |z_{\mathrm{cf}}|\, ,\quad 	\del_{a} |z_{\mathrm{cf}}|=-\frac{2\pi}{n^0_{\mathcal{C}}M^2}\delta_a |z_{\mathrm{cf}}|\, ,\\
	\del_a \del_b |z_{\mathrm{cf}}|&=\left(\left(\frac{2\pi}{n^0_{\mathcal{C}}M^2}\right)^2 \delta_a \delta_b-\frac{2\pi}{n_\mathcal{C}^0M^2}\delta_{ab}\right)|z_{\mathrm{cf}}|\, ,\quad \del_a \overline{\del_b}|z_{\mathrm{cf}}|=\left(\frac{2\pi}{n_\mathcal{C}^0M^2}\right)^2 \delta_a \overline{\delta_b}\, ,
\end{align}
and we define the low energy effective theory via imposing these relations onto the fields entering the K\"ahler potential and superpotential. We will work in a regime where $z_{\mathrm{cf}}\ll W_0$, so we may approximate our effective theory by sending $z_{\mathrm{cf}}\rightarrow 0$.

We now would like to address the backreaction of an anti-D3-brane uplift onto the light moduli. For this we assume an uplift term
\begin{equation}
    V_{\overline{D3}}=  \eta\,\frac{\left(z_{\mathrm{cf}}/\sqrt{\tilde{\calV}}\right)^{p}}{g_sM^2  \mathcal{V}_E^{4/3}}
    \, ,
\end{equation}
and we will set $p=\frac{4}{3}$, thus matching \eqref{eq:anti-D3-potential0}, where it is understood that $z_{\mathrm{cf}}$ is a function of the light moduli, and not itself an independent dynamical degree of freedom.

We find for the gradient
\begin{align}
	\del_{\tau}V_{\overline{D3}}&=\left(-\frac{i}{2\text{Im}(\tau)}-p\frac{2\pi}{n_{\mathcal{C}}^0 M^2}\delta_\tau\right)V_{\overline{D3}}=:k_{\tau}V_{\overline{D3}}\, ,\\
	\del_a V_{\overline{D3}} &= p\left(\frac{i\mathcal{T}_a}{4\tilde{\mathcal{V}}}-\frac{2\pi}{n_{\mathcal{C}}^0 M^2}\delta_{a}\right)V_{\overline{D3}}=:k_a V_{\overline{D3}}\, ,
\end{align}
and for the Hessian
\begin{align}
	\del_{\tau}^2V_{\overline{D3}}&=\left[\frac{1}{4\text{Im}(\tau)^2}+k_\tau^2\right]V_{\overline{D3}}\\
	\del_{\tau}\del_{\bar{\tau}}V_{\overline{D3}}&=\left[-\frac{1}{4\text{Im}(\tau)^2}+\left|k_\tau\right|^2\right]V_{\overline{D3}}\, ,\\
	\del_\tau \del_a V_{\overline{D3}}&=k_\tau k_a V_{\overline{D3}}\, ,\\
	\del_\tau \del_{\bar{a}} V_{\overline{D3}}&=k_\tau \overline{k_a} V_{\overline{D3}}\, ,\\
	\del_a \del_b V_{\overline{D3}} &=\left[ p\left(-\frac{1}{2}g^{hol}_{ab}-\frac{2\pi}{n^0_{\mathcal{C}}M^2}\delta_{ab}\right)+k_a k_b\right]V_{\overline{D3}}\, ,\\
	\del_a \del_{\overline{b}} V_{\overline{D3}} &=\left[ \frac{p}{2}g_{a\bar{b}}+ k_a \overline{k_b}\right]V_{\overline{D3}}	\, .
\end{align}
Evaluating the derivatives with respect to the K\"ahler moduli only involves derivatives of $\mathcal{V}_E$, which is straightforward.

\bibliographystyle{utphys}
\bibliography{refs}

\end{appendix}
\end{document}